\documentstyle[12pt,preprint,tighten,floats,aps,psfig]{revtex}
\def\simlt{\stackrel{<}{{}_\sim}}
\def\simgt{\stackrel{>}{{}_\sim}}
\tighten
\begin{document}
\draft
\preprint{UPR-0737-T, IEM-FT-150/97}
\title{Electroweak Breaking and the $\mu$ problem in Supergravity Models with 
an Additional $U(1)$ }
\author{M. Cveti\v c, D.A. Demir, J.R. Espinosa, L. Everett,  P. Langacker}
\address{Department of Physics, University of Pennsylvania, Philadelphia,
PA 19104-6396}
\date{\today}
\maketitle
\begin{abstract}
We consider electroweak symmetry breaking in supersymmetric
models with an extra non-anomalous $U(1)'$ gauge symmetry and
an extra standard-model singlet scalar $S$.
For appropriate charges the $U(1)'$ forbids an elementary
$\mu$ term, but an effective $\mu$ is generated by the VEV of $S$,
leading to a natural solution to the $\mu$ problem.
There are a variety of scenarios leading to acceptably small $Z-Z'$ 
mixing and other phenomenological consequences, all of which involve
some but not excessive fine tuning. One class, driven by
a large trilinear soft supersymmetry breaking term, implies
small mixing, a light $Z'$ (e.g., 200 GeV), and an
electroweak phase transition that may be first order at tree level.
In another class, with $m_S^2 < 0$ (radiative breaking), 
the typical scale of dimensional parameters, including
$M_{Z'}$ and the effective $\mu$, is
$O(1\ {\rm TeV})$, but the electroweak scale is smaller due
to cancellations. We relate the soft supersymmetry breaking parameters at the
electroweak scale to those at the string scale, choosing Yukawa couplings as
determined within a class of string models. We find that one does not obtain
either scenario for universal soft supersymmetry breaking mass parameters at the
string scale and no exotic multiplets contributing to the renormalization group
equations.
However, either scenario is possible when the assumption
of universal soft breaking is relaxed. Radiative breaking
can also be generated by exotics, which are expected in most 
string models.
\end{abstract}
\section{Introduction}
The simplest gauge extension of the standard model
involves one or more additional $U(1)$
symmetries and their associated extra $Z$ bosons.
Such $U(1)$'s often emerge in the breaking of grand unified
theories (GUT) or in string compactifications, for example.

There has been much phenomenological work on the implications of
such heavy $Z$'s for precision electroweak observables and for
future hadron and $e^+ e^-$ colliders.
Present \cite{PRESENTZ} and future \cite{FUTUREZ} limits as well as search 
and diagnostic capabilities depend on the $Z'$ mass, mixing with the $Z$, gauge
couplings, and chiral charges of the ordinary quarks and leptons, and are thus
very model dependent.
For many typical (especially GUT-motivated) models
the limits on the $Z-Z'$ mixing are around a few $\times 10^{-3}$.
The lower limits on the $Z'$ mass are typically
around 500 GeV, usually dominated by direct searches
at the Tevatron ($p\bar{p} \rightarrow Z' \rightarrow \ell^+ \ell^-$)  
\cite{CDFLIM}, but
with constraints from precision electroweak tests often competitive.
Recently, a number of authors \cite{LEPTOPHOBIC} have postulated that a possible 
excess of $Z \rightarrow b \bar{b}$ events at LEP could be accounted for
by the mixing between the $Z$ and a leptophobic (hadrophilic)
$Z'$ which mainly couples to quarks, but the most recent
LEP data, especially from ALEPH, have considerably weakened the
case that there is an excess \cite{LEPRB}. In the future it should be possible
to discover a heavy $Z'$ at the LHC for masses up to around 10 TeV.
Diagnostics of its couplings at the LHC or NLC
(which have complementary capabilities) should be possible up to
a few TeV \cite{FUTUREZ}.

In addition  to being a useful signature of the underlying theory,
an additional $U(1)'$ would have important theoretical implications.
For example, an extra $U(1)'$ breaking at the electroweak scale 
in a supersymmetric extension of the standard model could
solve the $\mu$ problem \cite{MUPROB,SY,CL,shrock}, by forbidding an elementary
$\mu$
term but inducing an effective $\mu$ at the electroweak scale
by the $U(1)'$ breaking. This possibility is one
of the major motivations of this paper. There are also implications
for baryogenesis. One popular scenario is that a lepton asymmetry 
\cite{LEPTONASY} (or an asymmetry in some other quantum number) was 
created by the out of equilibrium decay of a superheavy particle 
(e.g., a heavy Majorana neutrino) long before
the electroweak transition, and then converted to a baryon
asymmetry by sphaleron effects. Such a mechanism would 
not be consistent with an additional $U(1)'$ at the TeV or electroweak scale
unless  the Majorana neutrino were neutral under the $U(1)'$.
On the other hand, an extra $U(1)'$ might be useful for electroweak
baryogenesis, with cosmic strings providing the needed
``out of equilibrium'' ingredient \cite{COSTRINGS}.

Much of the phenomenological work on extra $Z'$s has been of the lamppost
variety, i.e., there was no strong motivation to think that
an extra $Z'$ would actually be light enough to observe.
Certainly, in ordinary GUTs there is no robust prediction for the
mass scale of the $U(1)'$ breaking. In supersymmetric models
there are constraints on the breaking scale, which are
usually of order a TeV, because the
$U(1)'$ D term may induce masses of order of the breaking 
for all scalars which carry the $U(1)'$ charge \cite{LYKKEN}. However, 
that is more a phenomenological constraint than a theoretical prediction,
and it can be evaded if the breaking occurs along a D--flat direction.

However, it was recently argued \cite{CL} that for a large class of string
models with extra $U(1)$'s, the breaking should be at the electroweak
scale and certainly not larger than a TeV.  The string models considered 
in \cite{CL} are based on $N=1$ supersymmetric  string models with  
the  standard model (SM) gauge group  $SU(2)_L\times U(1)_Y\times SU(3)_C$,  
three families,  and at least two standard model (SM)
doublets, {i.e.},  models with at least 
the particle content of the minimal supersymmetric standard model (MSSM). 
A number of such models are based on fermionic 
($Z_2\times Z_2$) orbifold  constructions \cite{ABK,NAHE,FARA,CHL} at a particular
point in moduli space.

Such models suffer from a number of  phenomenological problems
(see Section~II in \cite{CL} for a detailed discussion), and many such models
are already excluded experimentally.  Nevertheless, there
is a strong motivation to search for such $Z'$ bosons and also
for the exotic (vector under $SU(2)$) supermultiplets
with which they are usually associated.  In addition, they provide a useful testing
ground to address  the issues of $U(1)'$ breaking within  a large class of
string models.

The relevant models are those in which: (a) there is a non-anomalous
$U(1)'$ which does not acquire a large mass from string or shadow
sector dynamics, so that its mass must come from symmetry breaking
in the observable sector. (b) The soft supersymmetry breaking is such
that all scalar mass-squared terms are positive and of the same
order of magnitude at the string scale, which is the case for
most gravity mediated hidden sector models (but not 
necessarily for the gauge mediated supersymmetry breaking 
models that have been of recent interest).

Under these assumptions, the $U(1)'$ breaking may be radiative \cite{CL}. It can
take place if there are Yukawa couplings of order 1 of a  scalar which is a
standard model singlet (but which carries a $U(1')$ charge) to exotic particles. 
This is expected in many
string models, for which all non-zero Yukawas are typically
of the same magnitude, i.e., they are the same as the gauge coupling at the
string scale up to a coefficient of order unity.
These can drive the scalar mass-squared to a
negative value at low energies, which is typically of the same order
as the Higgs mass-squared, so that the electroweak and $U(1)'$ breaking scales
are comparable, both being controlled by the same soft
supersymmetry breaking scale\footnote{In some cases the
breaking will be at an intermediate scale if there is a D-flat
direction involving two scalars both of which have large
Yukawas.}.

In \cite{CL}, a model was considered in which {\it only one} (e.g., $H_2$) 
of the two SM Higgs doublets has non-zero couplings in the superpotential 
and contributes to the electroweak breaking; i.e., this model roughly 
corresponds to the large $\tan \beta$ scenario in the MSSM. 
The radiative symmetry breaking can take place
with  $M_{Z'}\sim 1\ TeV$,  and  sufficiently small  $Z-Z'$ mixing angle  (not yet
excluded by the direct and indirect heavy $Z'$ constraints), provided the
$U(1)'$ charge  assignments for the  
the  $H_2$ and the SM singlet $S $ (responsible for the symmetry breaking of
$U(1)'$) have the same sign. 

In this paper we  consider the more general case with the two SM doublets 
$H_{1,2}$  now coupled to the SM singlet $\hat{S}$  in the superpotential 
with the term $h_s{\hat S}{\hat H}_1\cdot{\hat H}_2$.   In this case, the 
$U(1)'$ charges of $\hat{H}_{1,2}$ and $\hat{S}$  must  sum to zero. This 
term provides an  effective  $\mu$ term   $h_s\langle S\rangle$, once $S$ 
acquires a non-zero vacuum expectation value. 

Due to this additional term  in the superpotential, a rich spectrum of possible
symmetry breaking scenarios emerges. In particular, we concentrate on a set
of phenomenologically viable scenarios with small $Z-Z'$ mixing ($\le {\cal
O}(10^{-3})$) and $M_{Z'}$ in the range $\le {\cal O}(1$ TeV$)$. We also
insist on no dangerous color breaking minimum, e.g., no negative squark  
mass-squared parameters or large trilinear soft supersymmetry breaking terms
that involve squarks. We find various ranges of parameters that allow for such
symmetry breaking scenarios. However,  all these cases involve some degree of 
fine-tuning of  parameters, either at the electroweak scale or  at the string 
scale\footnote{However, the tuning involved is no worse than that in the 
MSSM in the case for which the electroweak scale $M_Z$ is small compared to $\mu$,
e.g., for $\mu = {\cal O}(1\ TeV )$.}.  A few percent of the parameter space
gives
a phenomenologically acceptable  $U(1)'$ symmetry breaking scenario. This fact 
is important since it implies that in this class of string models there is
a reasonable probability that the heavy $Z'$ is in the experimentally observable
region (and not required to become massive at the string scale).  
In addition, these models provide an elegant solution to the $\mu$ problem,
complementary to that of the Giudice-Masiero mechanism \cite{GM}\footnote{With
additional $U(1)'s$ the required terms in the K\" ahler potential
are absent; thus the Giudice-Masiero mechanism is not
applicable. Other possible solutions are surveyed in \cite{CL}.}.

In Sec. II we give explicit expressions for the scalar potential, vector 
boson masses, scalar masses and related sparticle masses, and 
introduce certain definitions and conventions that will be used throughout the work.

In Sec. III, we present scenarios to obtain a small $Z-Z'$ mixing angle based on 
that portion of parameter space in which the trilinear coupling is much greater
than the soft mass parameters. In this case $M_{Z'}$ is typically comparable to
$M_Z$ (e.g., $200\ GeV$) and $\tan\beta\sim 1$. This scenario is only viable for 
certain (e.g., leptophobic) couplings. One version of the model has a first order
electroweak phase transition at tree level and thus has potentially interesting
cosmological consequences.

In Sec. IV, we present a scenario in which the singlet acquires a large VEV so
that $M_{Z'}={\cal O}(1\ TeV)$. In this case, all of the dimensional parameters in
the scalar potential are of ${\cal O}(1\ TeV)$ and the smaller electroweak scale
is due to a cancellation of parameters.

In Section V, we use the renormalization group to relate the electroweak scale
supersymmetry breaking parameters to those at the string scale. 
We first assume the minimal particle content, consisting of the MSSM particles, 
the additional singlet, and the $Z'$. We present the results of the numerical
integration of the renormalization group equations (RGEs) for the parameters of
the model as a function 
of their  boundary conditions at the string scale.  With the minimal particle
content, we conclude that it is necessary to invoke nonuniversal values of the
soft supersymmetry breaking parameters at the string scale to reach the desired
low energy region of parameter space.  Several examples of boundary conditions
at the string scale are presented which lead to the phenomenologically
acceptable scenarios of Sec.~III and IV.  We also discuss the implications of
additional exotic matter in the RGEs, and conclude that with additional $SU(3)$
triplets, for example, the large singlet VEV scenario is possible with universal
boundary conditions.  

The RGEs are presented in Appendix A.  In Appendix B, we present the
details of the numerical results, and we give semi-analytic solutions of the
RGEs.  Finally, in Appendix C we present examples of models with anomaly-free 
$U(1)'$.

Our goal is to explore the general features of electroweak breaking
in a class of string models, not to construct a specific model. We
therefore focus on the gauge and symmetry breaking sectors of the
theory and only specify the $U(1)'$ charges when we present
concrete numerical examples.

\section{Electroweak Symmetry Breaking}
The gauge group is extended to 
$G=SU(3)_{c}\times SU(2)_{L}\times U(1)_{Y} \times U(1)_{Y'}$ with the couplings
$g_{3}$, $g_{2}$, $g_Y$, $g_1'$, respectively\footnote{Here 
$g_Y=\sqrt{\frac{3}{5}}g_1$, where $g_1$ is the GUT normalized coupling. That
is, $g_Y$ is the coupling usually called $g'$ in the Standard Model.}. 
The particle content is given by the left-handed chiral superfields $\hat{L}_i 
\sim (1,2,-1/2, Q_{L})$, $\hat{E}_i^{c} \sim (1, 1, 1, Q_{E})$, $\hat{Q}_i \sim
(3, 2, 1/6, Q_{Q})$, $\hat{U}_i^{c} \sim (\bar{3}, 1, -2/3, Q_{U})$,
$\hat{D}_i^{c} \sim (\bar{3}, 1, 1/3, Q_{D})$, $\hat{H}_{1} \sim (1, 2, -1/2,
Q_{1})$, $\hat{H}_{2} \sim (1, 2, 1/2, Q_{2})$, $\hat{S}\sim (1, 1, 0, Q_{S})$,
where the subscript $i$ is the family index. 

The superpotential for our model is\footnote{The $U(1)'$ forbids not only an
elementary $\mu\hat{H}_{1}\cdot\hat{H}_{2}$ term in the superpotential, 
but also a term $\hat{S}^3$. Such a term is needed in the NMSSM \cite{NMSSM} 
to avoid the appearance of an axion after symmetry breaking. In our model, 
this massless pseudoscalar is eaten by the $Z'$. Also, unlike in the NMSSM the
discrete symmetry is embedded in the gauge symmetry and thus there is no domain
wall problem.} 
\begin{eqnarray} 
W=h_{s}\hat{S}\hat{H}_{1}\cdot\hat{H}_{2} + h_Q \hat{U}_3^c\hat{Q}_3\cdot\hat{H}_2. 
\label{superpot}
\end{eqnarray}
The form of (\ref{superpot}) is motivated by string models \cite{Faraggi}, in
which a given Higgs doublet (i.e., $\hat{H}_2$) only has Yukawa couplings to a
single (third) family. This family
index will not be displayed in the rest of the paper.

Gauge invariance of $W$ under $U(1)'$ requires $Q_{1}+Q_{2}+Q_{S} = 0$. The 
effective $\mu$ parameter is generated by the VEV $\langle S \rangle =
s/\sqrt{2}$,
and will then be given by $\mu_s=h_ss/\sqrt{2}$.

Within string models there is no mechanism for supersymmetry breaking with
quantitative predictive power. We thus parameterize supersymmetry breaking with
the most general soft supersymmetry
breaking mass parameters. 
The soft supersymmetry
breaking lagrangian takes the form \begin{eqnarray}
\label{soft}
{\cal L}_{SB}&=&(-\sum_{i}M_i\lambda_i\lambda_i+
   Ah_{s}SH_{1}\cdot H_{2}+A_Qh_QU^cQ\cdot H_2+h.c.)
   -m_{1}^{2}|H_{1}|^2-  m_{2}^{2}|H_{2}|^2\nonumber\\
  &-&m_{S}^{2}|S|^2-m_Q^2|Q|^2-m_U^2|U|^2
  -m_D^2|D|^2-m_E^2|E|^2-m_L^2|L|^2,
\end{eqnarray}
where the $\lambda_i$ are gauginos, and the other fields are the scalar
components of the corresponding supermultiplets.
Gauge symmetry breaking is now driven by the vacuum expectation values 
of the doublets $H_1$, $H_2$ and the singlet $S$. The Higgs potential 
is the sum of three pieces:
\begin{eqnarray}
V=V_F+V_D+V_{soft},
\label{pot}
\end{eqnarray}
with
\begin{eqnarray}
\label{vf}
V_F&=&
   |h_{s}|^2\left[ |H_{1}\cdot H_{2}|^2+ |S|^2 ( |H_{1}|^2+|H_{2}|^2)\right],
\end{eqnarray}
\begin{eqnarray}
\label{vd}
V_D&=&\frac{G^2}{8}\left(|H_{2}|^2-|H_{1}|^2\right)^2+ 
   \frac{g_{2}^2}{2}|H_{1}^{\dagger}H_{2}|^2+
   \frac{g_{1}'^2}{2}\left(Q_{1}|H_{1}|^2+Q_{2}|H_{2}|^2+
   Q_{S}|S|^2\right)^2, 
\end{eqnarray}
\begin{eqnarray}
V_{soft}&=&m_{1}^{2}|H_{1}|^2+
  m_{2}^{2}|H_{2}|^2+
  m_{S}^{2}|S|^2 - (Ah_{s}SH_{1}\cdot H_{2}+h.c.),
\end{eqnarray}
where $G^2=g_{Y}^2+g_{2}^{2}$, and
\begin{eqnarray}
H_{1}=\left(\begin{array}{c c} H_{1}^0\\H_1^-\end{array}\right),\;\;
H_{2}=\left(\begin{array}{c c} H_2^+\\H_2^0\end{array}\right).
\end{eqnarray}
By an appropriate choice of the global phases of the fields, we can take $Ah_s$ real 
and positive without loss of generality. By a suitable 
gauge rotation we can also make $\langle H_2^+\rangle=0$ and take $\langle 
H_2^0\rangle =v_2/\sqrt{2}$ and $\langle S\rangle=s/ \sqrt{2}$ real and positive.
The requirement $\langle H_1^- \rangle=0$ in the vacuum is equivalent to requiring 
the squared mass of the physical charged scalar to be positive and imposes some 
constraint on the parameter space of the model, as will be shown later. There is no 
room for explicit or spontaneous CP violation in the potential (\ref{pot}) so
that $\langle H_1^0\rangle=v_1/\sqrt{2}$ is 
real. Furthermore, with our choice $Ah_s>0$ one has $v_1>0$ in the true minimum.

Even after the replacement of $h_sS$ by $h_s\langle S \rangle=\mu_s\sqrt{2}$,  
$V$ differs from the MSSM by additional terms quadratic in the $H_i$ in $V_F$ and 
$V_D$. The minimization conditions when all VEVs are non-zero give\footnote{For a 
more precise analysis of the model, beyond the scope of this paper, it would be 
necessary to include one-loop 
corrections, which can have a non-negligible effect \cite{oneloopmin}.} 
\begin{eqnarray}
\label{min1}
m_1^2&=&m_3^2\tan\beta-\frac{1}{8}G^2v^2\cos 2\beta -\frac{1}{2}{g'}_1^{2}Q_1
({\overline Q}_Hv^2+Q_Ss^2)-\frac{1}{2}h_s^2(v^2\sin^2\beta+s^2),\\
\label{min2}
m_2^2&=&m_3^2\cot\beta+\frac{1}{8}G^2v^2\cos 2\beta -\frac{1}{2}{g'}_1^{2}Q_2
({\overline Q}_Hv^2+Q_Ss^2)-\frac{1}{2}h_s^2(v^2\cos^2\beta+s^2),\\
\label{min3}
m_S^2&=&m_3^2\frac{v^2}{s^2}\sin\beta\cos\beta-\frac{1}{2}{g'}_1^{2}Q_S({\overline 
Q}_Hv^2+Q_Ss^2)-\frac{1}{2}h_s^2v^2, 
\end{eqnarray}
where $m_3^2=(h_s/\sqrt{2})As$, ${\overline Q}_H=Q_1\cos^2\beta+Q_2\sin^2\beta$,
$v^2=v_1^2+v_2^2$ and $\tan\beta=v_2/v_1$.

To ensure that the extremum at $(v_1,v_2,s)$ is a minimum of the potential, the 
squared masses of scalar Higgses should be positive. In addition, $V(v_1,v_2,s)<0$
should also hold for the minimum to be acceptable. Even if all these conditions 
are satisfied, the minimum is not guaranteed to be the global minimum of the 
potential. Whether it is still acceptable will depend on the location and 
depth of the other possible minima and of the barrier height and width 
between the minima \cite{KLS}.

Letting $Z'$ be the gauge boson associated with $U(1)'$, the $Z-Z'$ mass-squared
matrix is given by  
\begin{eqnarray}
\label{Zmatrix}
(M^{2})_{Z-Z'}=\left (\begin{array}{c c} 
M_{Z}^{2}&\Delta^{2}\\\Delta^{2}&M_{Z'}^{2}\end{array}\right),
\end{eqnarray}
where 
\begin{eqnarray}
M_{Z}^{2}&=&\frac{1}{4}G^2(v_{1}^2+v_{2}^2),\\
M_{Z'}^{2}&=&{g'}_{1}^{2}(v_{1}^{2}Q_{1}^{2}+v_{2}^{2}Q_{2}^{2}+s^{2}Q_{S}^{2}),\\
\label{mix}
\Delta^{2}&=&\frac{1}{2}g_{1}'\,G(v_{1}^2Q_{1}-v_{2}^2Q_{2}).
\end{eqnarray}
The eigenvalues of this matrix are 
\begin{eqnarray}
M^{2}_{Z_{1},Z_{2}}&=&\frac{1}{2}\left[M^{2}_{Z}+M^{2}_{Z'}\mp
   \sqrt{(M^{2}_{Z}-M^{2}_{Z'})^{2}+4\Delta^4}\right].
\end{eqnarray}
The $Z-Z'$ mixing angle $\alpha_{Z-Z'}$ is given by
\begin{eqnarray}
\alpha_{Z-Z'}=\frac{1}{2}\arctan\left(\frac{2\Delta^2}{M^{2}_{Z'}-M^{2}_{Z}}
\right).
\end{eqnarray}
Phenomenological constraints typically require this mixing angle to be less than a 
few times $10^{-3}$ \cite{PRESENTZ}, although values as much as ten times larger
may be possible in some models with a light $Z'$ (e.g., $M_{Z_2}/M_{Z_1}\sim 
{\cal O}(2)$)
and certain (e.g., leptophobic) couplings. Then, with good precision
$M_{Z_1}^2=G^2v^2/4$ so that $v=246\,GeV$ is fixed.

The spectrum of physical Higgses after symmetry breaking consists of three 
neutral CP even scalars ($h^0_i$, $i=1,2,3$), one CP odd pseudoscalar ($A^0$) and 
a pair of charged Higgses ($H^\pm$), that is, one scalar more than in the 
MSSM. The tree-level masses of the Higgs bosons are 
\begin{equation}
\label{ma}
m_{A^0}^2=\frac{\sqrt{2}Ah_ss}{\sin 
2\beta}\left[1+\frac{v^2}{4s^2}\sin^22\beta \right],
\end{equation}
which is never negative, and
\begin{equation}
m_{H^\pm}^2=M_W^2+\frac{\sqrt{2}Ah_ss}{\sin 2\beta}-\frac{1}{2}h_s^2v^2.
\label{charged}
\end{equation}
$m_{H^\pm}^2$ could be lighter than the W boson due to the negative 
third contribution. It could even be negative for some choices 
of the parameters.

Masses for the three neutral scalars can be obtained by diagonalizing the
corresponding $3\times 3$ mass matrix, which, in the basis 
$\{H_1^{0r}\equiv Re (H_1^0) \sqrt{2},H_2^{0r},S^{0r}\}$, reads: 
\begin{eqnarray}
(M^{2})_{h^0}=\left(\begin{array}{c c c} 
\kappa_1^2v_1^2+m_3^2\tan\beta &
\kappa_{12}v_1v_2-m_3^2 &
\kappa_{1s}v_1s-m_3^2{\displaystyle\frac{v_2}{s}}\vspace{0.1cm}\\
\kappa_{12}v_1v_2-m_3^2 &
\kappa_2^2v_2^2+m_3^2\cot\beta &
\kappa_{2s}v_2s-m_3^2{\displaystyle\frac{v_1}{s}}\vspace{0.1cm}\\
\kappa_{1s}v_1s-m_3^2{\displaystyle\frac{v_2}{s}} &
\kappa_{2s}v_2s-m_3^2{\displaystyle\frac{v_1}{s}} &
\kappa_{s}^{2}s^2+m_3^2{\displaystyle\frac{v_1v_2}{s^2}}\end{array}\right), 
\label{neutral}
\end{eqnarray}
with $\kappa_i^2=G^2/4+{g'}_1^2Q_i^2$, $\kappa_{12}=h_s^2+{g'}_1^2Q_1Q_2-G^2/4$, 
$\kappa_{is}=h_s^2+{g'}_1^2Q_iQ_S$ and $\kappa_{s}^{2}={g'}_1^2Q_S^2$.

It is simple to obtain some useful information from the structure of this matrix.
The tree level mass of the lightest scalar $h_1^0$ satisfies the bound
\begin{equation}
\label{bound}
m_{h_1^0}^2\leq M_Z^2\cos^22\beta+\frac{1}{2}h_s^2v^2\sin^2 
2\beta+{g'}_1^2{\overline Q}_H^2v^2.
\end{equation}
The first term is the usual MSSM tree level bound. The second contribution comes 
from F-terms and appears also in the NMSSM \cite{NMSSM}, while the third is a D-term 
contribution from the $U(1)'$ and thus is a particular feature of this type of 
models \cite{HABSH,XG}. In contrast to the MSSM, $h_1^0$ can be 
heavier than $M_Z$ at tree level. 
In addition, radiative corrections \cite{radcor} will also be sizeable. 
This indicates that $h_1^0$ can easily escape detection at LEPII. 
For $m_{h_1^0}$ within the kinematical reach the composition of $h_1^0$ will 
determine its production cross sections (e.g., through $Z\rightarrow Z^*h_1^0 $). In 
particular, the $h_1^0ZZ$ coupling, and thus the cross section, are reduced
if $h_1^0$ has a significant singlet admixture. However, when that suppression 
takes place $h_2^0$ also tends to be light \cite{NTL}. Actually, in the limit of 
$h_1^0\rightarrow S^{0r}$ the mass of $h_2^0$ satisfies the limit (\ref{bound}).
In the event that both $h_1^0$ and $h_2^0$ have a substantial singlet
component, $h_3^0$ will also tend to be light. 

In the general case, when the masses governing
the scalar mass matrix $(m_{A^0}, M_Z, M_{Z'})$ have comparable magnitudes, 
the scalar states $h_i^0$ will be complicated mixtures of the interaction eigenstates. 
When there is some hierarchy in those masses, it is possible to make definite 
statements about the composition of the mass eigenstates:
\begin{itemize}
\item {\em H1)} If $M_{Z'}\gg m_A\sim M_Z$ the heavier scalar is singlet dominated 
($h_3^0\sim S^{0r}$) with mass $m_{h_3^0}\sim M_{Z'}$. The two lighter states
are mixtures of $H_1^{0r}$ and $H_2^{0r}$ (with some mixing angle much
like in the MSSM, although with masses in a different range) with  masses
around $m_A\sim M_Z$. More precisely, the lightest scalar $h_1^0$ satisfies
the (approximate) mass bound

\begin{equation}
\label{asympto}
m_{h_1^0}^2\simlt M_Z^2\cos^22\beta+
h_s^2v^2\left[\frac{1}{2}\sin^2 
2\beta-\frac{h_s^2}{{g'}_1^2Q_S^2}-2\frac{{\overline Q}_H}{Q_S}\right].
\end{equation}

\item {\em H2)} When $M_{Z'}\gg m_A\gg M_Z$ the two lighter mixed states of case
{\em H1)}
have a definite composition: $h_2^0\sim H_1^{0r}\sin\beta-H_2^{0r}\cos\beta$
with mass $\sim m_A$ and $h_1^0\simeq H_1^{0r}\cos\beta+H_2^{0r}\sin\beta$ 
with mass saturating the bound (\ref{asympto}). In this limit $h_1^0$ has Standard
Model couplings.
 
\item {\em H3)} If $m_A\gg M_{Z'}, M_Z$ then $m_{h_1^0}^2$ goes to negative
values. 
This means that the electroweak vacuum ceases to be a minimum and turns into a 
saddle point; the minimum of the potential lies at some other point in field 
space and the symmetry breaking is not in accord with the observed values
of the gauge boson masses.
\end{itemize}
More details about the Higgs spectrum in particular scenarios will be given in 
the next sections.

The parameter $\mu_s$ also plays an important role in the chargino-neutralino 
sector. Remembering that $\mu_s=h_ss/\sqrt{2}$, the masses 
for the two charginos $\tilde{\chi}_{1,2}^\pm$ are given by the MSSM formula
\begin{equation}
\label{charg0}
m^2_{\tilde{\chi}_{1,2}^\pm}=\frac{1}{2}\left\{
M_2^2+\mu_s^2+2M_W^2\pm\sqrt{
(M_2^2+\mu_s^2+2M_W^2)^2-4(M_2\mu_s-M_W^2\sin 2\beta)^2}
\right\},
\end{equation}
where $M_2$ is the $SU(2)$ gaugino mass. The following bounds result from
(\ref{charg0})
\begin{equation}
\label{charg1}
m^2_{\tilde{\chi}_{1}^\pm}\leq
\left\{
\begin{array}{l}
 \mu_s^2+2M_W^2\cos^2\beta \vspace{0.1cm}\\
 M_2^2+2M_W^2\sin^2\beta, 
\end{array}
\right.
\end{equation}
and the following limiting cases hold
\begin{equation}
\label{charg2}
m^2_{\tilde{\chi}_{1}^\pm}\rightarrow
\left\{
\begin{array}{ll} 
 \mu_s^2 &(M_2^2\gg \mu_s^2,2M_W^2\cos^2\beta) 
\vspace{0.1cm}\\
 M_2^2 &(\mu_s^2\gg M_2^2,2M_W^2\sin^2\beta).
\end{array}
\right.
\end{equation}
In the first (second) case, the lightest chargino is predominantly a higgsino
(gaugino).

Preliminary LEP results, including data collected at $\sqrt{s}=172\ GeV$
set a $95\%$ CL lower limit on the chargino mass of about $70-85$ GeV
\cite{lep}.
The weaker value corresponds to light enough $\tilde{e}^\pm, \tilde{\nu}_e$,
which can interfere destructively in the $e^+e^-\rightarrow \chi^+\chi^-$
cross-section. For definiteness we impose in our analysis $m_{\tilde{\chi}_1^\pm}
>80\ GeV$. Eqs.~(\ref{charg1},\ref{charg2}) imply that this lower bound 
puts a significant constraint on the parameter space of the model if 
$h_ss$ is relatively small (roughly $h_ss<M_Z$). In general, some parameter  
region around $M_2\mu_s=M_W^2\sin 2\beta$ will always be excluded (for parameter 
values satisfying exactly that condition, $m_{\tilde{\chi}^\pm_1}^2=0$).

In the neutralino sector, there is an extra $U(1)'$ zino and the 
higgsino ${\tilde S}$ as well as the four MSSM neutralinos. The
$6\times 6$ mass matrix reads
(in the basis $\{\tilde{B}',\tilde{B},\tilde{W_3},\tilde{H}_1^0,\tilde{H}_2^0,
\tilde{S}\}$):
\begin{eqnarray}
\label{neutralinos}
M_{\tilde{\chi}^0}=\left(\begin{array}{c c c c c c} 
M'_1 & 0 & 0 &
 g'_1Q_1v_1 &
 g'_1Q_2v_2 &
 g'_1Q_Ss \vspace{0.1cm}\\
0 & M_1 & 0 &
-{\displaystyle\frac{1}{2}}g_Yv_1 &
{\displaystyle\frac{1}{2}}g_Yv_2 & 0\vspace{0.1cm}\\
0 & 0 & M_2 &
{\displaystyle\frac{1}{2}}gv_1 &
-{\displaystyle\frac{1}{2}}gv_2 & 0\vspace{0.1cm}\\
g'_1Q_1v_1 &
-{\displaystyle\frac{1}{2}}g_Yv_1 &
{\displaystyle\frac{1}{2}}gv_1 & 0 &
-\mu_s &
-\mu_s{\displaystyle\frac{v_2}{s}}\vspace{0.1cm}\\
 g'_1Q_2v_2 &
{\displaystyle\frac{1}{2}}g_Yv_2 &
-{\displaystyle\frac{1}{2}}gv_2 &
-\mu_s & 0 &
-\mu_s{\displaystyle\frac{v_1}{s}}\vspace{0.1cm}\\
 g'_1Q_S s & 0 & 0 &
-\mu_s{\displaystyle\frac{v_2}{s}}&
-\mu_s{\displaystyle\frac{v_1}{s}}& 0
\end{array}\right), 
\end{eqnarray}
where $M_1$ and $M'_1$ are the gaugino masses associated with $U(1)$ and $U(1)'$, 
respectively. 

For general values of the parameters in this matrix, the mass eigenstates will be 
complicated mixtures of higgsinos and gauginos. It is useful to consider some 
limiting cases:

\begin{itemize}

\item {\em N1)} If $M_1'=M_1=M_2=\mu_s=0$ there are two massless neutralinos. One
is a pure photino ($\tilde{\chi}^0_1=\tilde{\gamma}=\cos\theta_W\tilde{B}+
\sin\theta_W\tilde{W}_3$) and the other a pure higgsino $\tilde{\chi}_2^0=
(\tilde{H}_1^0\sin\beta+\tilde{H}_2^0\cos\beta)\cos\alpha+\tilde{S}\sin\alpha$
with $\tan\alpha=(v/s)\sin\beta\cos\beta$. The rest of the neutralinos will have 
masses controlled by $M_Z$ and $M_{Z'}$.

\item {\em N2)} If $M_i^2,\mu_s^2\gg M_Z^2$, two of the eigenstates are just
$\tilde{B}$
and $\tilde{W_3}$ with masses $|M_1|$ and $|M_2|$, respectively. Next, two
higgsinos:
$\tilde{H}_1^0\sin\beta+\tilde{H}_2^0\cos\beta$ and 
$\tilde{H}_2^0\sin\beta-\tilde{H}_1^0\cos\beta$ are nearly degenerate with mass
$|\mu_s|$. The remaining two neutralinos are mixtures of $\tilde{B}'$ and 
$\tilde{S}$, and we can consider two different simple situations; first, if
${M_1'}^2\gg  {g_1'}^2Q_S^2s^2$, then $\tilde{B}$ has mass $|M_1'|$ while
$\tilde{S}$ is light, 
with mass controlled by $M_Z$. In the other case, with ${M_1'}^2\ll
{g_1'}^2Q_S^2s^2$,
they have masses $m^2_{\tilde{\chi}^0}={g_1'}^2Q_S^2s^2\pm g_1'Q_S M_1' s$.

\item {\em N3)} If $\mu_s^2\gg M_i^2,M_Z^2$ (which naturally requires $s\gg v$,
hence  $M_{Z'}^2\gg  M_i^2,M_Z^2$), the approximate eigenstates are:
 $(\tilde{B}'\pm
\tilde{S})/\sqrt{2}$ with
mass $M_{Z'}$; $\tilde{B}$, $\tilde{W}_3$, with masses
$|M_1|,|M_2|$ respectively, and $(\tilde{H}_1^0\pm\tilde{H}_2^0)/\sqrt{2}$
with mass $|\mu_s|$.
\end{itemize}

In the next sections we will give numerical examples of the pattern 
expected for charginos and neutralinos in different scenarios.

Masses for the rest of the MSSM particles (squarks and sleptons) can be  obtained 
directly from the MSSM formulae by setting $\mu=\mu_s=h_ss/\sqrt{2}$ and adding the 
pertinent D-term diagonal contributions from the $U(1)'$ \cite{LYKKEN}:
\begin{equation}
\label{sqdterm}
\delta m_i^2 = \frac{1}{2}{g'}_1^2Q_i(Q_1v_1^2+Q_2v_2^2+Q_Ss^2),
\end{equation}
where $Q_i$ is the $U(1)'$ charge of the corresponding particle. 
This extra term can produce significant mass deviations with 
respect to the minimal model and plays an important role
in the connection between parameters at the electroweak and string scales. 
However, in the low energy analysis,
its effect can always be absorbed in the unknown soft supersymmetry breaking mass
squared parameters. 

Before proceeding with the analysis of different scenarios it is useful to 
compare the present model with the simplified version discussed in ref.~\cite{CL}. 
That version contained one
Higgs doublet and one singlet, with $U(1)'$ charges $Q_H$ and $Q_S$ respectively.
It was shown that a sufficiently heavy $Z'$
(with mass up to $\sim 1\ TeV$) with small mixing to the $Z$ could be obtained 
for the case $Q_HQ_S>0$, which would allow cancellations so that $M_Z$ and $v$ 
can be small compared to $|m_H|$, $|m_S|$ and $s$. 
The more realistic case with two Higgs doublets offers several 
advantages. First, there can be a cancellation in the off-diagonal 
$Z-Z'$ mass matrix element (\ref{mix}) if $Q_1Q_2>0$. In addition, the presence 
of a trilinear 
coupling in the superpotential (forbidden by $SU(2)$ in the model of \cite{CL}) 
qualitatively changes the Higgs potential, allowing for a richer pattern of 
symmetry breaking mechanisms. In particular, the condition 
$Q_HQ_S>0$ (that in our model would be generalized to ${\overline Q}_HQ_S>0$) is 
no longer necessary.

We can classify the symmetry breaking scenarios in three different categories 
according to the value of the singlet VEV:

\begin{itemize}
\item (i) $s=0$.

This corresponds to the case of the breaking driven only by the two Higgs 
doublets (this would be the typical case if the soft mass of the singlet remains 
positive). The $Z'$ boson would acquire mass of the same order as the $Z$, and 
many other particles (Higgses, charginos and neutralinos) would tend to be 
dangerously light ($\mu_s=0$ now). There is in principle the possibility of a
small $Z-Z'$ mixing due to the cancellation mechanism described and one could
arrange the parameters to barely satisfy experimental constraints. However, this
requires considerable fine-tuning, and we do not pursue this singular scenario
further.

\item (ii) $s\sim v$.

This case would naturally give $M_{Z'}\simgt M_Z$ (if $g_1'Q$ is not too small) 
and a small $\mu_s$ (thus some sparticles will be expected to be light).  
One requires $Q_1=Q_2$ to have negligible $Z-Z'$ mixing. Such models are allowed
for leptophobic couplings \cite{LEPTOPHOBIC}. Particularly interesting examples 
of this type of scenario will be presented in the next section.

\item (iii) $s\gg v$.

In this case $M_{Z'}\gg M_Z$ and $\mu_s$ and $m_3^2$ are  naturally large. 
The $Z-Z'$ mixing is suppressed by the large mass $M_{Z'}$
(in addition to any accidental cancellation for particular choices of charges), 
eventually relaxing the constraint $Q_1Q_2>0$. As $M_{Z'}$ increases 
more fine-tuning is needed to keep $M_Z$ light. This case will be studied in 
section IV.
\end{itemize}

\section{Large Trilinear Coupling  Scenario}
For the sake of simplifying the analysis, the soft supersymmetry breaking mass
parameters can be written in terms of dimensionless parameters $c_i$ and an
overall mass scale $M_0$:
\begin{eqnarray}
\label{rescalemass}
&&m_{1}^{2}=c^{2}_{1}M_{0}^{2}\;,\;m_{2}^{2}=c^{2}_{2}M_{0}^{2}\;,\;
m_{S}^{2}=c^{2}_{S}M_{0}^{2}\;,\;A=c_{A}\,M_{0}.
\end{eqnarray}
Since these are the only dimensional parameters in (\ref{pot}), one can
conveniently parameterize the VEVs as:
\begin{eqnarray}
\label{rescalefields}
v_{1}=f_{1}M_{0}\;,\;v_{2}=f_{2}M_{0}\;,\;s=f_{s}M_{0}\, .
\end{eqnarray}

We first minimize the potential (\ref{pot}) with respect to the 
dimensionless parameters $f_{i}$ 
defined through  (\ref{rescalemass}), (\ref{rescalefields}) and then go to 
physical shell by choosing
\begin{eqnarray}
\label{M3/2}
M_{0}=\frac{v}{\sqrt{f_{1}^2+f_{2}^2}},
\end{eqnarray}
where $v=246\,GeV$ sets the scale of electroweak symmetry breaking. 

In contrast to the usual MSSM potential, $V$ in (\ref{pot}) has an important
trilinear term 
involving only the Higgs fields. Therefore, one can consider a symmetry 
breaking scenario driven by this large trilinear term, as opposed to 
the more common situation in which the value of the minimum is determined mainly 
by the signs and magnitudes of the soft mass-squared parameters $c^{2}_{1}$, 
$c^{2}_{2}$ and $c^{2}_{S}$. If $c_{A}$ is sufficiently large compared to the 
soft mass-squared parameters, a $c_{A}$- induced minimum occurs with
\begin{eqnarray}
\label{largecAlimit}
f_{1}\sim f_{2}\sim f_{s}\sim \frac{c_{A}}{\sqrt{2}\,h_{s}},
\end{eqnarray}
where we have also assumed that $h_{s}$ is large enough so that $V_{F}$ 
dominates the $D$-terms. (\ref{largecAlimit}) corresponds to $v_{1}\sim 
v_{2}\sim s\sim 174\, GeV$.

In the limit of large $c_{A}$, the relative signs and the magnitudes of the 
soft mass-squared parameters are not important since they contribute negligibly to 
the location of the minimum.  However, if the values of the soft mass
squared parameters are nearly the same, (\ref{largecAlimit}) is reached for
intermediate values of $c_A$.  In the present low energy analysis, we assume
for definiteness that $|c_{1}^{2}|\sim|c_{2}^{2}|\sim |c_{S}^{2}|$.
This relation is very fine-tuned in the context of the renormalization
group analysis, as discussed in Section V.

From (\ref{Zmatrix})-(\ref{mix}) it is clear that $M_{Z'}$ will generally be 
comparable to 
$M_{Z}$ in the large $c_{A}$ case, with the exact value depending on 
$g'_1Q_{1,2,S}$ (which we assume are of the same order of magnitude as $G$). Thus, 
the only phenomenologically allowed possibility is to have negligible mixing (and 
then only for small couplings to the ordinary leptons). From (\ref{mix}), we see 
that this occurs for $Q_{1}=Q_{2}$, in which case $\Delta^{2}\rightarrow 0$ for 
$f_{1}\sim f_{2}$. Both $D$- terms in (\ref{vd}) vanish in this case for large 
$c_{A}$. Therefore, in what follows we choose $Q_{1}=Q_{2}$.  

In the large $c_{A}$ solution (\ref{largecAlimit}), $M_{0}$ in (\ref{M3/2}) becomes
\begin{eqnarray}
\label{yeniM3/2}
M_{0}=\frac{h_{s}v}{c_{A}}, 
\end{eqnarray}
and 
\begin{eqnarray}
\tan\beta = \frac{f_{2}}{f_{1}}\simeq 1. 
\end{eqnarray}
The $Z'$ mass is simply given by
\begin{equation}
M_{Z'}^2=3Q_1^2{g'}_1^2v^2,
\end{equation}
and
\begin{equation}
A=h_sv.
\end{equation}
Using (\ref{largecAlimit}), (\ref{yeniM3/2}) in the expressions for the 
Higgs masses in (\ref{ma})-(\ref{neutral}), the 
limiting values for the Higgs masses are 
\begin{eqnarray}
m_{A^{0}}&\simeq& \sqrt{\frac{3}{2}}\,h_{s}\,v\nonumber\\
m_{H^{\pm}}&\simeq&\frac{1}{2}\sqrt{g_{2}^{2}+2h_{s}^{2}}\,v\nonumber\\ 
\label{limitHiggs}
m_{h^{0}_{1}}&\simeq&\frac{h_{s}}{\sqrt{2}}\,v\\
m_{h^{0}_{2}}&\simeq&\frac{1}{2}\sqrt{G^{2}+2h_{s}^{2}}\,v\nonumber\\
m_{h^{0}_{3}}&\simeq&\sqrt{3g_{1'}^{2}Q_{1}^{2}+\frac{h^{2}_{s}}{2}}\,v\nonumber.
\end{eqnarray}
Only $m_{h^{0}_{3}}$ depends explicitly 
on the $U(1)'$ charges. If a particular model allows $h_{s}$ to be much smaller
than 
the gauge couplings, $A^{0}$ and $h^{0}_{1}$ become light and $m_{H^{\pm}}\simeq 
M_{W}$, $m_{h^{0}_{2}}\simeq M_{Z}$.  

Chargino and neutralino masses depend on the gaugino masses of the $SU(2)_{L}$, 
$U(1)_{Y}$ and $U(1)'$ groups, and we discuss their spectrum later in this
section. In the $c_{A}$- induced minimum the 
effective $\mu$ parameter takes the form 
\begin{eqnarray}
\mu_{s}=\frac{h_{s}s}{\sqrt{2}}\simeq \frac{h_{s}\,v}{2}
\end{eqnarray}
This produces a small  $\mu$ parameter, $\mu_{s}\simeq 86\,GeV$ for 
$h_{s}\simeq 0.7$. 

To illustrate this scenario we take\footnote{Models which differ by a 
simultaneous rescaling of all of the $c$'s are equivalent since $M_{0}$ is 
chosen to give the observed $v=246\ GeV$.}
\begin{eqnarray} \label{fixmasses} |c_{1}^{2}|=|c_{2}^{2}|=|c_{S}^{2}|=1\, ,
\end{eqnarray}
and let $c_{A}$ vary from 0 to 10. Motivated by the renormalization group 
analysis in section V, we take $h_{s}=0.7$. We also take 
$Q_{1}=Q_{2}=-1$ and ${g'}_{1}^{2}=\frac{5}{3}G^2\sin^{2}\theta_{W}$, as is
suggested 
by simple version of gauge unification, and remark occasionally on different 
choices. 
\subsubsection{Hybrid Minimum}
First, consider
\begin{eqnarray}
c_{1}^{2}=1\;,\;c_{2}^{2}=-1\;,\;c_{S}^{2}=-1
\label{hybrid}
\end{eqnarray}
with $c_{A}$ varying from 0 to 10. We call this choice ``hybrid", since for 
small 
$c_{A}$ the minimum will be determined by these soft mass-squared parameters, and 
for large $c_{A}$ their signs and magnitudes will be irrelevant and a minimum 
described by (\ref{largecAlimit}) will occur. Though we are ultimately interested 
in the large $c_{A}$ minimum, we describe the properties of physical 
quantities in the whole $c_{A}$ range.

\begin{figure}
\centerline{\hbox{
\psfig{figure=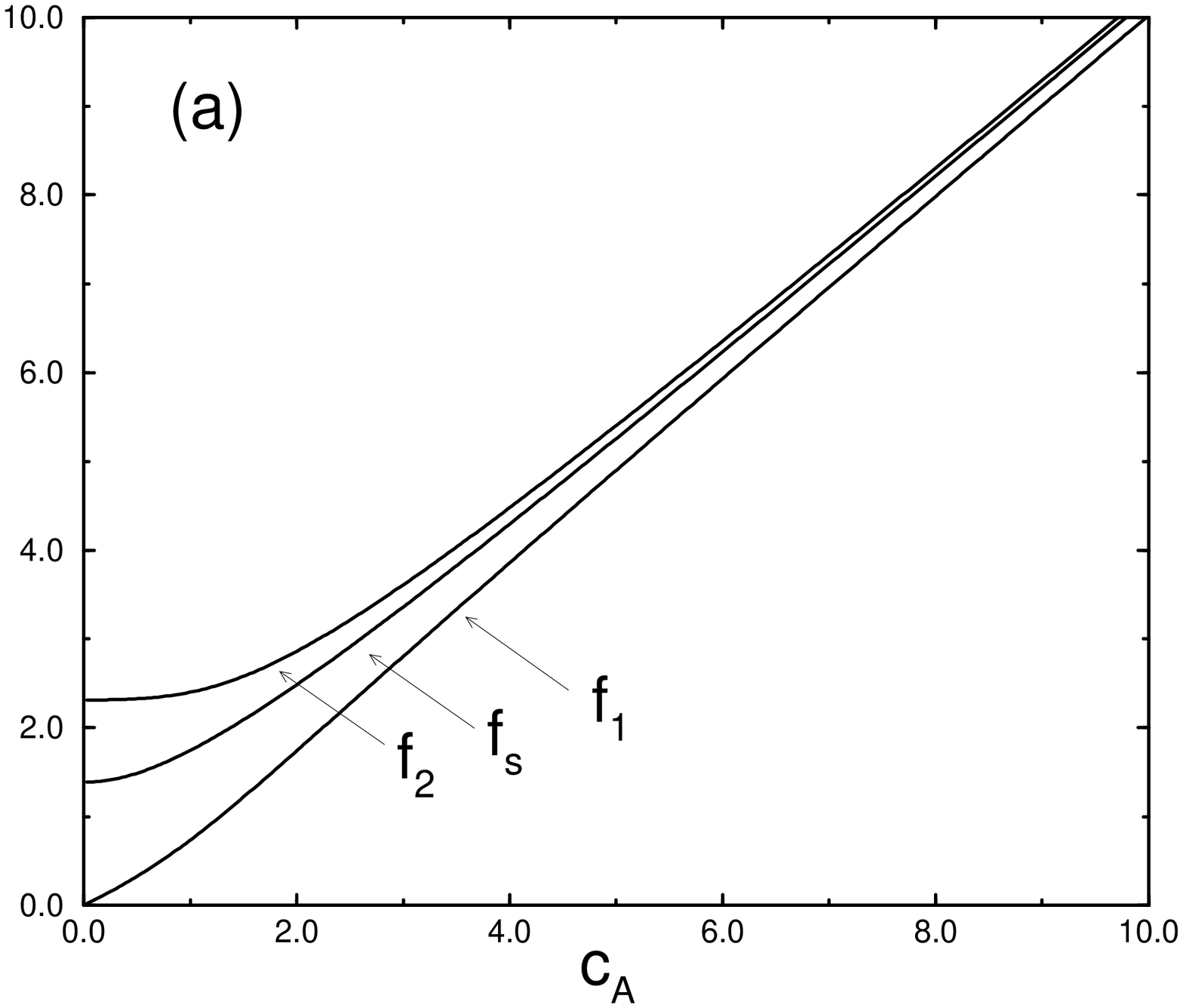,height=9cm,width=8cm,angle=-90,bbllx=3.cm,bblly=3.cm,bburx=21.cm,bbury=24.cm}
\psfig{figure=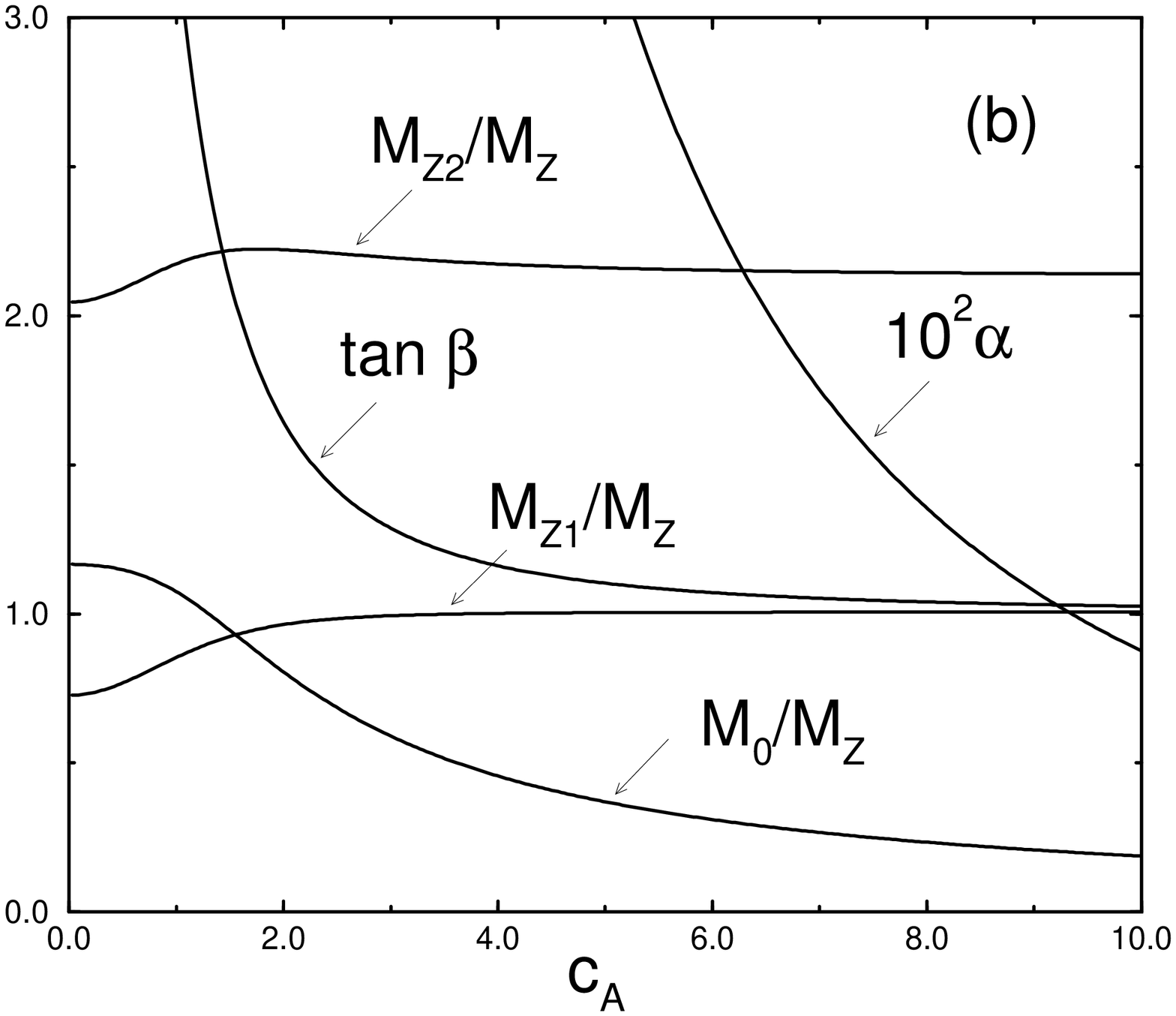,height=9cm,width=8cm,angle=-90,bbllx=3.cm,bblly=3.cm,bburx=21.cm,bbury=24.cm}}}
\caption{\footnotesize 
(a): $c_{A}$ dependence of the dimensionless field VEVs for the hybrid minimum.
(b): $c_{A}$ dependence of various dimensionless quantities for the hybrid minimum.} 
\end{figure} 

Fig.~1~(a) shows the variations of the dimensionless field VEVs with $c_{A}$. 
For large values of $c_{A}$, the effects of the quadratic mass 
parameters are unimportant and (\ref{largecAlimit}) becomes almost 
exact. It is mainly because of the biasing of the soft mass-squared parameters 
($c_{2}^{2}$ and $c_{S}^{2}$ are negative) that $f_{1}$, $f_{2}$ and 
$f_{s}$ approach their large $c_{A}$ character gradually. 

Taking $M_{Z}=91.19\,GeV$ the mass ratios $M_{Z_{1}}/M_{Z}$, 
$M_{Z_{2}}/M_{Z}$, $M_{0}/M_{Z}$, the $Z-Z'$ mixing angle $\alpha$, and $\tan\beta$
are shown as a function of $c_{A}$ in Fig.~1~(b) for the values of quadratic mass 
parameters in (\ref{hybrid}). 
We see that $M_{Z_{1}} \rightarrow M_{Z}$, $\tan\beta \rightarrow 1$, and $\alpha 
\rightarrow 0$ for large $c_{A}$; for example, $\tan\beta=1.03$ and
$\alpha=8.8\times 10^{-3}$ for $c_{A}=10$. 
With our specific $U(1)'$ charge assignments, $M_{Z_2}/M_Z\rightarrow 2.14$ 
($M_{Z_2}\simeq 196\ GeV$) for large $c_A$.
As we observe from Fig.~1~(a), the 
gap between $f_{1}$ and $f_{2}$ decreases rather gradually, and thus  
it is necessary to have larger values of $c_{A}$ to obtain a smaller
$Z-Z'$ mixing angle.  

Fig.~2 shows the variation of the scalar masses as a function of $c_{A}$ for the 
values of soft mass-squared parameters given by (\ref{hybrid}). 
For large enough $c_{A}$, all masses reach their 
asymptotic values given by (\ref{limitHiggs}): $m_{H^{\pm}}\simeq 146\,GeV\;,\; 
m_{A^{0}}\simeq 211\,GeV\;,\;m^{0}_{h_{3}}\simeq 230\,GeV\;,\;m^{0}_{h_{2}}\simeq 
152\,GeV\;,\; m^{0}_{h_{1}}\simeq 122\,GeV$. 

For the particular parameters in this example, the gauge symmetry is broken to 
$U(1)_{EM}$ for all values of $c_A$. However, for smaller $U(1)'$ couplings or
charges or larger values of $h_s$, the global minimum is $f_1=f_s=0$, $f_2\neq 0$
for values of 
$c_A$ smaller than some critical value, so that an additional $U(1)$ is unbroken. 
This is due to the positive quartic terms in $V_F$ (eq.~(\ref{vf})), which 
dominate the D-terms for large $h_s$ or small charges. The symmetry is broken to 
the desired $U(1)_{EM}$ as $c_A$ increases through 
this critical value, with the 
values of the $f_i$ varying continuously (as in a second order phase transition).
In the large $c_{A}$ limit, all 
quantities are controlled by (\ref{largecAlimit}).

\begin{figure}
\centerline{\hbox{
\psfig{figure=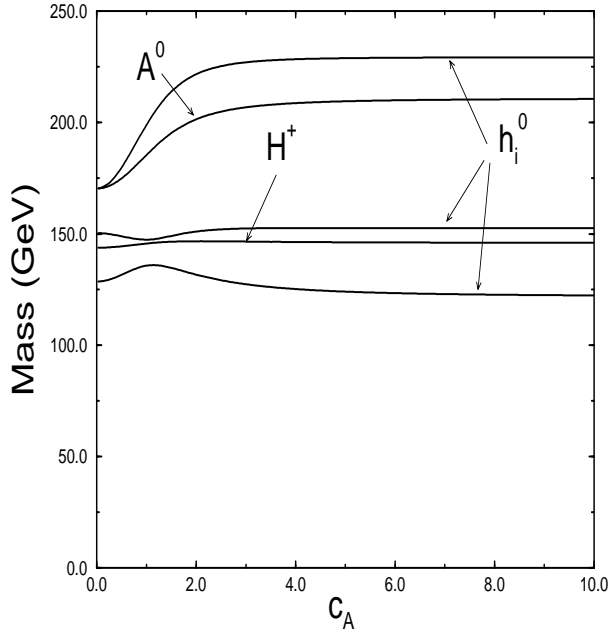,height=9cm,width=8cm,angle=-90,bbllx=3.cm,bblly=2.cm,bburx=21.cm,bbury=23.5cm}}}
\caption{\footnotesize 
$c_{A}$ dependence of the Higgs masses for the hybrid minimum.
}
\end{figure} 

\subsubsection{Pure Trilinear Coupling Minimum}
For a second example, we take 
\begin{eqnarray}
\label{posmass}
c_{1}^{2}=1\;,\;c_{2}^{2}=1\;,\;c_{S}^{2}=1
\end{eqnarray}
and vary $c_{A}$ from 0 to 10. The origin is a minimum, and a deeper minimum 
with nonvanishing fields can only be induced by $c_{A}$. 

Fig.~3~(a) shows the variations of the dimensionless field VEVs with $c_{A}$.
For  $c_{A}>c^{crit}_{A}=3$ all the fields are 
nonzero and identical, for our choices of the other parameters, approaching the 
values in (\ref{largecAlimit}) for large $c_{A}$.  

In Fig.~3~(b) we plot the dimensionless quantities $M_{Z_{1}}/M_{Z}$, 
$M_{Z_{2}}/M_{Z}$, $M_{0}/M_{Z}$, the $Z-Z'$ mixing angle $\alpha$, and $\tan\beta$
as a function of $c_{A}$ for the $c_{A}>c^{crit}_{A}$ portion of the total range.
In this minimum:
$M_{Z_{1}}=M_{Z}$, $M_{Z_{2}}=196\,GeV$, $\alpha=0$, $\tan\beta=1$, and 
$M_{0}=h_sv/c_{A}$. For other small positive values of the quadratic mass 
parameters, the minimum will again be induced by $c_{A}$, and the same values 
will be reached asymptotically. Fig.~4~(a) shows the variation of scalar masses as
a function of $c_{A}$ for the
soft mass-squared parameters of (\ref{posmass}). 

\begin{figure}
\centerline{\hbox{
\psfig{figure=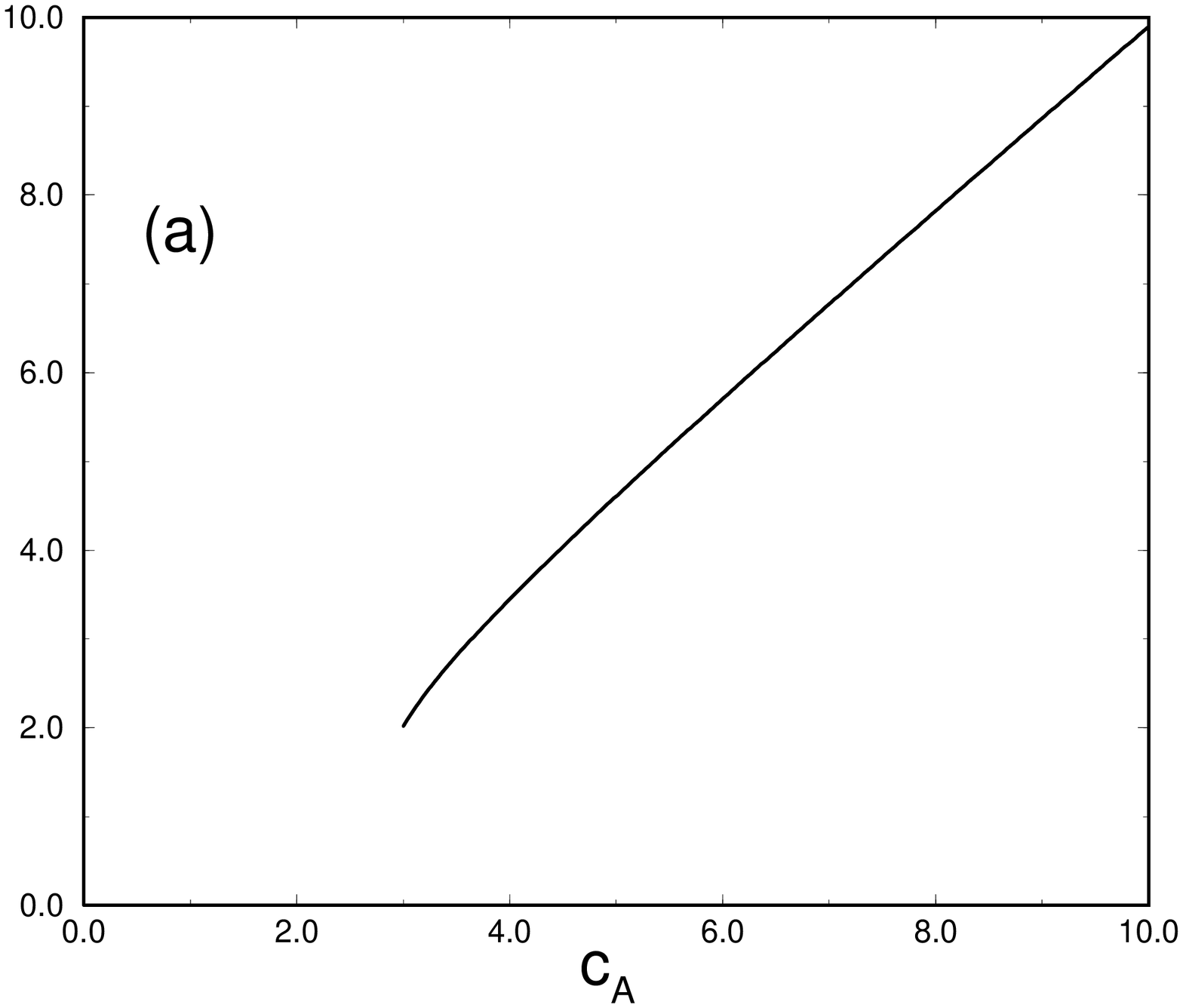,height=9cm,width=8cm,angle=-90,bbllx=3.cm,bblly=3.cm,bburx=21.cm,bbury=24.cm}
\psfig{figure=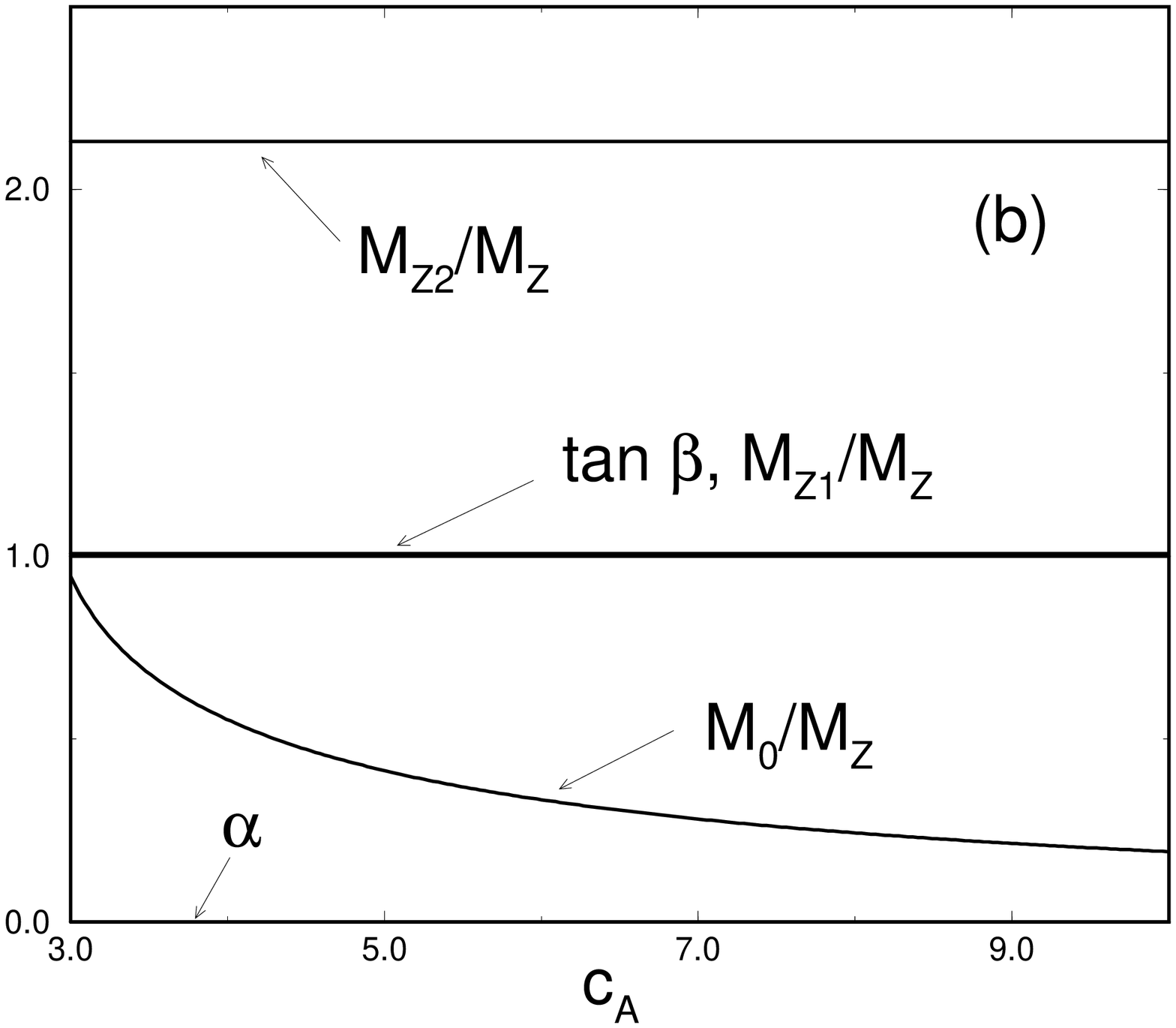,height=9cm,width=8cm,angle=-90,bbllx=3.cm,bblly=3.cm,bburx=21.cm,bbury=24.cm}}}
\caption{\footnotesize 
(a) $c_{A}$ dependence of the dimensionless VEVs, and (b) various 
dimensionless quantities for the pure 
trilinear coupling minimum.} 
\end{figure} 
\begin{figure}
\centerline{\hbox{
\psfig{figure=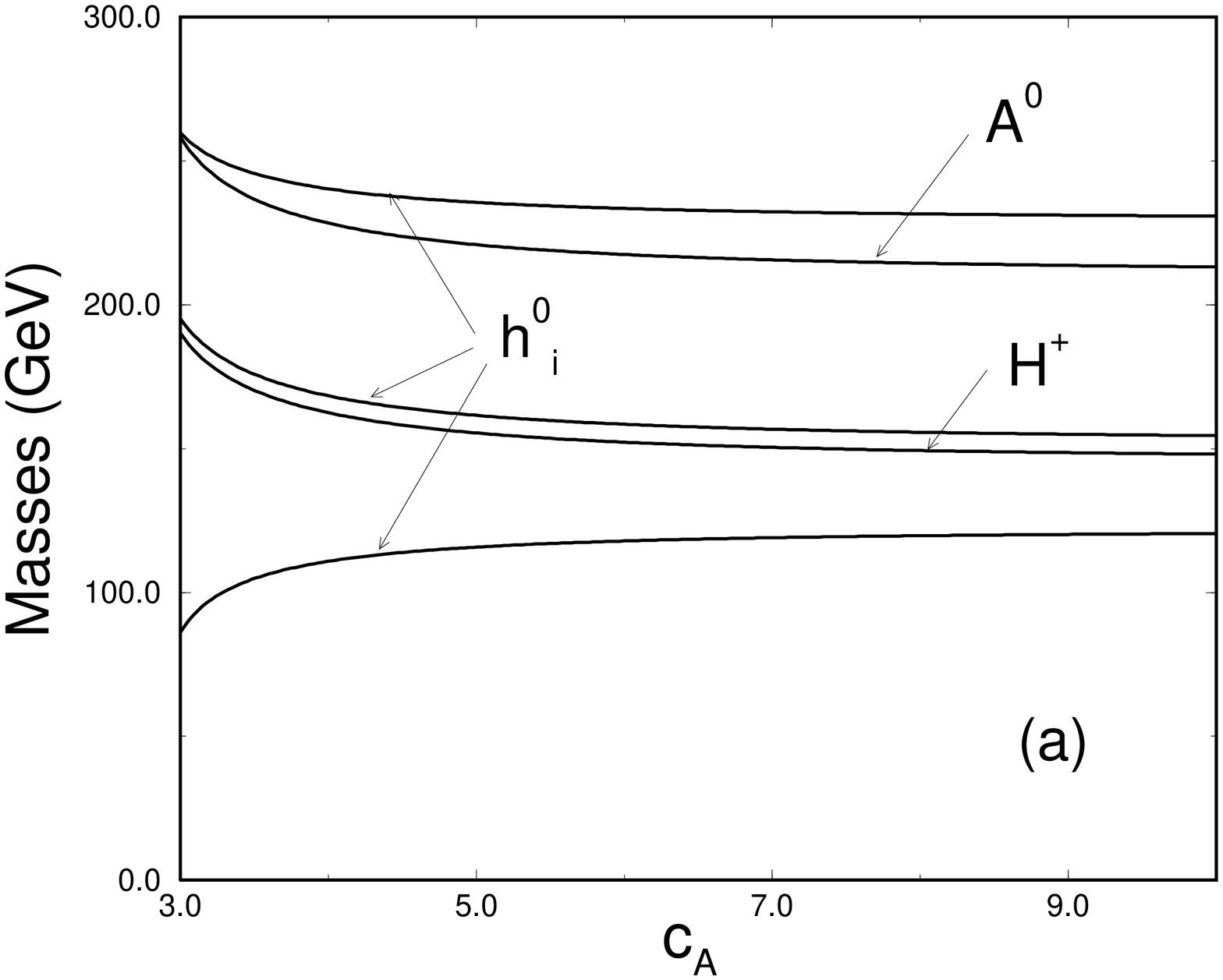,height=9cm,width=8cm,angle=-90,bbllx=3.cm,bblly=2.cm,bburx=21.cm,bbury=23.5cm}
\psfig{figure=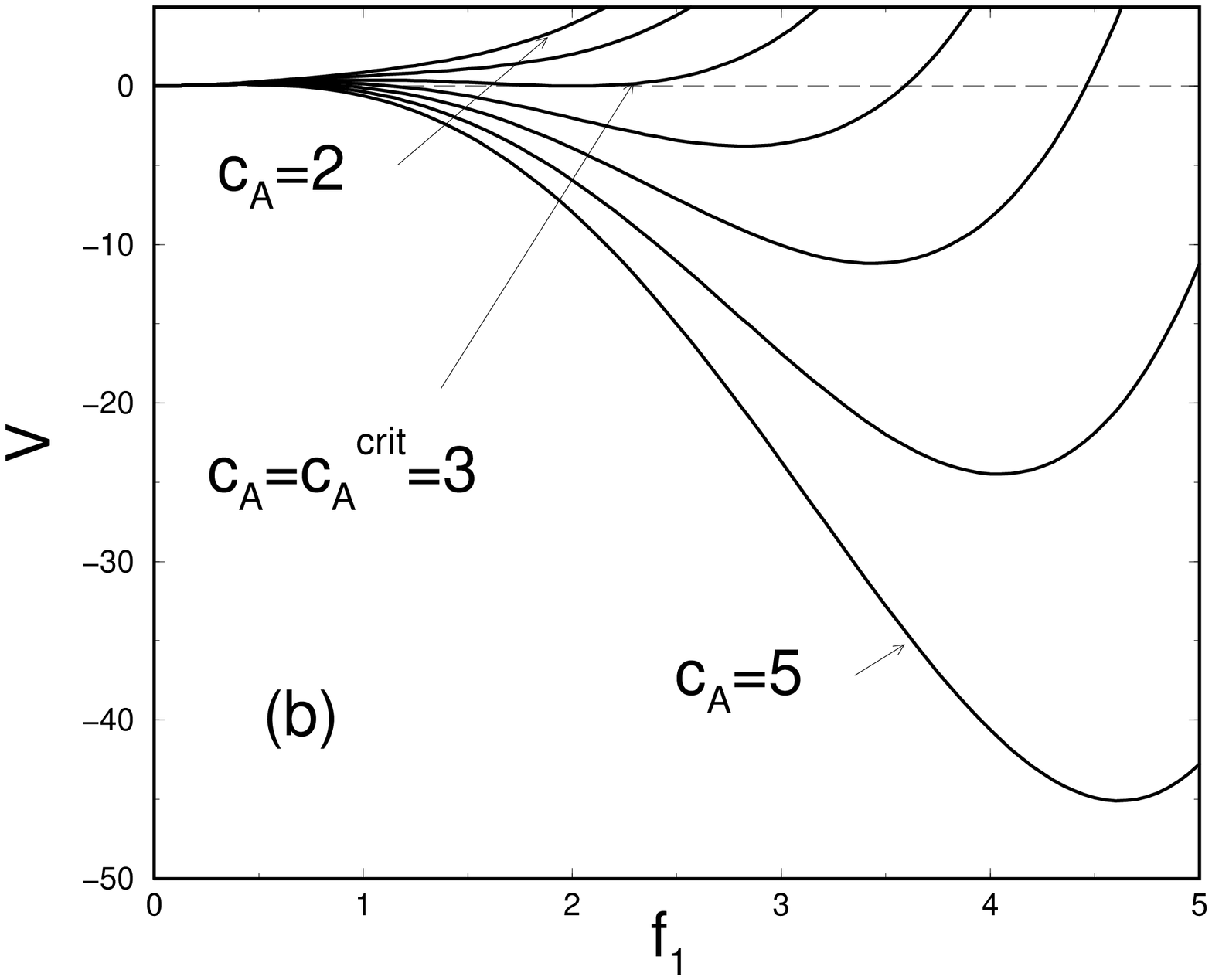,height=9cm,width=8cm,angle=-90,bbllx=3.cm,bblly=2.cm,bburx=21.cm,bbury=23.5cm}}}
\caption{\footnotesize 
(a): $c_{A}$ dependence of the Higgs masses for the pure $c_{A}$
minimum.
(b): Variation of the potential with $f_{1}$ for the parameter values in 
(\ref{posmass}) and different values of $c_{A}$ from 2 to 5 in steps of 0.5.}
\end{figure} 
\begin{figure}
\centerline{\hbox{
\psfig{figure=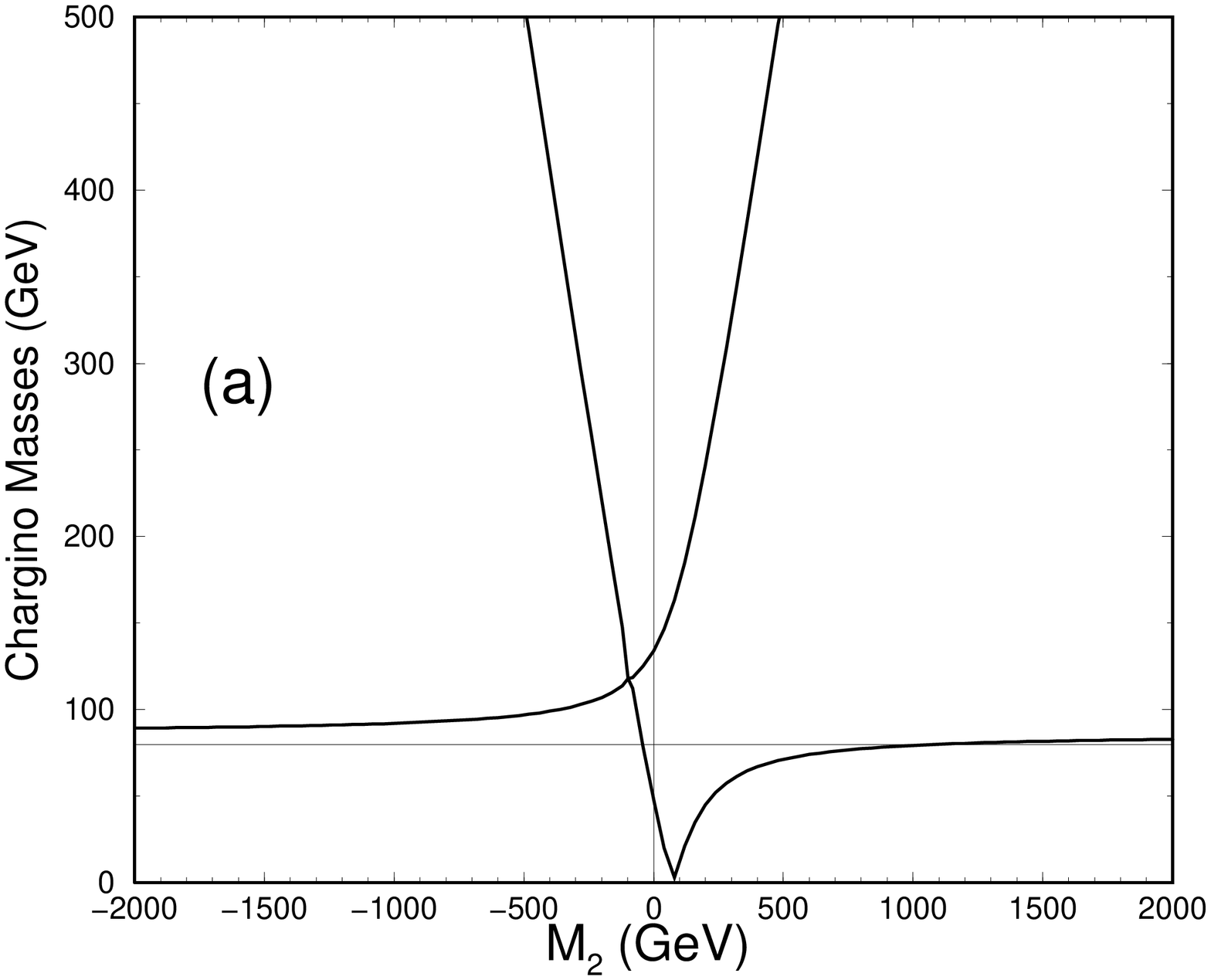,height=9cm,width=8cm,angle=-90,bbllx=3.cm,bblly=2.cm,bburx=21.cm,bbury=23.5cm}
\psfig{figure=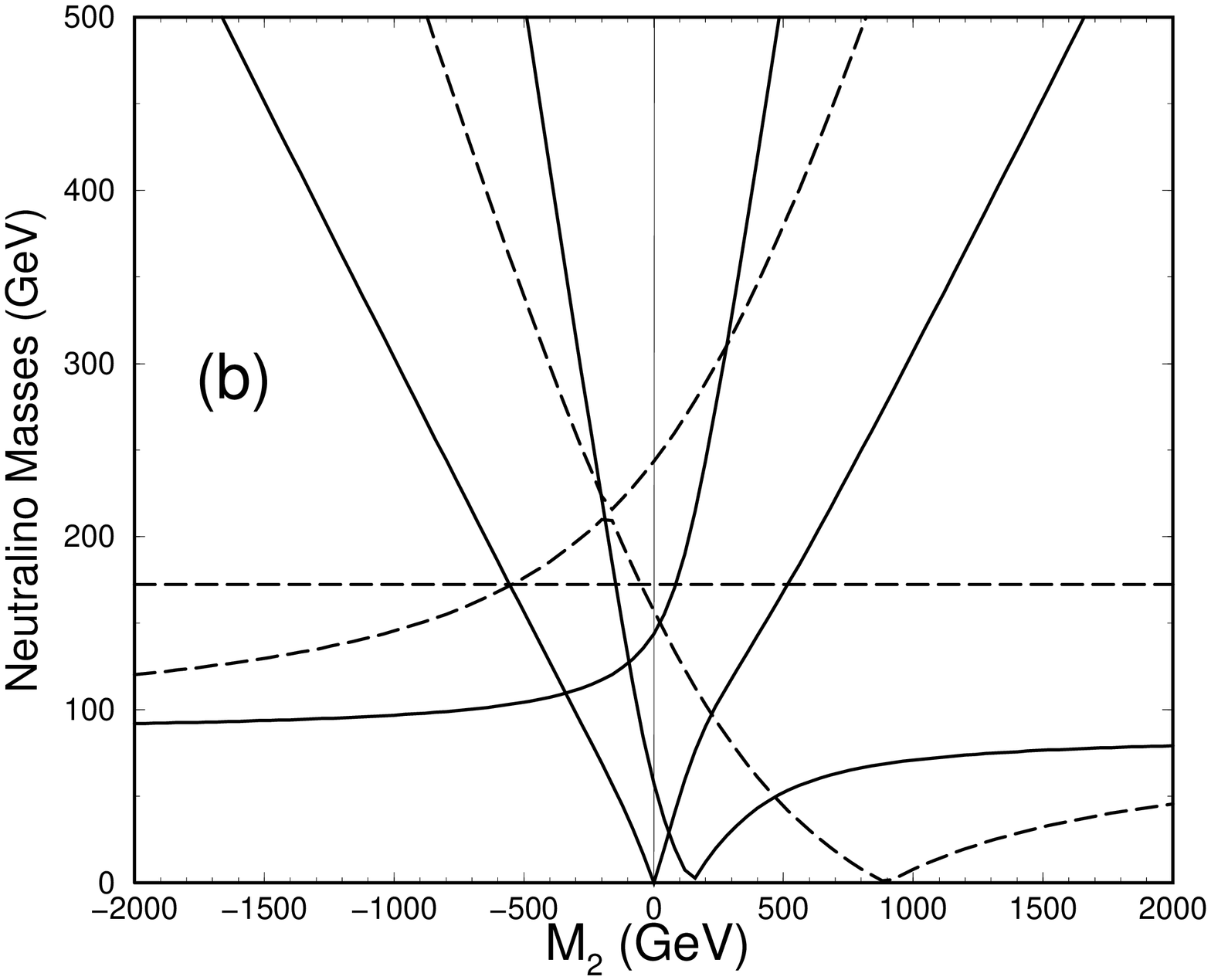,height=9cm,width=8cm,angle=-90,bbllx=3.cm,bblly=2.cm,bburx=21.cm,bbury=23.5cm}}}
\caption{\footnotesize 
Variation of the chargino masses (a) and the  neutralino masses (b) 
with the $SU(2)_{L}$ gaugino mass $M_{2}$ in the large $c_{A}$ minimum.}
\end{figure} 

In Fig.~4~(b) we investigate the $f_{1}$ dependence of the dimensionless potential   
for different values of $c_{A}$ and for the mass parameters in (\ref{posmass}). 
For each value of $f_1$, $V$ is minimized with respect to $f_2$ and $f_s$.
The straight dotted line at $V=0$ serves as a reference to separate the
two distinct minima. For all $c_{A}<c^{crit}_{A}$, the global minimum is at 
$f_{1}=f_2=f_s=0$.
 For $c_A>c_A^{crit}=3$ the minimum at $f_{1}\neq 0$ becomes the true minimum
and the gauge symmetry is broken. Passage of the system from one minimum to the
other requires quantum tunneling through the barrier. Presumably, as the universe
cooled it would have first been stuck in the local minimum, and could have
eventually tunnelled to the global minimum, with implications for baryogenesis
\cite{BG}. As is clear from Fig.~4~(b), the height of the barrier is very small
compared to the depth of the minimum for the large values of $c_A$ required to get
small enough $\alpha_{Z-Z'}$. In that case there is no danger of a large
supercooling and the transition can proceed without posing a cosmological
problem\footnote{We thank P.J. Steinhardt for a discussion on this point.}.
However, a detailed discussion of the cosmological implications of this model is
beyond the scope of this paper. 

In summary, the negative soft mass-squared parameters in the hybrid minimum
introduce a splitting among the fields for small $c_{A}$.  
The gap between $f_{1}$ and $f_{2}$ decreases gradually as a function of $c_A$,
which indicates that large values of $c_A$ are required to  
obtain a sufficiently small $Z-Z'$ mixing angle.
In the case of the pure trilinear coupling minimum, there is no bias from the 
soft mass-squared parameters and one can obtain a small mixing angle in a reasonable 
range of $c_{A}$ values. However, in the large $c_{A}$ limit the two minima have the 
same limiting properties solely determined by the value of the trilinear coupling.

In Fig.~5 we plot the chargino and neutralino masses as a function of the
$SU(2)_{L}$
gaugino mass $M_{2}$ (with $M_1$ and $M_1'$ as dictated by universality) in the 
large $c_{A}$ minimum (\ref{largecAlimit}). Fig.~5~(a) 
shows the chargino masses together with the LEP lower bound. If
$M_{2}\simgt 
1100\,GeV$ or $M_{2}\simlt  -40\,GeV$, $m_{\chi_{1}}$ is above the LEP bound. For 
$M_{2}\rightarrow\infty$, $m_{\chi_{1}}$ approaches $\mu_{s}$ from below, and for 
$M_{2}\rightarrow -\infty$, $m_{\chi_{1}}$ approaches $\mu_{s}$ from above.

In Fig.~5~(b), we show the $M_{2}$ variation of the neutralino masses in the large 
$c_{A}$ minimum. In this scenario, the neutralino 
mass matrix takes a simple form if $Q_1=Q_2\equiv Q$ and $\tan\beta=1$. 
The matrix decomposes in two $3\times 3$ matrices [in the basis
($\tilde{B},\tilde{W}_3,\tilde{H}_2^0\sin\beta-\tilde{H}_1^0\cos\beta$),
($\tilde{B}',\tilde{H}_1^0\sin\beta+\tilde{H}_2^0\cos\beta,\tilde{S}^0$)
].
The first of them has a 
$2\times 2$ submatrix identical to the chargino mass matrix. For $g_1=0$ the 
three eigenvalues are exactly equal to $M_1$ and $m_{\tilde{\chi}_{1,2}^\pm}$.
The presence of a non-zero $g_1$ slightly changes the picture, with the 
deviations largest when $M_1$ is close to $m_{\tilde{\chi}_{1,2}^\pm}$.
This behaviour is shown in Fig.~5~(b) where these particular three eigenvalues are 
singled out by solid lines. The second $3\times 3$ matrix 
has one eigenvalue equal to $2\mu_s$, independent of the gaugino masses. The other 
two eigenvalues are:
\begin{equation}
\label{zeroneut}
m_{\tilde{\chi}^0}=\frac{1}{2}\left[
M_1'+\mu_s\pm\sqrt{(M_1'-\mu_s)^2+12{g_1'}^2Q^2v^2}
\right].
\end{equation}
These three eigenvalues are plotted in Fig.~5~(b) with dashed lines.
For $M_1'\mu_s=3{g_1'}^2Q^2v^2$ one of the neutralino masses from (\ref{zeroneut})
goes to zero.
If the lightest chargino is to satisfy the LEP bound, the LSP is 
the $\chi^{0}_{1}$ neutralino.

\section{Large S Scenario}
Unless $g_1'Q_S$ is large, $M_{Z'}\gg M_Z$ requires $s \gg v$.
In that case it is convenient to examine the $U(1)'$ breaking 
first, separately from $SU(2)\times U(1)$ breaking, which will represent only a 
small correction. The breaking of the $U(1)'$ is triggered by the running of 
the soft mass $m_S^2$ towards negative values in the infrared. As a result the 
singlet gets a VEV [see eq.~(\ref{min3})]
\begin{equation}
\label{mzp}
s^2\simeq \frac{- 2 m_S^2}{{g_1'}^2Q_S^2}.
\end{equation}
That is, $M_{Z'}^2\sim -2 m_S^2(\mu=S)$.

The presence of this large singlet VEV influences, already at tree level,  
$SU(2)\times U(1)$ breaking, which is governed by the minimization conditions
(\ref{min1},\ref{min2}). Let us rewrite these conditions in a form that resembles 
the MSSM ones: 
\begin{eqnarray}
-m_3^2&=&\frac{1}{2}\left[
({\widetilde m}_1^2-{\widetilde m}_2^2)\tan 2\beta +
(M_Z^2-\frac{1}{2}h_s^2v^2)\sin 2\beta +\frac{1}{2} {g_1'}^2 (Q_1-Q_2) {\overline 
Q}_H v^2 \tan 2\beta\right],\nonumber\\ 
\label{mins2} 
\mu_s^2&=&\frac{{\widetilde m}_2^2\sin^2\beta
-{\widetilde m}_1^2\cos^2\beta}{\cos 2\beta}-\frac{M_Z^2}{2} 
-\frac{1}{2} {g_1'}^2 v^2 \frac{Q_1^2\cos^4\beta-Q_2^2\sin^4\beta}{\cos 2\beta},
\end{eqnarray}
where ${\widetilde m}_i^2=m_i^2+\frac{1}{2}{g_1'}^2Q_iQ_S s^2$ are the Higgs 
doublet soft 
masses corrected by the singlet VEV. The MSSM case would be recovered by setting 
$g_1'=h_s=0$ (but keeping $\mu_s$ fixed). The last term in (\ref{mins2}) is 
negligible if there is a 
cancellation in the off-diagonal $Z-Z'$ mass term (\ref{mix}). It is interesting to 
note that ${\widetilde m}_i^2+\mu_s^2$ (the effective Higgs mass terms in the 
potential) can be made negative by the $S$ contribution. Then
$SU(2)\times U(1)$ breaking can be triggered by the previous $U(1)'$ breaking. 
This is yet another alternative to the usual radiative breaking (although the 
breaking of the $U(1)'$ is itself radiative).

Turning back to the minimization equations (\ref{mins2}) one would
naturally expect $v^2\sim s^2$. The lightness of $M_Z$ 
compared to $M_{Z'}$ results from a cancellation of different mass terms of order 
$M_{Z'}$. The fine-tuning involved is then roughly given by the ratio $M_{Z'}/M_Z$.
It is illustrative to look at this cancellation in more detail. Consider first the
case of the MSSM. By naturalness one usually assumes that soft supersymmetry breaking 
mass parameters are 
at most of ${\cal O} (1\ TeV)$. If the soft mass parameters are as heavy as that
limit, then 
some fine-tuning is needed to get $M_Z$ one order of magnitude lower. We will 
take this as the limit of admissible (low-energy) fine-tuning. As already 
mentioned, the $\mu$ parameter in the MSSM does not naturally satisfy that 
constraint.
Consider next the simple model discussed in \cite{CL} with one single 
Higgs doublet. For large $s$, the cancellation to be enforced is
\begin{equation}
\label{cvla}
m_H^2 + \frac{1}{2}{g_1'}^2Q_HQ_Ss^2\sim {\cal O} (M_Z^2),
\end{equation}
where $m_H^2$ is the Higgs soft mass-squared parameter. One 
sees that $Q_HQ_S>0$ is needed for the 
cancellation to occur (note that, if $m_H^2>0$, corresponding to a non-radiative 
breaking of $SU(2)\times U(1)$, the opposite condition $Q_HQ_S<0$ would be required).
Substituting (\ref{mzp}) in (\ref{cvla}) and imposing $|m_H^2|\simlt{\cal O}(1\ 
TeV)^2$ one arrives at the condition
\begin{equation}
\left(\frac{M_{Z'}}{1\ TeV}\right)^2\simlt\frac{|Q_S|}{|Q_H|}.
\end{equation}
From this, it follows that the only possibility of having $M_{Z'}$ significantly 
heavier than $1\ TeV$ without excessive fine-tuning to keep $M_Z$ light is  
to have $|Q_H|\ll |Q_S|$. The natural possibility is to have $Q_H=0$; that would 
correspond to a $U(1)'$ trivially decoupled from electroweak breaking.

In the case of two Higgs doublets, we can similarly require that $\mu_s^2$, 
$m_3^2$ and ${\widetilde m}_{1,2}^2$ are at most ${\cal O} (1\ TeV)^2$. Then we 
arrive at the condition
\begin{equation}
\label{up1}
\left(\frac{M_{Z'}}{1\ TeV}\right)^2\simlt min\left\{
\frac{|Q_S|}{|Q_1|},\frac{|Q_S|}{|Q_2|},\frac{{g'}_1^2Q_S^2}{h_s^2}
\right\},
\end{equation}
and also
\begin{equation}
\label{up2}   
\left(\frac{A}{1\ TeV}\right)\left(\frac{M_{Z'}}{1\ 
TeV}\right)\simlt\frac{g'_1|Q_S|}{h_s}. \end{equation}
Consider first the case of $h_s^2$ small compared to ${g'}_1^2Q_S^2$. This means
$\mu_s$ is small compared to $M_{Z'}$ so that no restriction comes from the
$\mu_s$
condition in (\ref{up1}). In this case,
\begin{equation}
\left(\frac{M_{Z'}}{1\ TeV}\right)^2\simlt min\left\{
\frac{|Q_1+Q_2|}{|Q_1|},\frac{|Q_1+Q_2|}{|Q_2|}\right\}\equiv m \leq 2.
\end{equation}
There is a maximum value of $m$, ($m=2$, reached for $Q_1=Q_2$) and 
it is not possible to 
decouple the $Z'$ from electroweak breaking by a large hierarchy between the charges
because of the constraint $Q_1+Q_2+Q_S=0$. If $h_s^2$ is larger than
${g'}_1^2Q_S^2$ (that is, $\mu_s^2\gg M_{Z'}^2$), then 
the minimum in (\ref{up1}) goes to zero, which indicates that
$M_{Z'}\ll \mu_s\sim 1\ TeV$ to avoid a large fine-tuning.
We conclude that, to have $M_{Z'}\gg {\cal O}(1\ TeV)$ requires 
excessive fine-tuning in both cases. 
From (\ref{up2}) we also find a natural upper limit to 
impose on the $A$ parameter:
\begin{equation}
h_s A \simlt g'_1|Q_S|{\cal O}(1\ TeV).
\end{equation}
\begin{figure}
\centerline{\hbox{
\psfig{figure=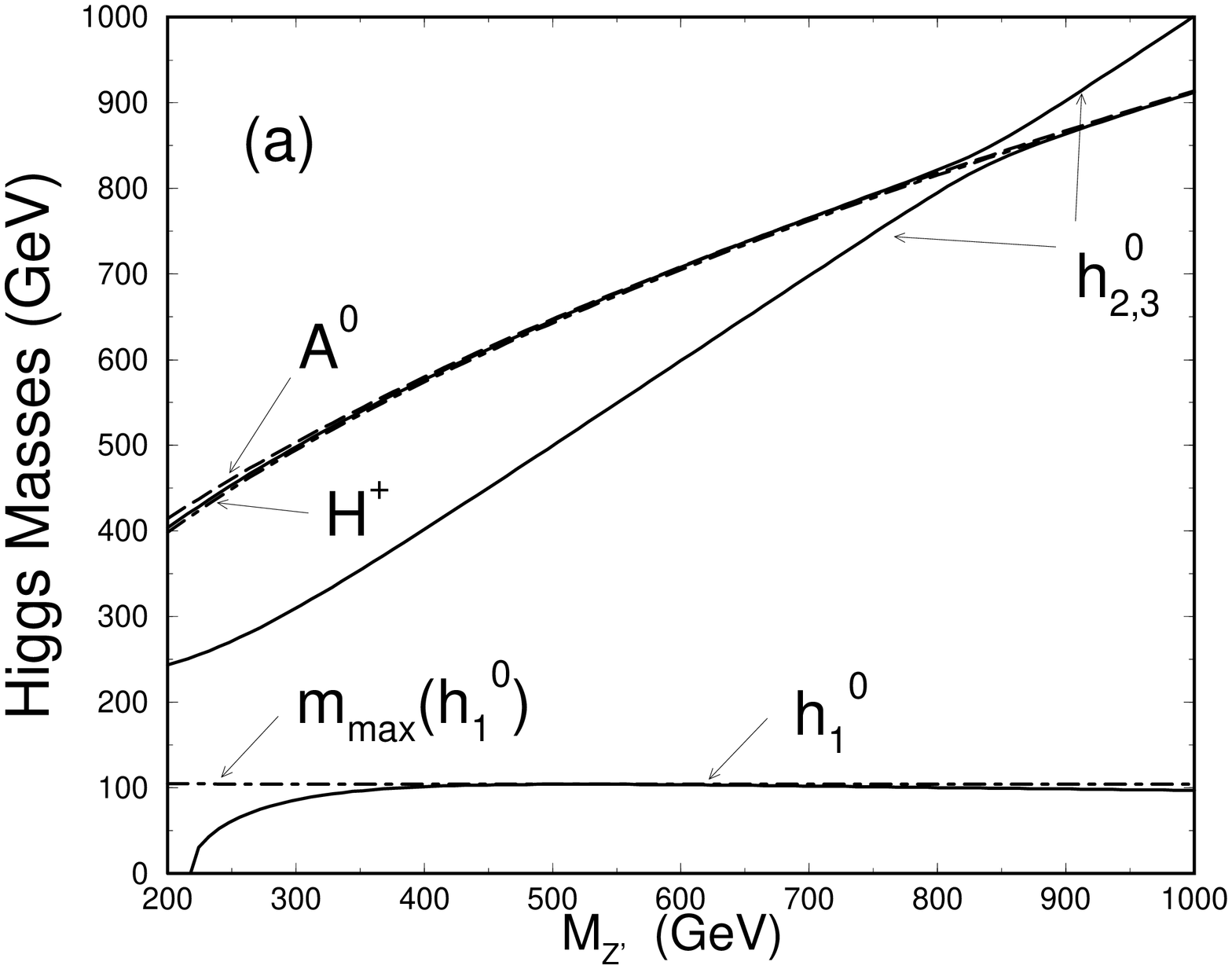,height=9cm,width=8cm,angle=-90,bbllx=3.cm,bblly=2.cm,bburx=21.cm,bbury=23.5cm}
\psfig{figure=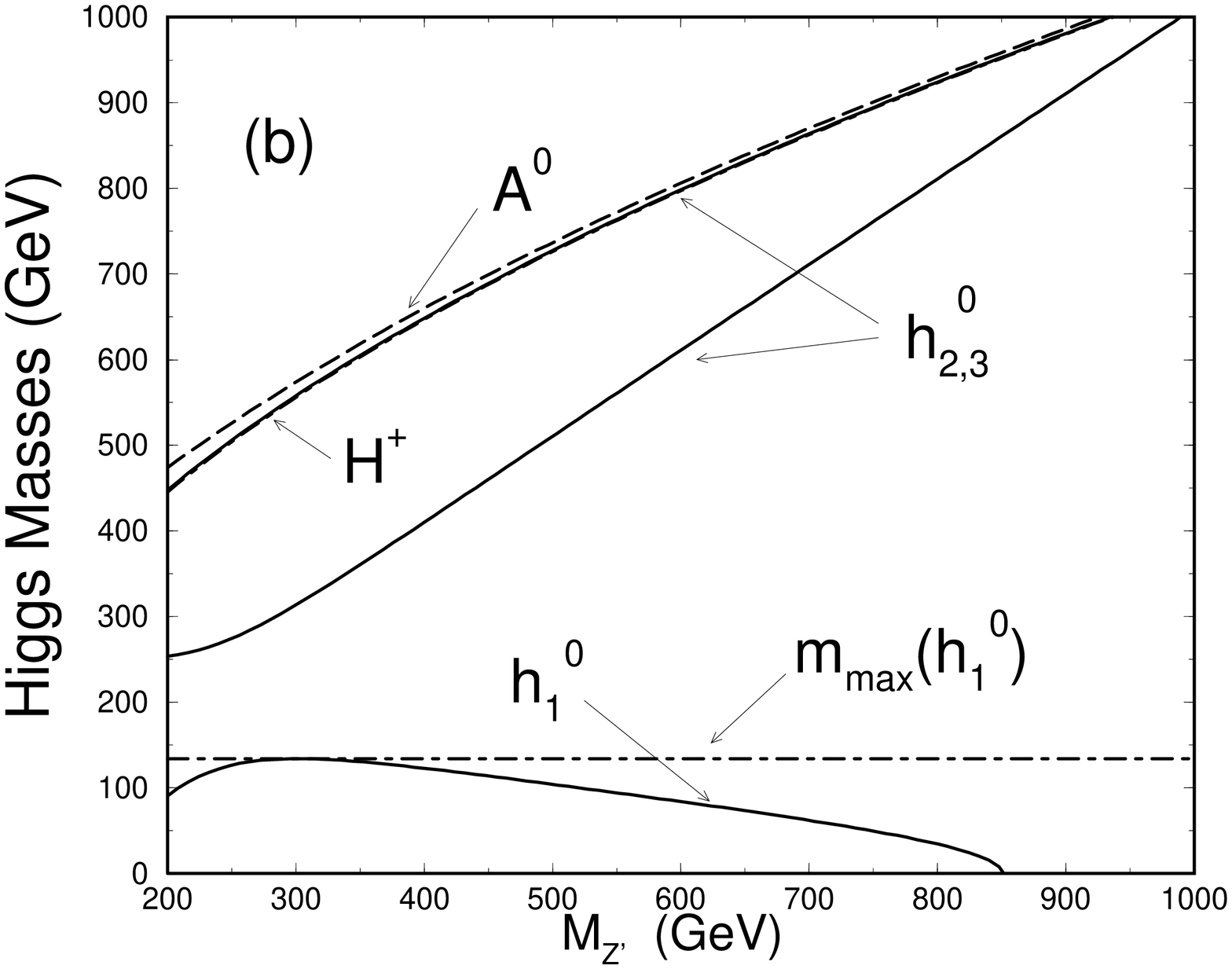,height=9cm,width=8cm,angle=-90,bbllx=3.cm,bblly=2.cm,bburx=21.cm,bbury=23.5cm}}}
\caption{\footnotesize Variation of the Higgs spectrum with $M_{Z'}$ for $A=500\ 
GeV$, $Q_1=Q_2=-1/2$ and (a) $h_s=0.5$ and $\tan\beta=1.5$; (b)  $h_s=0.7$ and 
$\tan\beta=1$. There are three scalars 
(solid lines) one pseudoscalar (dashed line) and one charged pair (dash-dotted). 
The horizontal dash-dotted line gives the bound (\ref{bound}).} 
\end{figure} 

In addition, the $Z-Z'$ mixing should be small enough. For moderate values of 
$M_{Z'}$ (say $500\ GeV$), small $Z-Z'$ mixing requires a 
small off-diagonal element in the $Z,Z'$ mass matrix. In fact,  
this matrix element vanishes for some value of $\tan\beta$ if $Q_1Q_2>0$. 
More precisely, 
$\theta_{Z-Z'}\leq 
\delta\theta$ if $\tan\beta$ is in the interval
\begin{equation}
\tan\beta\simeq \sqrt{Q_1/Q_2}\left[
1\pm\delta\theta\frac{G(Q_1+Q_2)}{4g'_1Q_1Q_2}\frac{M_{Z'}^2}{M_Z^2}
\right], \;\;\; (Q_1Q_2>0),
\label{ftb}
\end{equation}
(with $\theta_{Z-Z'}=0$ for the central value). This quantifies the 
fine-tuning required in $\tan\beta$.
This effect reduces the fraction of acceptable
parameter space for low values of $M_{Z'}$. The reduction is less 
important for a $Z'$ closer to the upper natural limit of $1\ TeV$, where a good 
cancellation in the off-diagonal $Z-Z'$ mass term is not required and eventually the 
condition $Q_1Q_2>0$ can be relaxed.

The pattern of the spectrum of physical Higgses in the large $s$ case is 
particularly simple. As discussed in Section~II, one neutral scalar $h_1^0$
remains
below 
the bound (\ref{bound}) and approaches the value (\ref{asympto}). The pseudoscalar 
$A^0$ mass, $m_{A^0}^2\simeq\sqrt{2}Ah_ss/\sin 2\beta$ is naturally expected to be 
large (unless $Ah_s$ is very small) and in that case, one of the neutral scalars
and the charged Higgs are approximately degenerate with $A^0$, completing a full
$SU(2)$ doublet $(H^0,A^0,H^\pm)$ not involved in $SU(2)$ breaking. The
lightest 
neutral scalar is 
basically the (real part of the) neutral component of the Higgs doublet which is
involved in the $SU(2)$ breaking and has then a very small singlet component. 
The third neutral scalar has mass controlled by $M_{Z'}$ and is basically the 
singlet. This mass pattern can be clearly seen in Fig.~6 for different 
choices of couplings and $U(1)'$ charges.

The mass of the lightest Higgs boson is of particular interest. 
The limiting value (\ref{asympto}) for $m_{h_1^0}$ can be bigger or smaller 
than the MSSM upper bound 
$M_Z^2\cos^22\beta$ depending on couplings and charge assignments. Note that the 
$D$-term contribution ${g'}_1^2{\overline Q}_H^2v^2$ in (\ref{bound}) is exactly 
compensated after integrating out $S$ and disappears in this decoupling limit. 
However, this exact cancellation does not take place for the $F$-term
contributions.
The behaviour of $m_{h_1^0}$ as a function of $M_{Z'}$ is shown in Fig.~7~(a)
for two 
different cases. Horizontal dash-dotted lines give the upper bound [eq.~(\ref{bound})], 
the MSSM bound $M_Z|\cos 2\beta|$ (which is zero in the figure), and the
asymptotic 
value eq.~(\ref{asympto}) [to make the figure simpler the parameters have been 
chosen such that (\ref{bound}) and (\ref{asympto}) are the same in both cases].
Fig.~7~(a) shows an example for which the asymptotic value is bigger than the
MSSM 
upper bound. This value is approached slowly. After including 
subdominant terms ${\cal O}(m_A^2/M_{Z'}^2)$ in eq.~(\ref{asympto}), one obtains
\begin{figure}
\centerline{\hbox{
\psfig{figure=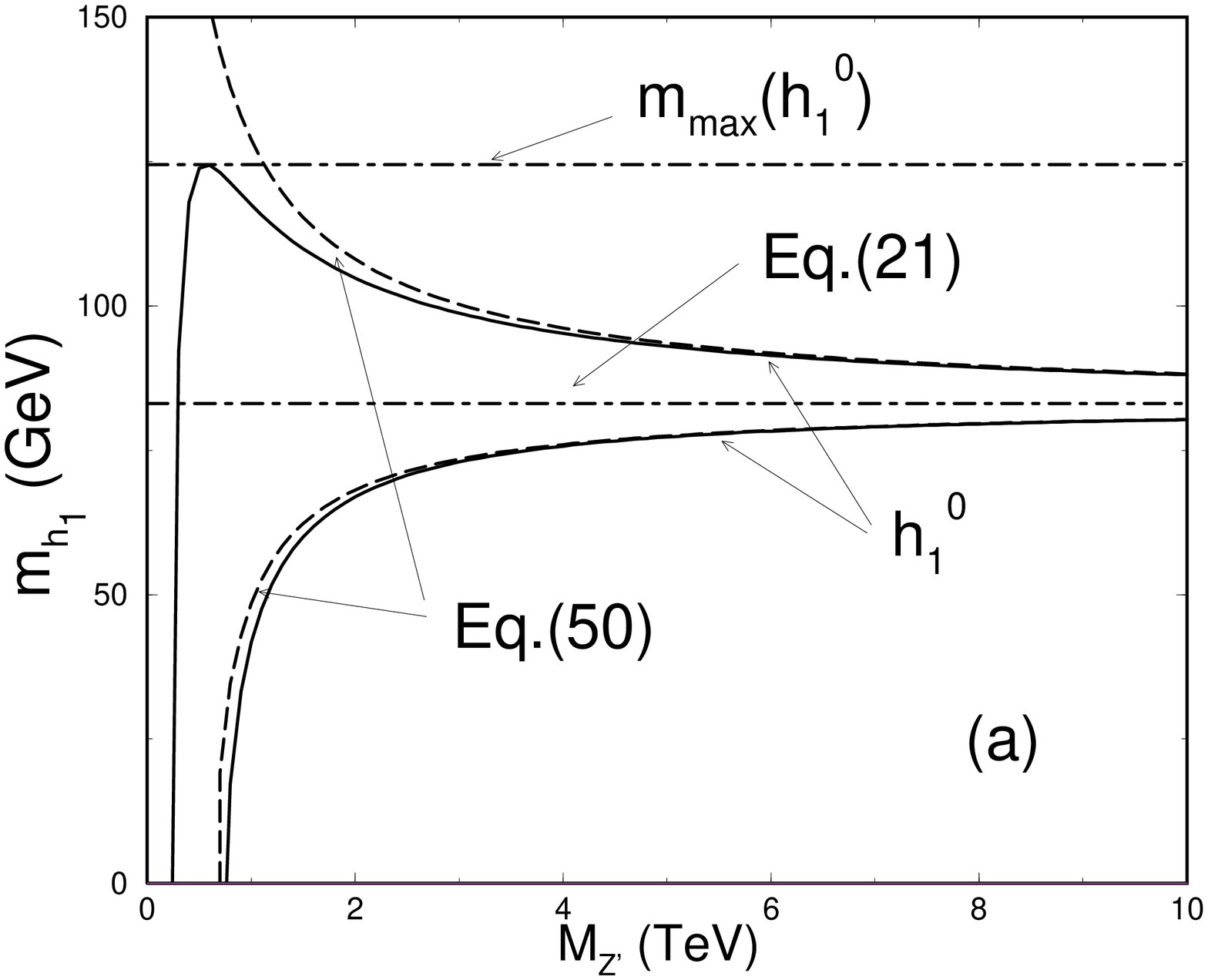,height=9cm,width=8cm,angle=-90,bbllx=3.cm,bblly=2.cm,bburx=21.cm,bbury=23.5cm}
\psfig{figure=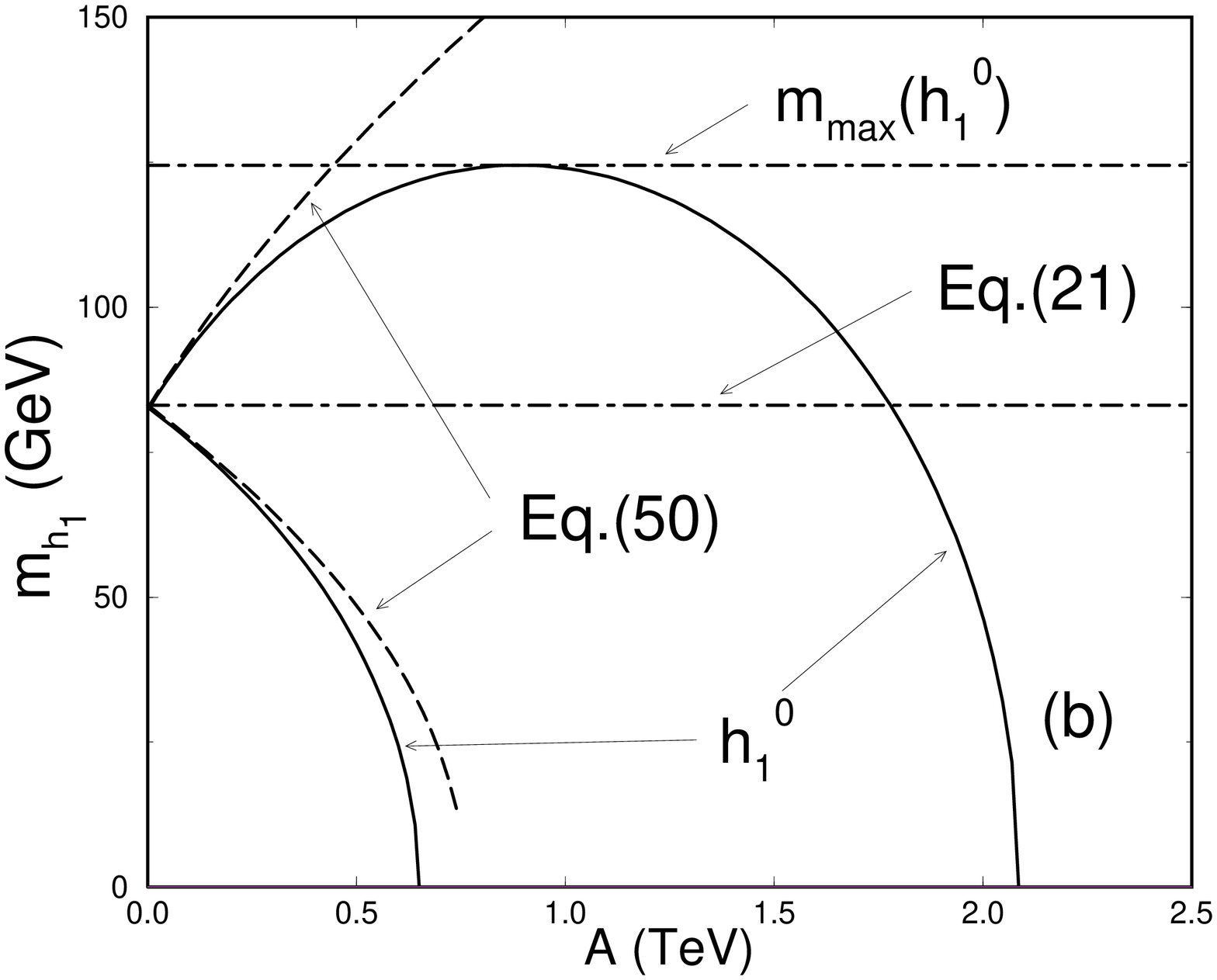,height=9cm,width=8cm,angle=-90,bbllx=3.cm,bblly=2.cm,bburx=21.cm,bbury=23.5cm}}}
\caption{\footnotesize 
Mass of the lightest Higgs scalar, $h_1^0$, (solid line): (a) as a function of 
$M_{Z'}$ showing the decoupling limit $s\gg v$ for two different cases with  
$A=500\ GeV$,  $\tan\beta=1$. The upper curve has $Q_1=Q_2= -3/5$,
$h_s\simeq 0.6$ and the lower $Q_1=Q_2=-1$ and $h_s\simeq 0.3$; (b)
as a function of $A$ for $M_{Z'}=1\ TeV$ in the two same cases. Dashed  
and dash-dotted lines give different mass bounds and limits as discussed in the 
text.} \end{figure}

\begin{eqnarray}
\label{asimpr}
m_{h_1^0}^2&\rightarrow & M_Z^2\cos^22\beta+
h_s^2v^2\left[\frac{1}{2}\sin^2 
2\beta-\frac{h_s^2}{{g'}_1^2Q_S^2}-2\frac{{\overline Q}_H}{Q_S}\right]\nonumber\\
&+&\sqrt{2}h_s\frac{A}{{g'}_1^2Q_S^2s}(h_s^2+{g'}_1^2Q_S\overline{Q}_H)v^2\sin2\beta.
\end{eqnarray}
This approximation is represented by dashed lines in Fig.~7~(a) and gives 
$m_{h_1^0}$ rather precisely for large $M_{Z'}$. The sign of  
$K=h_s^2+{g'}_1^2Q_S\overline{Q}_H$ determines whether the asymptotic value is 
reached from below ($K<0$) or above ($K>0$). 

\begin{figure}[tbp]
\centerline{\hbox{
\psfig{figure=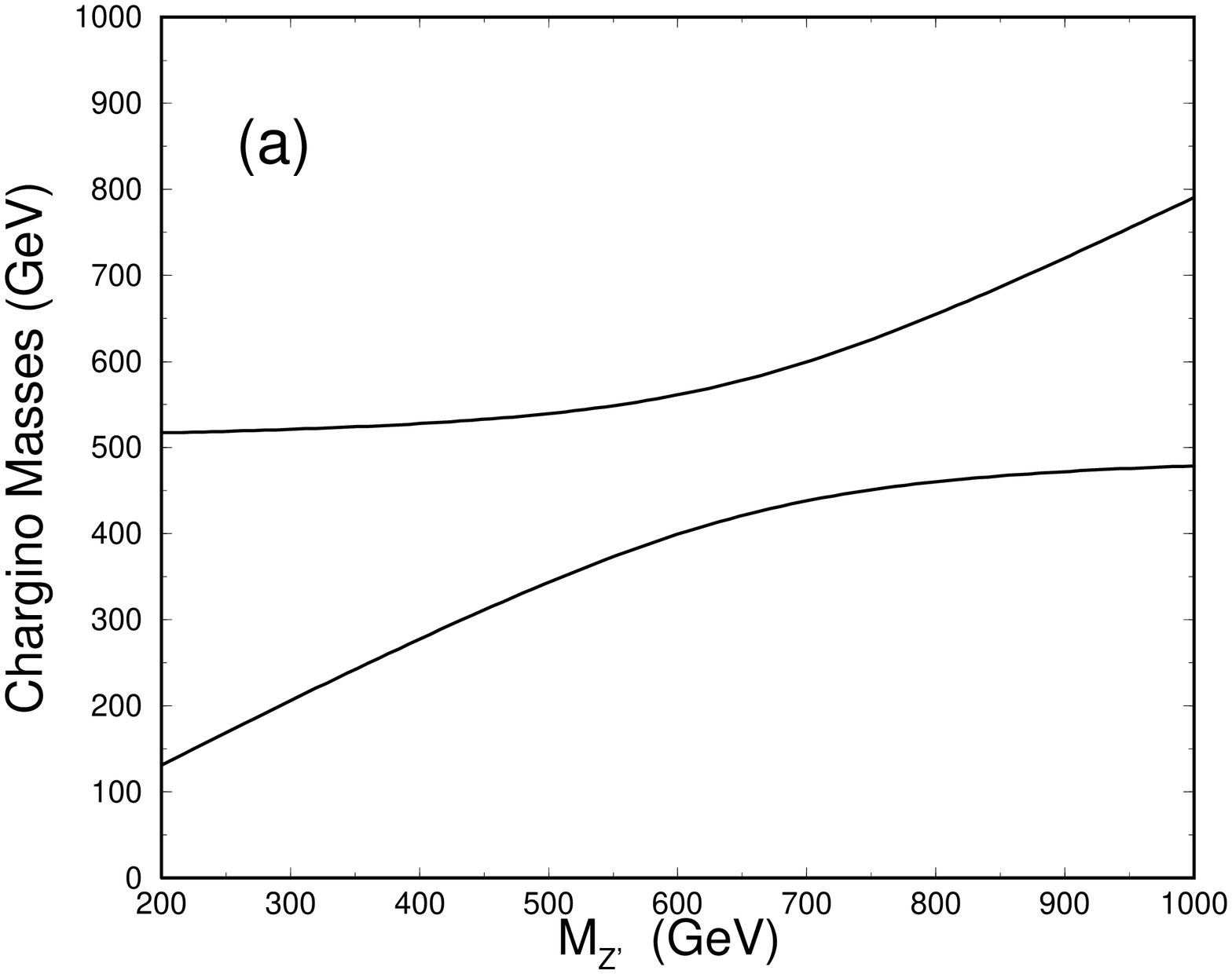,height=9cm,width=8cm,angle=-90,bbllx=3.cm,bblly=2.cm,bburx=21.cm,bbury=23.5cm}
\psfig{figure=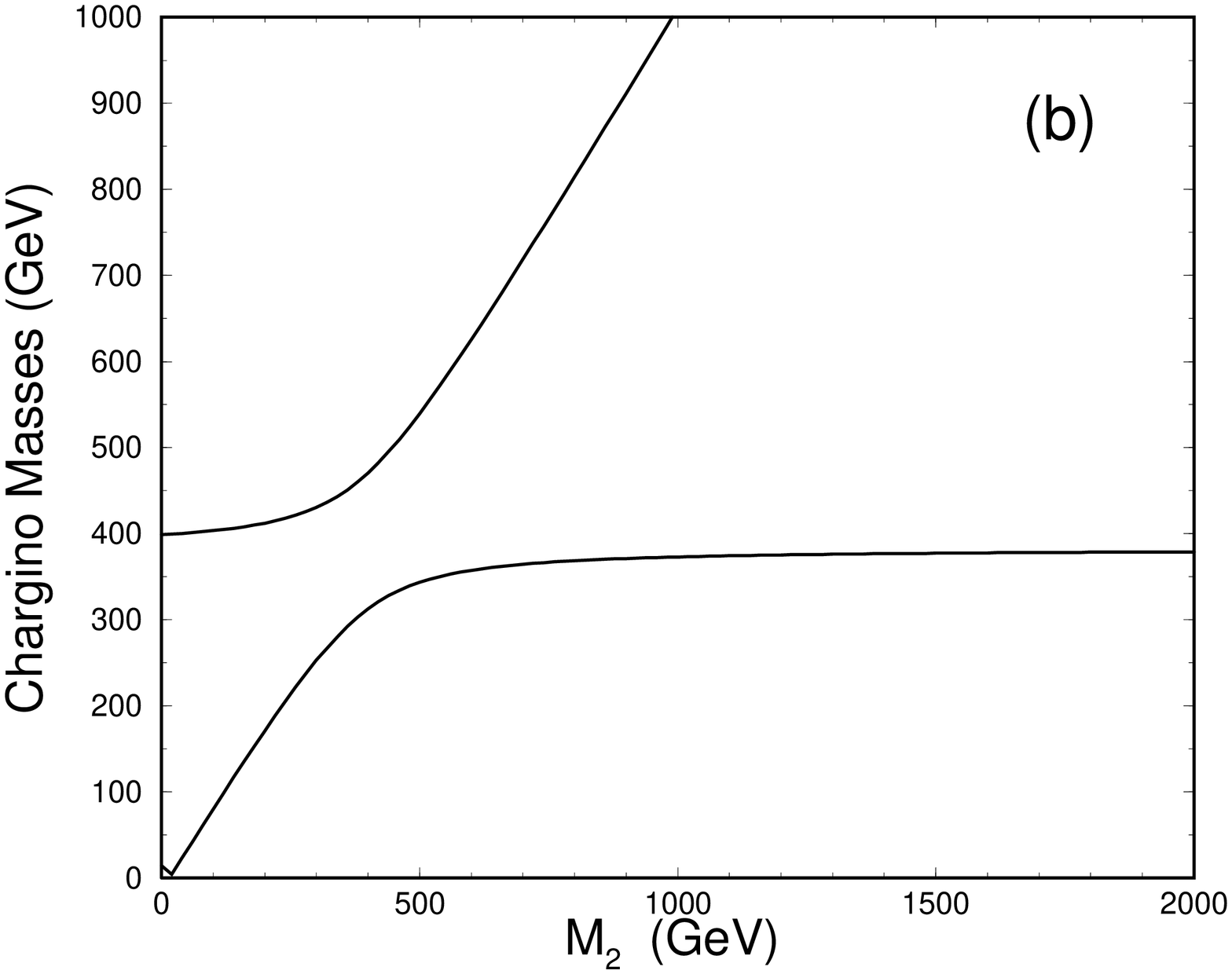,height=9cm,width=8cm,angle=-90,bbllx=3.cm,bblly=2.cm,bburx=21.cm,bbury=23.5cm}}}
\caption{\footnotesize 
Chargino masses (a) as a function of $M_{Z'}$ for $M_2=500\ GeV$;
(b) as a function of $M_2$ for $M_{Z'}=500\ GeV$.
(We fix $M_1$ and $M_1'$ by universality, $Q_1=Q_2=-1/2$, $h_s=0.5$ and
$\tan\beta=1.5$.).}
\end{figure}
\begin{figure}[tbp]
\centerline{\hbox{
\psfig{figure=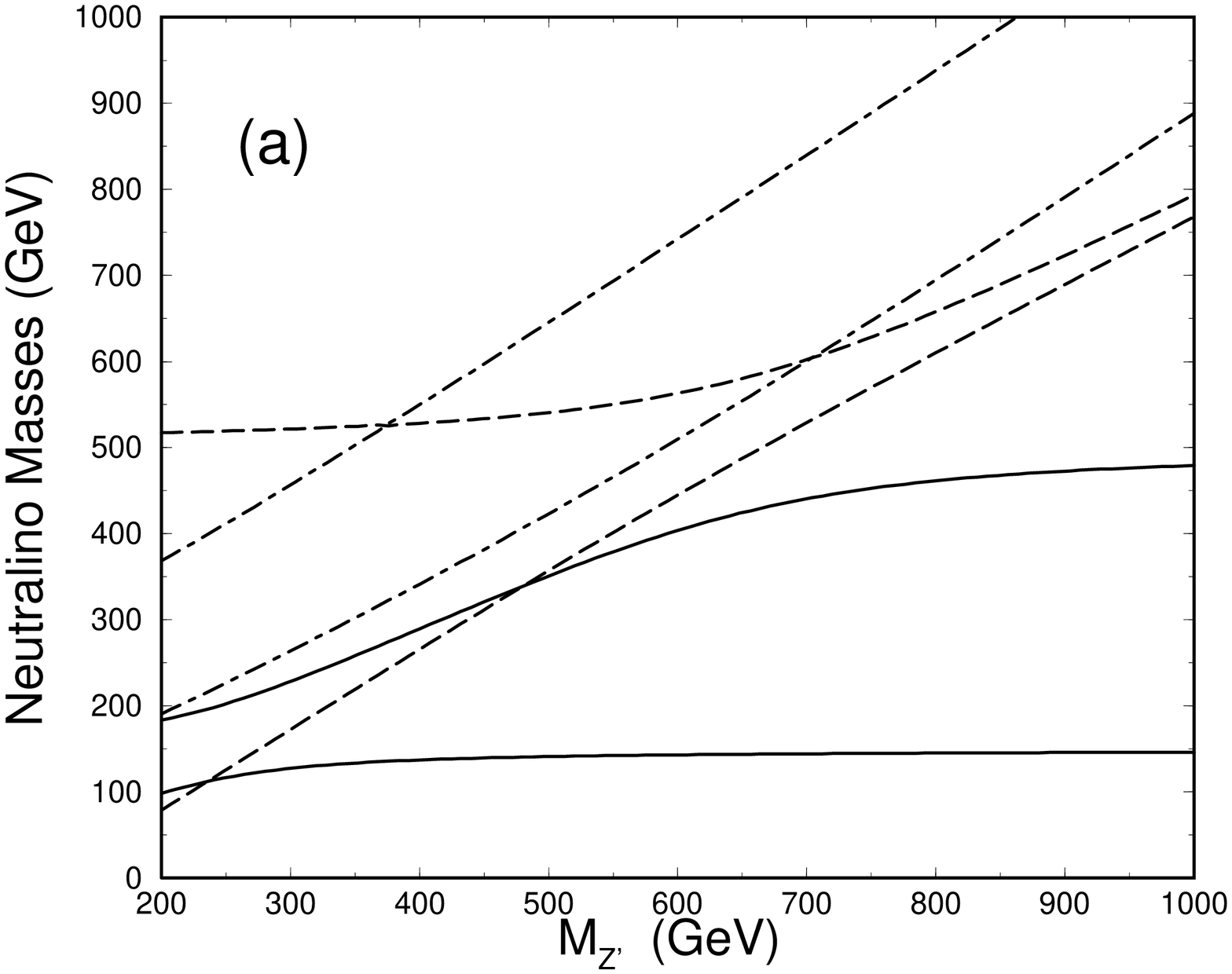,height=9cm,width=8cm,angle=-90,bbllx=3.cm,bblly=2.cm,bburx=21.cm,bbury=23.5cm}
\psfig{figure=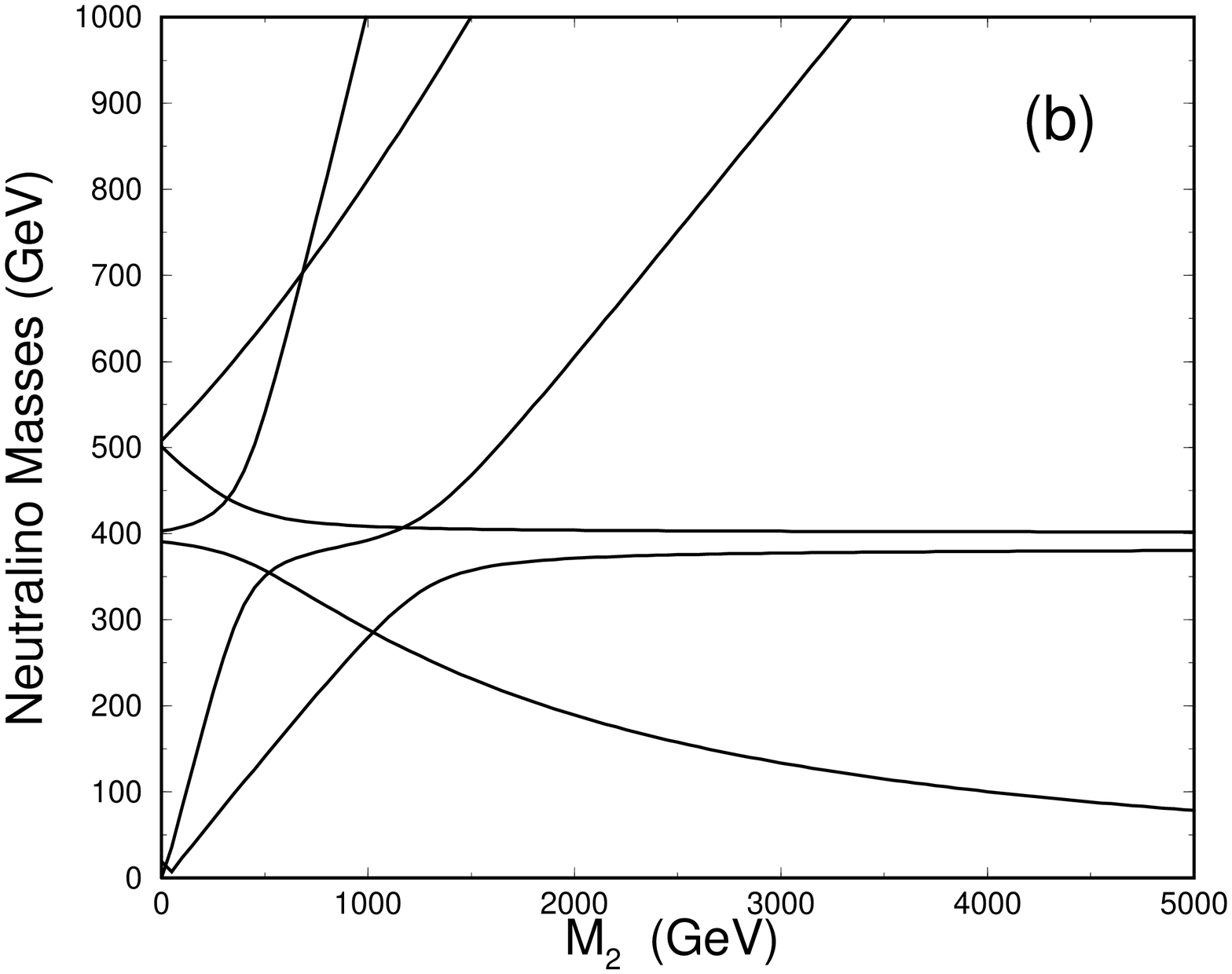,height=9cm,width=8cm,angle=-90,bbllx=3.cm,bblly=2.cm,bburx=21.cm,bbury=23.5cm}}}
\caption{\footnotesize 
Neutralino masses (a) as a function of $M_{Z'}$ for $M_2=500\ GeV$;
(b) as a function of $M_2$ for $M_{Z'}=500\ GeV$.
(We fix $M_1$ and $M_1'$ by universality, $Q_1=Q_2=-1/2$, $h_s=0.5$ and
$\tan\beta=1.5$).}
\end{figure}

In Fig.~7~(b), we show the dependence of $m_{h_1^0}$ on $A$ for fixed $M_{Z'}$ in 
the same two cases of Fig.~7~(a). For small $A$, we are in case {\em H1} of
Section~II
and the inequality (\ref{asympto}) holds (actually, it is saturated for the 
parameters chosen). The approximation (\ref{asimpr}) works well in that region. 
For larger $A$, $m_{h_1^0}$ increases or decreases depending on 
the sign of $K$. In both cases, when $A$ grows beyond $m_A\sim M_{Z'}$, 
$m_{h_1^0}^2$ drops to negative values (case {\em H3} in Section~II). The minimum
of the potential does not give a correct electroweak breaking; for 
sufficiently large $A$ the pattern of VEVs is similar to the one encountered
in the large $c_A$ case of the previous section, but the gauge boson masses 
would be much larger than the observed values. This 
behaviour differs from the MSSM where $m_{h_1^0}$ always increases with larger
$m_A$ until the upper bound is saturated. 

For some values of the parameters the large $s$ asymptotic value for $m_{h_1^0}^2$,
as computed from eq.~(\ref{asympto}), is negative. An example of this case is 
shown in Fig.~6~(b). In such cases there is an upper bound on $M_{Z'}$ beyond 
which the vacuum would be destabilized.

Next we show typical examples of the neutralino-chargino spectra. In Figs.~8~(a)
and 9~(a) we fix $M_2=500\ GeV$ (assuming that $M_1$ and $M_1'$ have values as 
dictated by universality) and show the dependence on the mass of the $Z'$ boson of 
the masses in the neutralino-chargino sector. Figs.~8~(b) and 9~(b) instead show 
the variation of the masses with $M_2$ for a fixed value of $M_{Z'}=500\ GeV$.
In Fig.~8~(a), we clearly see how the chargino masses are controlled by $M_2$
(fixed) and $\mu_s$ (growing linearly with $M_{Z'}$). For low $M_{Z'}$, meaning 
$\mu_s<M_2$, the lighter chargino mass follows $\mu_s$ and the heavier mass is 
nearly constant and equal to $M_2$. This role is interchanged after crossing
the $\mu_s\sim M_2$ region. The same behaviour is manifest in Fig.~8~(b), where
$\mu_s$ is kept constant and $M_2$ varies.

In Figs.~9~(a) and 9~(b) we plot the spectrum of neutralinos for the same two
cases. In Fig.~9~(a), for large $M_{Z'}$ we have $M_i^2,\mu_s^2\gg M_Z^2$ and 
the masses follow the pattern described in the discussion (case {\em N2}) after
eq.~(\ref{neutralinos}): the two lower (solid) curves asymptotically flattening 
approach $|M_1|$ and $|M_2|$ and correspond to $\tilde{B}$ and $\tilde{W_3}$ 
respectively. Then there are two (dashed) curves for the doublet Higgsinos 
tending to $|\mu_s|$ and finally two (dash-dotted) curves for two $\tilde{B}'-
\tilde{S}$ mixed states with masses $M_{Z'}\pm M'_1/2$. Also note that two
neutralino states follow closely the chargino pattern of Fig.~8~(a).

Concerning the nature of the LSP, the lightest neutralino is the natural 
candidate in these models. In particular, we see that the LSP is mostly $\tilde{B}$.
For large gaugino masses however, if ${M'}_1^2\gg M_{Z'}^2$, the lightest 
neutralino is the singlino $\tilde{S}$ whose mass is then of the order of $M_Z$. 
This possibility is realized in the case shown in Fig.~9~(b). 

\section{Renormalization Group Analysis}
We now turn to the renormalization group analysis 
of the model presented in Section~II to determine what boundary conditions 
at the string scale are required 
to reach the desired low energy parameter space as described in Sections~III and
IV.

As our model is motivated from string theory, we normalize the gauge 
couplings so that at the string scale 
\begin{eqnarray}
g^{0}_{3}=g^{0}_{2}=g^{0}_{1}=g_1'^{0}=g_{0}.
\label{unification}
\end{eqnarray}
In (weakly coupled) heterotic string 
theory this relation among the couplings is valid for the level one Ka\v c-Moody  
models\footnote{For the Ka\v c-Moody level $k \ne 1$ the relationship among the 
coupling constants is altered  by adding appropriate factors of $\sqrt k$  
in the equation.}. 
This is approximately consistent with the observed gauge coupling 
unification, which occurs at $M_G \simeq  3 \times 10^{16}\,GeV$, one order of
magnitude below $M_{String}\simeq 5 \times 10^{17}\,GeV$;  this difference
introduces a numerically small inconsistency in our analysis.

String models based on fermionic ($Z_2\times Z_2$) orbifold constructions 
\cite{ABK,NAHE,CHL} at a 
special point in moduli space possess the feature that the  couplings of the 
trilinear terms in the superpotential are equal for the fields whose 
string  vertex operators do not involve additional (real) world-sheet fermion
fields (with conformal dimension (1/2,1/2))\footnote{We 
thank G. Cleaver for a discussion on this point.}. In this case, the trilinear
coupling is $h^0_i=g_0{\sqrt 2 }$.  For a majority  of models all of the
observable fields are of that type. However, for fields whose string vertex
operators involve one such world-sheet fermion field the trilinear coupling is
$h^0=g_0$. Since in the vertex operator one can add at 
most {\it two} such world-sheet fermions (they now saturate (1,1) 
conformal dimension of the vertex operator), the trilinear 
coupling with one such field is $h^0=g_0/{\sqrt{2}}$ (which 
is then the smallest possible non-zero value of the Yukawa coupling).  
In the latter case, however, such fields usually correspond to exotics.

Thus, for  the sake of simplicity we assume that the boundary conditions for
the Yukawa couplings are given by
\begin{eqnarray}
\label{yukawainit}
h^{0}_{Q}=h^{0}_{S}=g_{0}\sqrt{2},
\end{eqnarray}
where $g_{0}$ is defined in (\ref{unification}).
Using the RGEs of the MSSM (i.e., in the absence of trilinear couplings of
$h_2$ to exotics), this  value of the Yukawa coupling $h^0_Q$ determines 
the value of $h_Q $ at $M_{Z}$.  When combined with the
VEV of $H_2$, which ensures the
correct electroweak symmetry  breaking vacuum, this result
 yields a  prediction for the top quark  mass  in the
range of $\sim 170-200$ GeV\cite{Faraggi}.

We first consider universal boundary conditions for the soft supersymmetry
breaking mass parameters at the string scale:  
\begin{itemize} 
\item Universal Scalar Soft Mass-Squared Parameters:
\begin{eqnarray}
m^{0\,2}_{1}=m^{0\,2}_{2}=m^{0\,2}_{S}=m^{0\,2}_{U}=m^{0\,2}_{Q}=M_{0}^{2}.
\label{univmass}
\end{eqnarray}
\item Universal Gaugino Masses:
\begin{eqnarray}
M^{0}_{3}=M^{0}_{2}=M^{0}_{1}=M_1'^0=M_{1/2}=C_{1/2}M_{0}.
\label{univgau}
\end{eqnarray}
\item Universal Trilinear Couplings:
\begin{eqnarray}
A^{0}=A^{0}_{Q}=C_{0}M_{0}.
\label{univtri}
\end{eqnarray}
\end{itemize}

As a second step, we will allow for nonuniversal initial conditions for the
trilinear 
couplings and the soft mass-squared parameters, such that in general
\begin{eqnarray}
\label{nonunivtri}
A^0_i&=&c^0_{A_{i}}M_{0},\\
\label{nonunivmass}
m^{0\,2}_i&=&c^{0\,2}_iM^2_0.
\end{eqnarray}

The one-loop RGEs for the parameters are 
presented in Appendix A.  We assume a minimal particle content, 
consistent with the superpotential (\ref{superpot}).  The renormalization 
group analysis of the model depends on 
the choice of $U(1)'$ charges of the theory, that enter the RGEs 
for the $U(1)'$ gauge coupling and gaugino.  In general, the $U(1)$ factors
have a small effect in the RGEs of the other
parameters due to the small magnitudes of the $U(1)$ gauge couplings and 
gaugino masses.  The $U(1)$ factors are neglected in the running of the 
parameters in the semi-analytic approach, which is often a good approximation.  
In the numerical analysis,
we choose for definiteness the $U(1)'$ charges $Q_{1}=Q_{2}=-1$, $Q_{L}=Q_{Q}=-1/2$, 
and most of the $U(1)$ factors are retained\footnote{The factors $S_1$ and 
$S_1'$ defined in Appendix A are not included in the numerical analysis of the 
RGEs, as discussed in Appendix C.}.  

We have solved the RGEs numerically, and
investigated the evolution of the parameters for a wide range of boundary
conditions.  With a
specific choice of the boundary conditions of the Yukawa couplings, we 
have obtained the numerical solutions for the
parameters at the electroweak scale as a function of the initial values of the
trilinear couplings and soft mass-squared parameters.  The results are 
qualitatively the same with other choices of initial values of 
the Yukawa couplings motivated by string theory; thus for definiteness we 
consider only the case with initial Yukawa couplings given by 
(\ref{yukawainit}).  To further 
our understanding of the evolution of these parameters, we have also 
derived semi-analytic solutions of the RGEs.  The numerical and  
semi-analytic solutions are presented and discussed in detail 
in Appendix B, and shown in some representative graphs.  With the numerical 
results (\ref{cAqsoln})-(\ref{cqsoln}), we are able to investigate
systematically the effect of the choice of boundary conditions 
on the evolution of the trilinear couplings and the soft mass-squared parameters. 

First, we consider the case of universal boundary conditions, as stated in
(\ref{univmass})-(\ref{univtri}), assuming that the only contributions to the
RGEs are from the MSSM supermultiplets, $\hat{S}$, and $Z'$ vector multiplet.  An
example of universal boundary conditions is presented in Figures
10-11, which show the scale dependence of the Yukawa couplings, the 
dimensionless trilinear
couplings, and the dimensionless soft mass-squared parameters, for $C_0=1.0$ 
and $C_{1/2}=0.1$. The dimensionless quantities are related to the physical
parameters by rescaling with $M_0$, which is defined in (\ref{M3/2}).

These graphs illustrate the general features of universal initial boundary
conditions: $h_Q(M_Z)$ 
is larger than $h_s(M_Z)$, $A_Q(M_Z)$ is larger than $A(M_Z)$ 
for $C_{1/2}\simgt 0.019\ C_0$,
and $m^2_2(M_Z)$ is negative while the other 
mass-squared parameters are positive at the electroweak scale.  This behaviour 
can be seen from the solutions (\ref{cAqsoln})-(\ref{cqsoln}), and 
the semi-analytic solutions discussed in 
Appendix B.  These solutions also demonstrate that the initial value of 
the gaugino mass parameter $M_{1/2}$ directly controls the splitting of 
the low energy values of the trilinear couplings and the 
mass-squared parameters.

These results indicate that the values of the low energy parameters 
obtained with universal boundary conditions at the string scale (and assuming
no exotic supermultiplets) do not 
lie within the phenomenologically acceptable region of parameter space.  
The large trilinear coupling scenario of Section~III requires $c_A \gg c^2_1 \sim
c^2_2 \sim 
c^2_S$ at the electroweak scale, which clearly does not follow from Figure 11.  
The scenario of Section~IV also does not result from universal initial 
conditions; Figure 11(b) demonstrates that while $m_2^2(M_Z)$ is negative,
$m_S^2(M_Z)$ is positive, so the singlet does not develop the 
large VEV necessary for this minimum.

\begin{figure}[tbp]
\centerline{\hbox{
\psfig{figure=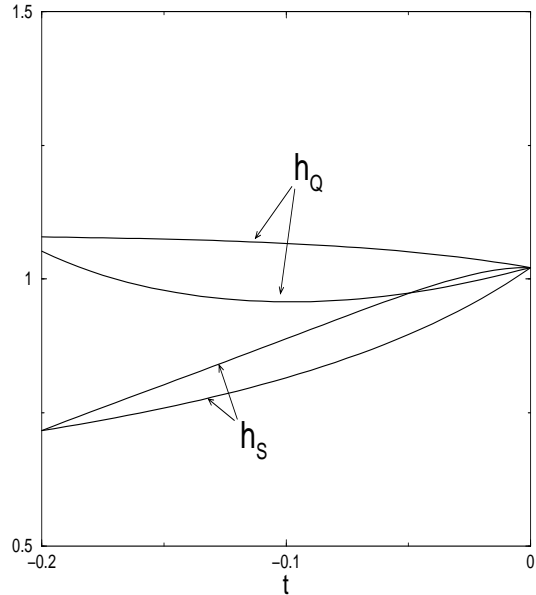,height=9cm,width=8cm,angle=-90,bbllx=3.cm,bblly=3.cm,bburx=21.cm,bbury=24.cm}}}
\caption{\footnotesize 
 Scale dependence of the Yukawa couplings (bold curves are for exact solutions). 
$t=\frac{1}{16\pi^{2}}\ln\frac{\mu}{M_{String}}$, such that $t\sim-0.2$ at the
electroweak scale.}
\end{figure} 
\begin{figure}[tbp]
\centerline{\hbox{
\psfig{figure=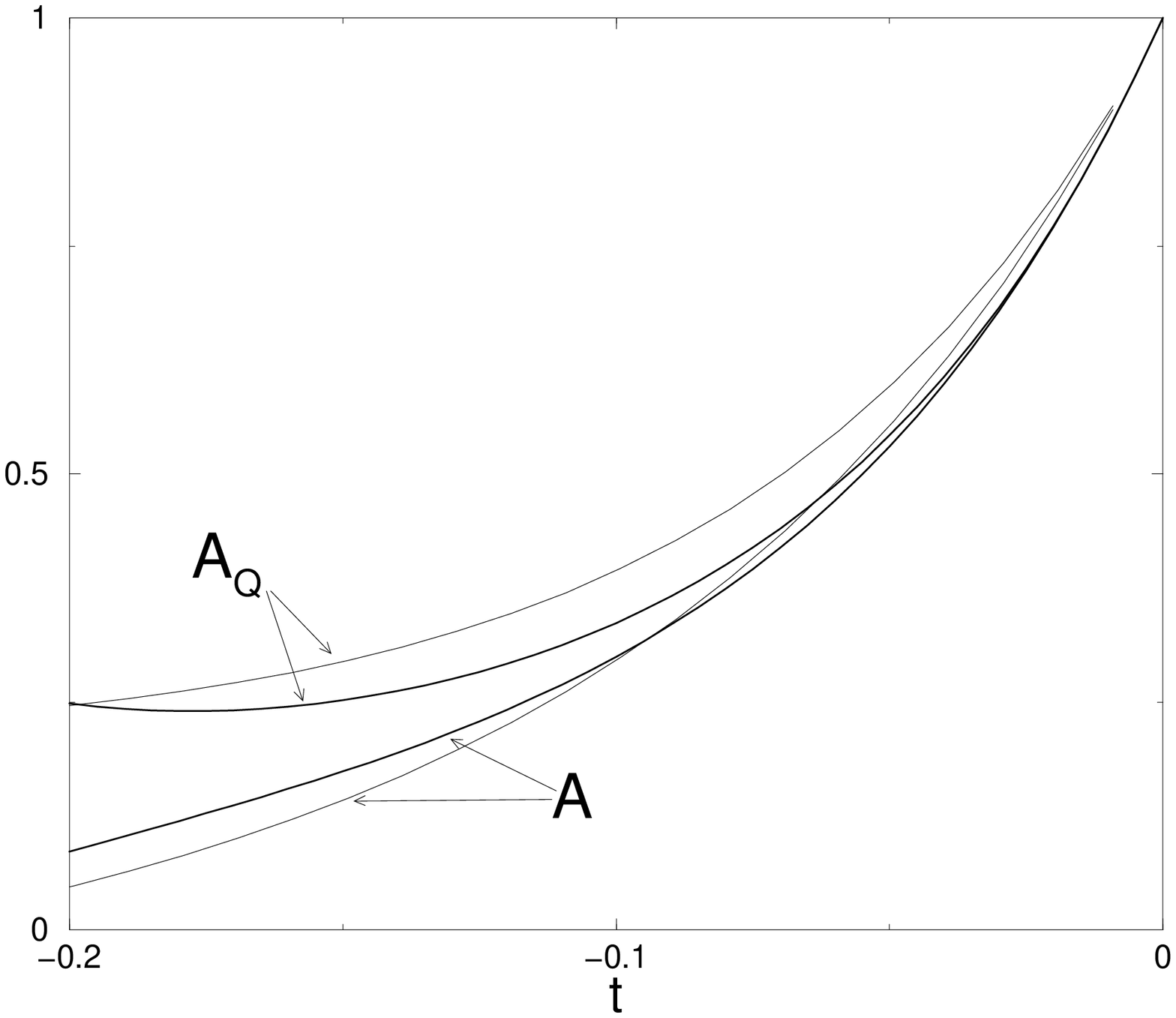,height=9cm,width=8cm,angle=-90,bbllx=3.cm,bblly=3.cm,bburx=21.cm,bbury=24.cm}
\psfig{figure=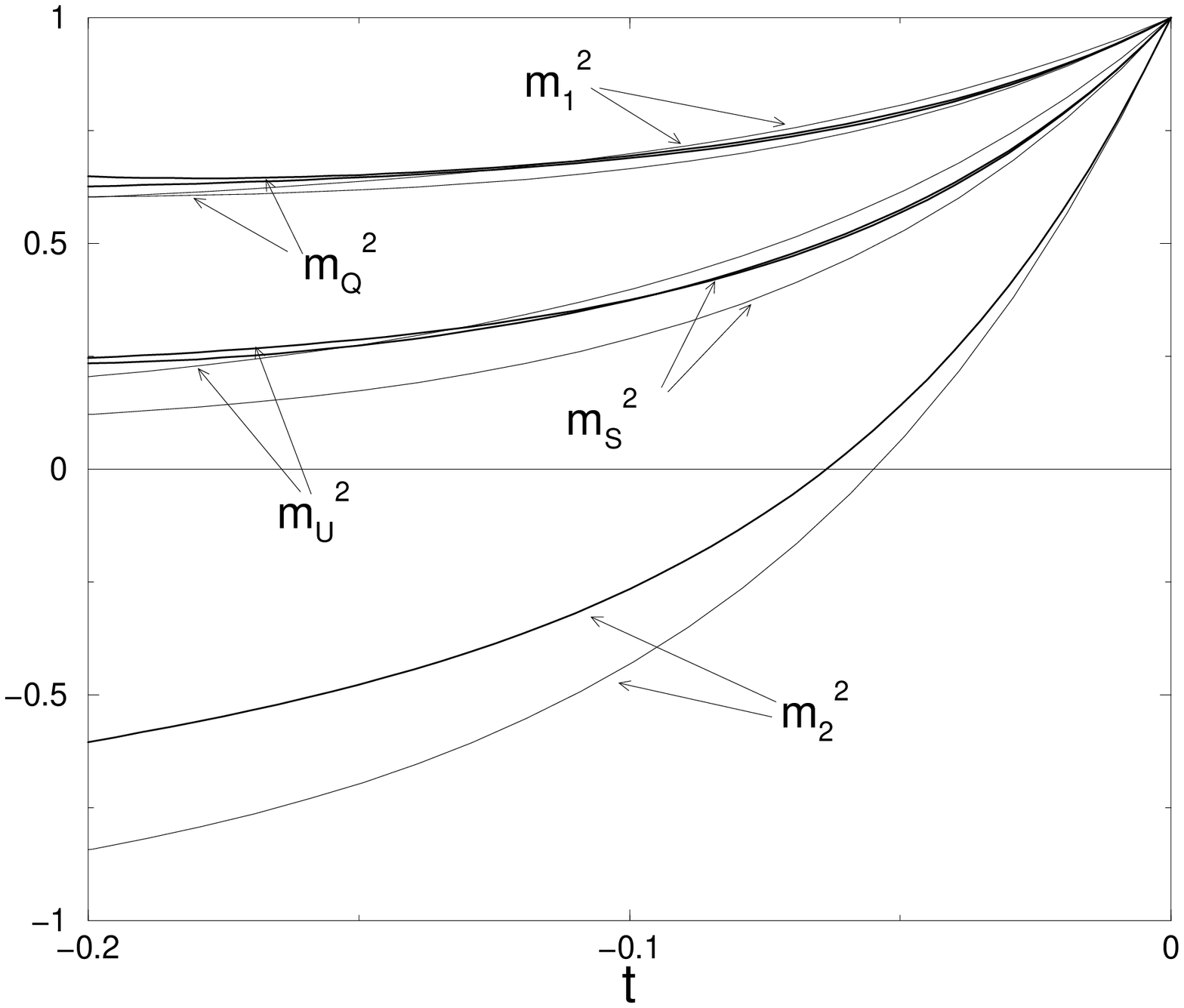,height=9cm,width=8cm,angle=-90,bbllx=3.cm,bblly=3.cm,bburx=21.cm,bbury=24.cm}}}
\caption{\footnotesize 
 Scale dependence of the dimensionless trilinear coupling parameters (a) and 
 soft mass-squared parameters (b). Bold curves
are for exact solutions.}
\end{figure} 

Therefore, we must relax our assumptions of universality and/or of no exotics 
to reach the desired low energy parameter space.  We first consider the
possibility of nonuniversal (but of the same order of magnitude) 
trilinear couplings and soft
 mass-squared parameters at the string scale.  In most cases, we must choose
$M_{1/2}$ small compared to other soft
masses at the string scale.  The value of $M_{1/2}$ must also be chosen to
satisfy the phenomenological bounds on the chargino masses and the gluino
masses at the electroweak scale. 
The boundary conditions are chosen to avoid a
dangerous color breaking minimum  \cite{CCB}, which could result from negative squark
mass-squares or large
values of $A_Q$ at the electroweak scale\footnote{Moderate trilinear terms
involving squarks
may be allowable because the charge-color breaking minimum may be only local
or, if global, may be separated from the standard-like minimum by a large
barrier. Whether a color and charge breaking global minimum  
is allowed depends on the cosmological history and on the tunneling rate from the 
standard-like minimum \cite{KLS}.}.  Negative squark mass-squares (including
both the
supersymmetric and soft breaking contributions) are always unacceptable, because
they imply that the standard-like minimum is an unstable saddle point.  
We present several illustrative examples of nonuniversal boundary
conditions, the resulting low energy parameters, and the relevant physical
quantities in Tables I-VI.  For economy of presentation, we display the low
energy values of the $SU(3)$ and $SU(2)$ gaugino masses ($M_3$, $M_2$) explicitly 
in the last line of each table, and do not present the values of $M_1$ and 
$M_1'$, which follow from the assumption of universal gaugino
masses [see (\ref{gauginosoln}) and (\ref{univgau})].

In Table I, we present a set of examples of boundary conditions that lead 
to the special case of the large trilinear coupling scenario in which 
$c^2_1(M_Z)=c^{2}_2(M_Z)=c^{2}_S(M_Z)=1.0$, and $c_A(M_Z)=5.0$.  This 
special case, chosen for definiteness to address the effect of the large 
value of $c_A$, has the $Z-Z'$ mixing angle identically zero, as 
discussed in Section~III.  In each case, the initial values of the gaugino 
mass and the trilinear couplings must be chosen such that
$A$ takes a large value  compared to the soft mass-squared parameters at the
electroweak 
scale. This can be obtained 
either by choosing $A^0_Q$ negative, choosing $A^0$ much larger than
$A^0_Q$, or taking $M_{1/2}$ negative.  The initial
values of the soft mass-squared parameters also must be chosen carefully so that
$m_1^2=m_2^2=m_S^2$ at the electroweak scale, which clearly is very fine-tuned. 
In each example, the 
initial values of the parameters are much larger than the low energy values of 
the soft mass-squared parameters of the singlet and the Higgses.  

In Table II, we present examples of the more general case of the 
large $c_A$ minimum in which the
 magnitudes of the soft supersymmetry breaking mass-squared parameters are not
exactly equal at the electroweak scale.  The
first example II(a) has the values $c_A(M_Z)=5.0$, $c_1^2(M_Z)=1.1$, 
$c_2^2(M_Z)=0.9$, and $c_S^2(M_Z)=1.0$; these small deviations in the 
low energy values of the soft mass-squared parameters yield a mixing angle around
$10^{-2}$, which may be barely allowable for $M_{Z'} \sim 200\,GeV$. 
Smaller mixing may be obtained for larger values of $c_A$, such as $c_A(M_Z)=10.0$, as 
shown in II(b), with the
values of the dimensionless low energy soft mass-squared parameters as above. 
Example II(c) also
has $c_A(M_Z)=10.0$, but $c_1^2(M_Z)=1.5$, $c_2^2(M_Z)=0.5$, and
$c_S^2(M_Z)=1.0$, and the mixing angle is again of ${\cal{ O}}(10^{-2})$. 
These examples are presented to emphasize the 
increase in the hierarchy between the values of the parameters at the 
string scale and the low energy values with the increasing value of $c_A$. 
The comparison of examples II(b) and II(c) demonstrates the fine-tuning required
at the string
scale (as well as at the electroweak scale) for this scenario.  The values
of the soft mass-squared parameters at the string scale are very similar, 
yet the resulting low energy parameters yield quite different values for the
mixing angle.  

Table III shows examples that yield the hybrid minimum of the 
large $c_A$ scenario discussed in
Section~III, for $c_A(M_Z)=5.0$, $c_A(M_Z)=8.0$, and $c_A(M_Z)=10.0$ (cases
(a), (b), and (c), respectively). Large values of $c_A(M_Z)$ are needed to obtain
a small enough mixing
angle when the low energy soft mass-squared parameters differ in magnitude or sign.  
This in turn causes the values of the parameters at the string scale 
to be much larger in magnitude than those at the electroweak scale, similar to 
the results presented in Table~II.

In Table IV, we present examples of boundary conditions that lead to the 
case (large $s$ scenario) described in Section~IV.  The initial values of
the parameters are chosen to lead to the negative value of $m^2_S$ at the electroweak
scale required for this scenario.  In addition, we choose values of the squark 
soft mass-squared parameters such
that the masses of the squarks will not be made negative when adding the
large $U(1)'$ D-term contribution (\ref{sqdterm}).
In this case, $M_{Z_2}=1\,TeV$, $\tan\beta=1$, and the $Z-Z'$ mixing angle is 
zero; the last two results are due to our assumption that
$c^2_1(M_Z)=c^2_2(M_Z)$, which requires fine-tuned boundary conditions. 
In addition, $m^2_S$ is negative at low energies, while the other 
soft mass-squared parameters are positive.  This requires taking the initial 
values of the parameters very large relative to the low energy values, 
and choosing $m^{0\,2}_2$ large compared to the initial values of 
the other soft mass-squared parameters. 
In this minimum, the chargino mass constraint is satisfied as 
long as $|M_{1/2}|$ is 
chosen large enough.

Table V presents more typical examples of boundary conditions which lead to
the large $s$  minimum with $M_{Z_2}=1\,TeV$, $\tan\beta=2$ and a nonzero
mixing angle.  
In each example, the initial values of the mass parameters are larger (by a factor
5-10) than 
the low energy values.  In comparison with the results of Table IV, in most
cases the 
magnitude of $m^{0\,2}_2$ need not be taken as large relative to 
the other soft mass-squared parameters, because in this case $m^2_2$ is allowed to 
be negative at the electroweak scale.

In Table VI, we present examples which lead to a case of the large $s$ 
minimum with a lighter $Z'$ mass ($\sim 700\,GeV$), a nonzero mixing
angle and $\tan\beta=1.4$.  This case has a different choice of $U(1)'$ 
charges $Q_1=-1$, $Q_2=-1/2$.  Once again, the initial values of the
parameters are larger than the values of the parameters at the electroweak scale.
As in Table V, $m^2_2(M_Z)$ is negative, so in most of the examples the
magnitude of $m^{0\,2}_2$ is comparable to the initial values of the other
soft mass-squared parameters.

\begin{table}[tbp]
\begin{center}
\begin{tabular}{||c||c|c||c|c||c|c||}
 & (a) $M_Z$ & $M_{String}$ & (b) $M_Z$ & $M_{String}$& (c) $M_Z$ &
$M_{String}$ \\ \hline
$m_1^2$ $(GeV)^2$& $(40.8)^2$ & $(710)^2$ & $(40.8)^2$ & $(413)^2$&$(40.8)^2$
&$(586)^2$ \\
$m_2^2$ $(GeV)^2$& $(40.8)^2$ & $(1460)^2$ 
&$(40.8)^2$&$(667)^2$&$(40.8)^2$ & $(986)^2$ \\
$m_S^2$ $(GeV)^2$& $(40.8)^2$ & $(1020)^2$&$(40.8)^2$&$(590)^2$&$(40.8)^2$ &
$(847)^2$ \\
$m_U^2$ $(GeV)^2$& $(450)^2$ & $(1010)^2$&$(150)^2$&$(373)^2$&$(180)^2$
&$(431)^2$ \\
$m_Q^2$ $(GeV)^2$& $(500)^2$ & $(714)^2$ &$(250)^2$&$(289)^2$&$(300)^2$ &
$(129)^2$\\
$A$ $(GeV)$& 204& $-$4230 &204&1080&204&2190\\
$A_Q$ $(GeV)$& 405 & 634&$-$175&667&$-$250&2270 \\
$M_{1/2}$ $(GeV)$&(578, 164) & 200 &($-$289, $-$82) &$-$100 &
($-$578, $-$164)&$-$200 
\end{tabular}
\end{center}
\caption{ \footnotesize Large $c_A$ minimum: $M_{Z_2}=196\,GeV$,
$\alpha_{Z-Z'}=0.0$, $\tan\beta=1.0$, $m_h=126\,GeV$, and $Q_1=Q_2=-1$. We
present the values of ($M_3$, $M_2$) at the electroweak scale.  The gluino
mass is $\mid\!M_3\!\mid$.}
\end{table}
\begin{table}[tbp]
\begin{center}
\begin{tabular}{||c||c|c||c|c||c|c||}
 & (a) $M_Z$ & $M_{String}$ & (b) $M_Z$ 
& $M_{String}$& (c) $M_Z$ & $M_{String}$ \\ \hline
$m_1^2$ $(GeV)^2$& $(42.8)^2$ & $(443)^2$ & $(19.9)^2$ & $(390)^2$&$(23.4)^2$
&$(392)^2$ \\
$m_2^2$ $(GeV)^2$& $(38.7)^2$ & $(819)^2$&$(18.0)^2$&$(775)^2$&$(13.5)^2$ &
$(776)^2$ \\
$m_S^2$ $(GeV)^2$& $(40.8)^2$ & $(640)^2$&$(19.0)^2$&$(568)^2$&$(19.1)^2$ &
$(569)^2$ \\
$m_U^2$ $(GeV)^2$& $(201)^2$ & $(456)^2$&$(245)^2$&$(458)^2$&$(245)^2$
&$(457)^2$ \\
$m_Q^2$ $(GeV)^2$& $(400)^2$ & $(399)^2$ &$(380)^2$&$(367)^2$&$(380)^2$ &
$(367)^2$\\
$A$ $(GeV)$& 204& 1070 &190&814&191&818\\
$A_Q$ $(GeV)$& $-$270 & 695&$-$290&398&$-$290&400\\
$M_{1/2}$ $(GeV)$&($-$434, $-$123)& $-$150 &($-$434, $-$123) &$-$150 & ($-$434,
$-$123)&$-$150
\end{tabular}
\end{center}
\caption{ \footnotesize  Large $c_A$ minimum: (a) $M_{Z_2}=196\,GeV$,
$\alpha_{Z-Z'}=7.8 \times 10^{-3}$, $\tan\beta=1.02$, $m_h=125\,GeV$; 
(b) $M_{Z_2}=196\,GeV$, $\alpha_{Z-Z'}=1.9 \times 10^{-3}$, $\tan\beta=1.01$, 
$m_h=130\,GeV$; (c) $M_{Z_2}=197\,GeV$,
$\alpha_{Z-Z'}=9.3 \times 10^{-3}$, $\tan\beta=1.03$,  $m_h=131\,GeV$.  
In all cases $Q_1=Q_2=-1$.}
\end{table}
\begin{table}[tbp]
\begin{center}
\begin{tabular}{||c||c|c||c|c||c|c||}
 & (a) $M_Z$ & $M_{String}$ & (b) $M_Z$ & $M_{String}$& (c) $M_Z$ & $M_{String}$ 
\\ \hline
$m_1^2$ $(GeV)^2$& $(36.5)^2$ & $(366)^2$ & $(23.1)^2$ & $(372)^2$&$(18.5)^2$
&$(350)^2$ \\
$m_2^2$ $(GeV)^2$& $-(36.5)^2$ & 
$(789)^2$&$-(23.1)^2$&$(809)^2$&$-(18.5)^2$ & $(702)^2$ \\
$m_S^2$ $(GeV)^2$& $-(36.5)^2$ & 
$(531)^2$&$-(23.1)^2$&$(542)^2$&$-(18.5)^2$ & $(512)^2$ \\
$m_U^2$ $(GeV)^2$& $(210)^2$ & $(471)^2$&$(208)^2$&$(489)^2$&$(180)^2$
&$(360)^2$ \\
$m_Q^2$ $(GeV)^2$& $(405)^2$ & $(410)^2$ &$(410)^2$&$(426)^2$&$(397)^2$ &
$(350)^2$\\
$A$ $(GeV)$& 183& 586 &184&581&185&683\\
$A_Q$ $(GeV)$& $-$310 & 116&$-$311&104&$-$301&240 \\
$M_{1/2}$ $(GeV)$&($-$434, $-$123)& $-$150 &($-$434, $-$123) &$-$150 & ($-$434,
$-$123)&$-$150
\end{tabular}
\end{center}
\caption{ \footnotesize Hybrid minimum: (a) $M_{Z_2}=200\,GeV$,
$\alpha_{Z-Z'}=3.4 \times 10^{-2}$, $\tan\beta=1.11$, $m_h=135\,GeV$; (b)
$M_{Z_2}=198\,GeV$,
$\alpha_{Z-Z'}=1.4 \times 10^{-2}$, $\tan\beta=1.04$, $m_h=134\,GeV$; (c)
$M_{Z_2}=197\,GeV$,
$\alpha_{Z-Z'}=9.3 \times 10^{-3}$, $\tan\beta=1.03$,  $m_h=133\,GeV$.  In all 
cases $Q_1=Q_2=-1$.}
\end{table}
\begin{table}[tbp]
\begin{center}
\begin{tabular}{||c||c|c||c|c||c|c||}
 & (a) $M_Z$ & $M_{String}$ & (b) $M_Z$ & $M_{String}$& (c) $M_Z$ & 
$M_{String}$ \\ \hline
$m_1^2$ $(GeV)^2$& $(430)^2$ & $(1260)^2$ & $(430)^2$ & $(804)^2$&$(430)^2$
&$(679)^2$ \\
$m_2^2$ $(GeV)^2$& $(430)^2$ & $(2350)^2$ &$(430)^2$&$(1400)^2$&$(430)^2$ &
$(1270)^2$ \\
$m_S^2$ $(GeV)^2$& $-(701)^2$ & $(1520)^2$&$-(701)^2$&$(696)^2$&$-(701)^2$ &
$(363)^2$ \\
$m_U^2$ $(GeV)^2$& $(450)^2$ & $(1670)^2$&$(425)^2$&$(786)^2$&$(425)^2$
&$(602)^2$ \\
$m_Q^2$ $(GeV)^2$& $(511)^2$ & $(1230)^2$ &$(475)^2$&$(441)^2$&$(495)^2$ &
$(100)^2$\\
$A$ $(GeV)$& 500& 2320 &500&863&500&416\\
$A_Q$ $(GeV)$& 190 &948& 339&$-$1170&363&$-$1810 \\
$M_{1/2}$ $(GeV)$&(289, 82) & 100 &(746, 212) &258 &(853, 242) &295
\end{tabular}
\end{center}
\caption{ \footnotesize Large $s$ minimum: $M_{Z_2}=1\,TeV$,
$\alpha_{Z-Z'}=0.0$, $\tan\beta=1.0$, $m_h=172\,GeV$, and $Q_1=Q_2=-1$.}
\end{table}
\begin{table}[tbp]
\begin{center}
\begin{tabular}{||c||c|c||c|c||c|c||}
 & (a) $M_Z$ & $M_{String}$ & (b) $M_Z$ & $M_{String}$& (c) $M_Z$ & 
$M_{String}$ \\ \hline
$m_1^2$ $(GeV)^2$& $(427)^2$ & $(670)^2$ & $(427)^2$ & $(670)^2$&$(427)^2$
&$(806)^2$ \\
$m_2^2$ $(GeV)^2$& $-(173)^2$ & $(1180)^2$
&$-(173)^2$&$(2280)^2$&$-(173)^2$ &
$(1480)^2$ \\
$m_S^2$ $(GeV)^2$& $-(704)^2$ &
$(210)^2$&$-(704)^2$&$(296)^2$&$-(704)^2$ &
$(669)^2$ \\
$m_U^2$ $(GeV)^2$& $(200)^2$ & $(861)^2$&$(310)^2$&$(1730)^2$&$(262)^2$
&$(1090)^2$ \\
$m_Q^2$ $(GeV)^2$& $(380)^2$ & $(676)^2$ &$(400)^2$&$(1170)^2$&$(362)^2$ &
$(806)^2$\\
$A$ $(GeV)$& 250& 1940 &250&$-$2230&250&2020\\
$A_Q$ $(GeV)$& $-$109 & 1640&125&$-$4330&278&$1630$ \\
$M_{1/2}$ $(GeV)$& ($-$289, $-$82)& $-$100 &(723, 205) &250 &(289, 82) &100 
\end{tabular}
\end{center}
\caption{ \footnotesize Large $s$ minimum: $M_{Z_2}=1\,TeV$,
$\alpha_{Z-Z'}=6.3\times 10^{-3}$, $\tan\beta=2.0$, 
$m_h=163\, GeV$, and $Q_1=Q_2=-1$.}
\end{table}
\begin{table}[tbp]
\begin{center}
\begin{tabular}{||c||c|c||c|c||c|c||}
 & (a) $M_Z$ & $M_{String}$ & (b) $M_Z$ & $M_{String}$& (c) $M_Z$ &
$M_{String}$ 
\\ \hline
$m_1^2$ $(GeV)^2$& $(207)^2$ & $(425)^2$ & $(207)^2$ & $(540)^2$&$(207)^2$
&$(510)^2$ \\
$m_2^2$ $(GeV)^2$& $-(354)^2$ & $(605)^2$ &$-(354)^2$&$(944)^2$&$-(354)^2$ &
$(711)^2$ \\
$m_S^2$ $(GeV)^2$& $-(499)^2$ &
$(212)^2$&$-(499)^2$&$(510)^2$&$-(499)^2$ &
$(455)^2$ \\
$m_U^2$ $(GeV)^2$& $(200)^2$ & $(331)^2$&$(262)^2$&$(692)^2$&$(242)^2$
&$(367)^2$ \\
$m_Q^2$ $(GeV)^2$& $(350)^2$ & $(246)^2$ &$(362)^2$&$(526)^2$&$(384)^2$ &
$(264)^2$\\
$A$ $(GeV)$& 250& 2180 &250&1800&250&1840\\
$A_Q$ $(GeV)$& $-$190 & 2140&303&1370&400&1380 \\
$M_{1/2}$ $(GeV)$&($-$463, $-$132)& $-$160 & (361, 103)&125 &(506, 144) &175 
\end{tabular}
\end{center}
\caption{ \footnotesize Large $s$ minimum: $M_{Z_2}=700\,GeV$,
$\alpha_{Z-Z'}=1.4\times 10^{-4}$, $\tan\beta=1.4$, $m_h=120\,GeV$, 
and $Q_1=-1$, $Q_2=-1/2$.}
\end{table}

In summary, without exotic particles it is necessary to invoke 
nonuniversal trilinear couplings and soft
mass-squared parameters at the string scale to reach either scenario.  In most
 cases, small initial values of the gaugino masses relative to the soft
mass-squared parameters 
are required, such that $M_{1/2} \ll m^0_i$.  It is also necessary
to have $  m^{0\,2}_i \gg m^2_i(M_Z) $ for the large trilinear coupling
scenario,
and for many of the examples that lead to the large $s$ minimum.  With these generic
features of the values of the parameters at the string scale, it is possible to 
reach the phenomenologically viable low energy parameter space with the minimal 
particle content. 

Another possibility is to add  to our 
model by considering exotic particles, as are expected in many string models.
One example involves color triplets 
$\hat{D}_{1}\sim (3, 1, Y_{D_{1}}, Q_{D_{1}})$ and $\hat{D}_{2}\sim (\bar{3}, 1,
Y_{D_{2}}, Q_{D_{2}})$ 
which couple to the singlet through the additional term in the superpotential
\begin{eqnarray}
\label{superpotex}
W=h_{D}\hat{S}\hat{D}_{1}\hat{D}_{2}.
\end{eqnarray}
The presence of these exotics affects the running of the $SU(3)$ and $U(1)$
gauge couplings. Taken by themselves they would destroy the gauge coupling 
unification\footnote{Small ${\cal O}(10\%)$ corrections to the RGE predictions,
which could be due to exotics, may even be desirable, due to the values of the
predicted unification scale and $\alpha_3$.}. Thus, one must assume that
$\hat{D}_{1,2}$ are associated with other exotics so that the gauge
unification is restored.  One example would be for $\hat{D}_i$ to be part of a
complete GUT supermultiplet.  Examples of anomaly-free models consistent with
gauge unification are given in Appendix C.  Clearly, the implications are very
model dependent. A precise numerical analysis of the associated renormalization
group equations of such models is beyond the scope of this paper.  However, it
is useful to consider the consequences of these exotics on the low energy
parameter space using a semi-analytic approach.  With the additional color
triplets, a large singlet VEV  is guaranteed with {\it universal boundary
conditions},
as $m_S^2$ is negative at the electroweak scale.  This was shown in \cite{CL}
in the limit in which the gaugino masses and trilinear couplings can be 
neglected.  The additional coupling of the singlet to the exotic triplets 
increases the overall weight driving $m_S^2$ negative in its RGE in analogy
with $m_2^2$, as discussed in Appendix B.  In contrast, the large trilinear
coupling scenario is more difficult to obtain in this case.  The presence of
the new trilinear coupling $A_D$ acts to lower the fixed point value of $A$
further, such that at low energies $A_D(M_Z) \sim A_Q(M_Z) \gg A(M_Z)$. 
Universal boundary conditions would not lead to this minimum; the
initial values of the trilinear couplings and the soft supersymmetry breaking
mass-squared parameters would have to be
chosen to invert this hierarchy and obtain similar values of $m_1^2$, $m_2^2$,
and $m_S^2$ at the electroweak scale.

\section{Conclusions}

In this paper, we  explored the  features of the
supersymmetric standard model with an additional non-anomalous $U(1)'$ 
gauge symmetry.  The
model is a ``minimal'' extension of the 
Minimal Supersymmetric Standard Model (MSSM), with  one  standard model singlet
chiral superfield $\hat S$
added to
the MSSM particle content.  
The  $U(1)'$ charges  are chosen  
to allow 
the  trilinear coupling  of $\hat S$ to the MSSM doublet
chiral superfields ${\hat H}_{1,2}$ in the superpotential. 
This choice of $U(1)'$ charges implies that the  bilinear coupling
 of the two doublets  ${\hat H}_{1,2}$ 
is absent; hence, there is no elementary $\mu$  parameter in the
superpotential. However, when $S$ (scalar component of $\hat{S}$) acquires a
nonzero vacuum expectation value  (VEV), this trilinear term generates 
an effective $\mu$  term, which leads to a natural solution of the $\mu$
problem.

The gauge structure, particle content, and
nature of the couplings of this type of model are  
key ingredients of a large class of  $N=1$ supersymmetric  string models
based on fermionic constructions ({e.g.}, $Z_2\times Z_2$ asymmetric orbifolds)
 at a particular point 
in moduli space. Within this  approach, we identified the minimal  particle
content and their couplings in the supersymmetric part of the
theory which are necessary to address the symmetry breaking patterns.
Thus,
we ignored the difficult problems associated with the couplings of
additional exotic particles in such string  models. Another  difficulty of 
this class 
of 
string models is the absence of a  mechanism
for supersymmetry breaking with unique quantitative predictions.  We chose
to parameterize the
supersymmetry breaking with a general set
 of soft supersymmetry breaking  mass parameters.

The analysis given in this paper generalizes the  work of 
\cite{CL}, which investigated the  gauge symmetry breaking pattern of
the above class of string models in the limit of a large $\tan
\beta$ scenario.
We have addressed the  nature of  phenomenologically acceptable
electroweak symmetry breaking scenarios  and the resulting  particle
spectrum in detail. In addition, we have analyzed the RGEs of the model to
explore the range of parameters at the string scale which leads to the
phenomenologically viable low energy parameter space.

We summarize the main results of the analysis as follows:

\noindent{\it   Gauge Symmetry Breaking Scenarios}

We found a rich structure of  phenomenologically acceptable 
gauge symmetry breaking patterns, which involved a certain but not
excessive amount of 
fine tuning of the parameters.  The symmetry breaking necessarily takes
place 
in the electroweak energy range\footnote{The scale of $U(1)'$ symmetry 
breaking can be in the $10^{10}-10^{14}$ GeV range for the case of
more than
one SM singlet and the appropriate choices of their $U(1)'$ charges 
\cite{CL}.}. 
 For a range of the parameters which comprises
a few percent of the  full parameter  space, the 
  $Z-Z'$ mixing  is acceptably small  and the $Z'$ mass is sufficiently
large. 
 The symmetry breaking patterns fall
into two characteristic  classes:
\begin{itemize}
\item   {\it Large  trilinear  coupling scenario}

 The symmetry breaking is driven by 
a large value of the soft supersymmetry breaking trilinear coupling.  When
the trilinear coupling is larger than the scalar soft mass parameters by a
factor of 5 to 10, the VEVs   of  $ H_{1,2}$, and $S$ are approximately equal. 
For equal $U(1)'$ charges for ${\hat H}_1$
 and ${\hat H}_2$,  the $Z-Z'$ mixing  is suppressed; it can be
 easily ensured to be $<10^{-3}$.  The  $Z'$  is light,
with  mass $\sim  200$ GeV. In this scenario, the electroweak phase
transition may be first order with
potentially interesting cosmological implications.

\item {\it  Large singlet VEV scenario}

In this case, the  symmetry breaking is driven
by a negative mass squared term for $S$. 
Its absolute magnitude is in
general  larger than that of the mass squared terms  for $H_{1,2}$. 
A certain fine tuning of the soft mass parameters is needed
to ensure  acceptably small $Z-Z'$ mixing.  This scenario
is viable (for different ranges of parameters)  without imposing additional
constraints on the $U(1)'$ charges of  the Higgs fields. 
The $Z'$ mass is typically in the range of $1$ TeV.
It is interesting to note that the range  of mass parameters for this
scenario 
is similar to that  of the MSSM.

\end{itemize}

\noindent{\it Renormalization Group Analysis}

We have also explored the relationship between the values of the soft supersymmetry breaking
mass parameters at the electroweak scale and the values at the string scale by 
analyzing the RGEs of the model.  We have solved the RGEs numerically as a
function of the boundary conditions at the string scale.  We have also 
derived semi-analytic solutions of the RGEs to further our understanding
of the evolution of the parameters. In the analysis, we  chose the 
initial values of the Yukawa couplings (of the
Higgs fields to the singlet and of the Higgs field to
the third quark family) to be of  the order of magnitude of the gauge
coupling, as determined in a class of string models based on
the fermionic construction.  These couplings provide a dominant
contribution to the RGEs of the soft mass parameters.  

We found that with the minimal particle content, universal soft
supersymmetry breaking mass parameters at the string scale do not yield
the phenomenologically acceptable range of parameters at the electroweak
scale.  The  results  which lead to 
the phenomenologically
acceptable low energy parameter space can be classified as follows:


\begin{itemize}
\item {\it Nonuniversal boundary conditions}


 With the minimal particle content, nonuniversal soft
supersymmetry breaking mass parameters are required at the string scale
to obtain the viable gauge symmetry breaking scenarios previously described.
In most cases, the gaugino
masses at the string scale must be  chosen
small relative to the other soft supersymmetry breaking mass parameters.
For the large trilinear coupling scenario,
the soft mass-squared parameters at the string scale are about a factor
of ten larger than their values at the electroweak scale\footnote{In a
large class of models for supersymmetry breaking, the
values of these mass parameters at the string scale 
are closely related  to the value of the 
gravitino mass.}. 

\item {\it Additional exotics}

Many string models predict the existence of  additional exotic particles, 
such as
additional $SU(3)$ triplets which couple to $\hat S$  with
Yukawa couplings of the order of the gauge couplings.
 The presence of such exotic particles can modify the RGE analysis
significantly\footnote{Since such exotics destroy the
gauge coupling 
unification,  one has to assume that
there are additional exotics (that, however, do not couple to $\hat S$),
so that the gauge coupling
unification is restored.}. Using the semi-analytic approach,
we determined that, for example, additional color
triplets ensure a large value of the singlet VEV even with universal
boundary conditions.  This indicates that the latter
scenario is obtainable for universal soft mass parameters at the
string scale when such exotics are present. In the limit of small
gaugino masses 
and trilinear couplings,  this result was exhibited
numerically in \cite{CL}.
In contrast, the large trilinear
coupling scenario is more difficult to obtain with additional exotic particles.  
We found that  nonuniversal boundary
conditions for  the soft supersymmetry breaking trilinear couplings are
required to reach this scenario.

\end{itemize}

The  analysis presented in this paper   exhibits  the viability and
predictive 
power of  supersymmetric models with an additional  $U(1)'$, whose 
 gauge structure, particle content, and
nature of couplings  are
key ingredients of a large class of string vacua. For a range of
soft supersymmetry breaking parameters at the string scale,  such models
allow for interesting gauge symmetry breaking scenarios,
 which can be tested  at  future colliders.

\acknowledgments
We thank Jing Wang for useful suggestions.

The work was supported by the School of Arts and Sciences of the University of
Pennsylvania, the U.S. Department of Energy Grant No. DOE-EY-76-02-3071, and
the Scientific and Technical Research Council of Turkey. 

\newpage
\appendix
\section{Renormalization Group Equations}

We present the renormalization group equations for the gauge couplings, gaugino
masses, Yukawa couplings, trilinear couplings, and soft mass-squared parameters 
for the model\footnote{We do not present the
renormalization group equations for the soft mass-squared parameters of the staus and
the sbottoms, as they do not influence directly the symmetry breaking pattern. 
These terms are included in the definitions (\ref{sumhyp}) and
(\ref{sumpr}).}.  In the following equations, $S_{1}$
and $S_1'$ are defined to be
\begin{eqnarray}
\label{sumhyp}
S_{1}=\sum_aY_am_a^2&=&
\sum_{i=1}^{N_F}({m_{E}}_i^2-{m_L}_i^2+{m_Q}_i^2+{m_D}_i^2-2{m_U}_i^2)-m_1^2
+m_2^2\\
\label{sumpr}
S_1'=\sum_aQ_am_a^2&=&\sum_{i=1}^{N_F}({Q_{E}}_i{m_{E}}_i^2 
+2{Q_{L}}_i{m_{L}}_i^2+6{Q_{Q}}_i{m_{Q}}_i^2+3{Q_{D}}_i{m_{D}}_i^2\nonumber\\
&+&3{Q_{U}}_i{m_{U}}_i^2)+2Q_{1}m_{1}^2+2Q_{2}m_{2}^2+Q_{S}m_{S}^2,
\end{eqnarray} 
$N_F$ denotes the number of families, and the scale variable is given
by
\begin{eqnarray}
t=\frac{1}{16\pi^{2}}\ln\frac{\mu}{M_{String}}.
\label{tvar}
\end{eqnarray}
The normalization of the $U(1)'$ gauge coupling is model dependent.  For
definiteness, we choose to normalize the gauge couplings by requiring that 
the gauge couplings and charges satisfy the constraint that $g_i^{0\,2}$
Tr$\,Q^2$ is constant, where the trace is evaluated over one family.  With the
choice of $U(1)'$ charges used in the renormalization group analysis, $g_1'(t)$
is numerically very similar to $g_1(t)$.

\subsubsection{Gauge Couplings}
\begin{eqnarray}
\label{gaugecoupl3}
\frac{d}{dt}g_{3}&=&(2N_F-9)g_{3}^{3}\\
\label{gaugecoupl2}
\frac{d}{dt}g_{2}&=&(2N_F-5)g_{2}^{3}\\
\label{gaugecoupl1}
\frac{d}{dt}g_{1}&=&(2N_F+\frac{3}{5})g_{1}^{3}\\
\label{gaugecoupl1pr}
\frac{d}{dt}g_{1}'&=&(2N_F+\rho (2Q_1^2+2Q_2^2+Q_S^2))g_{1}'^{3},
\end{eqnarray}
where 
\begin{eqnarray}
\label{rdef}
\rho=\frac{2}{6Q_{Q}^{2}+3(Q_{U}^{2}+Q_{D}^{2})+2Q_{L}^{2}+Q_{E}^{2}}.
\end{eqnarray}
\subsubsection{Gaugino Masses}
\begin{eqnarray}
\label{gaugn3}
\frac{d}{dt}M_{3}&=&2(2N_F-9)g_{3}^{2}M_{3}\\
\label{gaugn2}
\frac{d}{dt}M_{2}&=&2(2N_F-5)g_{2}^{2}M_{2}\\
\label{gaugn1}
\frac{d}{dt}M_{1}&=&2(2N_F+\frac{3}{5})g_{1}^{2}M_{1}\\
\label{gaugn1pr}
\frac{d}{dt}M_{1}'&=&2(2N_F+\rho(2Q_1^2+2Q_2^2+Q_S^2))g_{1}'^{2}M_{1}'
\end{eqnarray} 
\subsubsection{Yukawa Couplings}
\begin{eqnarray}
\label{heqn}
\frac{d}{dt}h_{s}&=&h_{s}\{4h_{s}^{2}+3h_{Q}^{2}-(3g_{2}^{2}+\frac{3}{5} g_{1}^{2} 
+2\rho g_{1}'^{2}(Q_{1}^{2}+Q_{2}^{2}+Q_{S}^{2}))\}\\
\label{hQeqn}
\frac{d}{dt}h_{Q}&=&h_{Q}\{6h_{Q}^{2}+h_{s}^{2}-(\frac{16}{3}g_{3}^{2}+3g_{2}^{2} 
+\frac{13}{15}g_{1}^{2}+2\rho g_{1}'^{2}(Q_{U}^{2}+Q_{Q}^{2}+Q_{2}^{2}))\}
\end{eqnarray}
\subsubsection{Trilinear Couplings}
\begin{eqnarray}
\label{cAseqn}
\frac{d}{dt}A&=&8h_{s}^{2}A 
+6h_{Q}^{2}A_{Q}- 2(3M_{2}g_{2}^{2}+ \frac{3}{5}M_{1}g_{1}^{2}
+2\rho M_{1}'g_{1}'^{2}(Q_{1}^{2}+Q_{2}^{2}+Q_{S}^{2}))\\
\label{cAqeqn}
\frac{d}{dt}A_{Q}&=&12h_{Q}^{2}A_{Q}+2h_{s}^{2}A 
-2(\frac{16}{3}M_{3}g_{3}^{2}+3M_{2}g_{2}^{2}+\frac{13}{15}M_{1}g_{1}^{2}
+2\rho M_{1}'g_{1}'^{2}(Q_{U}^{2}+Q_{Q}^{2}+Q_{2}^{2}))
\end{eqnarray}
\subsubsection{Soft Scalar Mass-Squared Parameters}
\begin{eqnarray}
\label{cSeqn}
\frac{d}{dt}m_{S}^{2}&=&4(m_{S}^{2}+m_{1}^{2}+m_{2}^{2} 
+A^{2})h_{s}^{2}
-8\rho M_{1}'^{2}g_{1}'^{2}Q_{S}^{2}+2\rho Q_{S}g_{1}'^{2}S_{1}'\\
\label{c1eqn}
\frac{d}{dt}m_{1}^{2}&=&2(m_{S}^{2}+m_{1}^{2}+m_{2}^{2}+A^{2})h_{s}^{2}\\
&-&8(\frac{3}{4}M_{2}^{2}g_{2}^{2}+\frac{3}{20}M_{1}^{2}g_{1}^{2}+ 
\rho M_{1}'^{2}g_{1}'^{2}Q_{1}^{2})-
\frac{3}{5}g_{1}^2S_{1}+2\rho g_{1}'^2Q_{1}S_{1}'\nonumber\\
\label{c2eqn}
\frac{d}{dt}m_{2}^{2}&=&2(m_{S}^{2}+m_{1}^{2}+m_{2}^{2} 
+A^{2})h_{s}^{2}+6(m_{2}^{2}+{m_{Q}}_3^{2}+{m_{U}}_3^{2}+A_{Q}^{2})h_{Q}^{2}\\
&-&8(\frac{3}{4}M_{2}^{2}g_{2}^{2}+\frac{3}{20}M_{1}^{2}g_{1}^{2}+
\rho M_{1}'^{2}g_{1}'^{2}Q_{2}^{2})+\frac{3}{5}g_{1}^2S_{1}
+2\rho g_{1}'^2Q_{2}S_{1}'\nonumber\\
\label{cUeqn}
\frac{d}{dt}{m_{U}}_3^{2}&=&4(m_{2}^{2}+{m_{Q}}_3^{2}+ 
{m_{U}}_3^{2}+A_{Q}^{2})h_{Q}^{2}\\
&-&8(\frac{4}{3}M_{3}^{2}g_{3}^{2}+\frac{4}{15}M_{1}^{2}g_{1}^{2}+
\rho M_{1}'^{2}g_{1}'^{2}Q_{U}^{2})
-\frac{4}{5}g_{1}^{2}S_{1}+2\rho g_{1}'^2Q_{U}S_{1}'\nonumber\\
\label{cQeqn}
\frac{d}{dt}{m_{Q}}_3^{2}&=&2(m_{2}^{2}+{m_{Q}}_3^{2}
+{m_{U}}_3^{2}+A_{Q}^{2})h_{Q}^{2}\\
&-&8(\frac{4}{3}M_{3}^{2}g_{3}^{2}+\frac{3}{4}M_{2}^{2}g_{2}^{2}+ 
\frac{1}{60}M_{1}^{2}g_{1}^{2}+ \rho M_{1}'^{2}g_{1}'^{2}Q_{Q}^{2})
+\frac{1}{5}g_{1}^{2}S_{1}+2\rho g_{1}'^2Q_{Q}S_{1}'
\end{eqnarray} 

\section{Solutions of RGEs}
\subsection{Numerical Results}
The RGEs for the gauge couplings and gaugino masses with the initial conditions
(\ref{unification}) and (\ref{univgau}) can be solved to yield
\begin{eqnarray}
\label{gaugecpls3}
g_{3}^{2}(t)=\frac{g_{0}^{2}}{1-2(2N_F-9)g_{0}^{2}t},\\
\label{gaugecpls2}
g_{2}^{2}(t)=\frac{g_{0}^{2}}{1-2(2N_F-5)g_{0}^{2}t},\\
\label{gaugecpls1}
g_{1}^{2}(t)=\frac{g_{0}^{2}}{1-2(2N_F+\frac{3}{5})g_{0}^{2}t},
\end{eqnarray}
\begin{eqnarray}
\label{gaugecpls1pr}
g_{1}'^{2}(t)=\frac{g_{0}^{2}}{1-2(2N_F+\rho (2Q_1^2+2Q_2^2+Q_S^2))g_{0}^{2}t},
\end{eqnarray}
where $\rho$ is defined in (\ref{rdef}), and 
\begin{eqnarray}
\label{gauginosoln}
M_i(t)=M^0_i\frac{g_{i}^{2}(t)}{g_0^2}.
\end{eqnarray}
These solutions are inserted in the RGEs for the other parameters, which we 
integrated numerically.  As a
concrete example, we choose the initial
values of the Yukawa couplings $h^{0}_{Q}=h^0_s=g_{0}\sqrt{2}$. 
  With the choice of charges $Q_1=Q_2=-1$,
$Q_S=2$, $Q_Q=Q_U=\frac{1}{2}$, and $Q_L=Q_E=Q_D=0$,  the results are as
follows:

\begin{itemize}

\item Yukawa couplings:
\begin{eqnarray}
h_{s}(M_{Z})=0.70\;,\;h_{Q}(M_Z)=1.074.
\label{yukawaMw}
\end{eqnarray}

\item Trilinear Couplings:
\begin{eqnarray}
\label{cAqsoln}
A_{Q}(M_{Z})&=&-0.047\,A^{0}+0.109\,A^{0}_{Q}+1.97\,M_{1/2},\\
\label{cAssoln}
A(M_{Z})&=&0.316\,A^{0}-0.230\,A^{0}_{Q}-0.162\,M_{1/2}.
\end{eqnarray}

\item Soft Mass-Squared Parameters:
\begin{eqnarray}
\label{c1soln}
m^{2}_{1}(M_{Z})&=&-0.13\,m^{0\,2}_{2}+0.8\,m^{0\,2}_{1}-0.2\,m^{0\,2}_{S}
+0.062\,m^{0\,2}_{U}+0.062\,m^{0\,2}_{Q}\nonumber\\&-&0.056\,A^{0\,2}
+0.0083\,A^{0\,2}_{Q}+0.61\,(M_{1/2})^{2}+
0.034\,A^{0}A^0_{Q}\nonumber\\&-&0.051\,A^{0}M_{1/2}
+0.044\,A^0_{Q}M_{1/2},\\
\label{c2soln}
m^{2}_{2}(M_{Z})&=&0.47\,m^{0\,2}_{2}-0.12\,m^{0\,2}_{1}-0.12\,m^{0\,2}_{S}
-0.41\,m^{0\,2}_{U}-0.41\,m^{0\,2}_{Q}\nonumber\\&-&0.031\,A^{0\,2}
-0.039\,A^{0\,2}_{Q}-3.21\,(M_{1/2})^{2}+
0.034\,A^{0}A^0_{Q}\nonumber\\&+&0.035\,A^{0}M_{1/2}
-0.18\,A^0_{Q}M_{1/2},\\
\label{cssoln}
m^{2}_{S}(M_{Z})&=&-0.25\,m^{0\,2}_{2}-0.38\,m^{0\,2}_{1}+0.62\,m^{0\,2}_{S}
+0.12\,m^{0\,2}_{U}+0.12\,m^{0\,2}_{Q}\nonumber\\&-&0.11\,A^{0\,2}
+0.017\,A^{0\,2}_{Q}+0.42\,(M_{1/2})^{2}+
0.068\,A^{0}A^0_{Q}\nonumber\\&-&0.1\,A^{0}M_{1/2}
+0.087\,A^0_{Q}M_{1/2},\\
\label{cusoln}
m^{2}_{U}(M_{Z})&=&-0.27\,m^{0\,2}_{2}+0.05\,m^{0\,2}_{1}+0.05\,m^{0\,2}_{S}
+0.68\,m^{0\,2}_{U}-0.32\,m^{0\,2}_{Q}\nonumber\\&+&0.017\,A^{0\,2}
-0.032\,A^{0\,2}_Q+4.1\,(M_{1/2})^{2}
+0.00\,A^{0}A^{0}_{Q}\nonumber\\&+&0.06\,A^{0}M_{1/2}
-0.15\,A^{0}_{Q}M_{1/2},\\
\label{cqsoln}
m^{2}_{Q}(M_{Z})&=&-0.14\,m^{0\,2}_{2}+0.024\,m^{0\,2}_{1}+0.024\,m^{0\,2}_{S}
-0.16\,m^{0\,2}_{U}+0.84\,m^{0\,2}_{Q}\nonumber\\&+&0.0084\,A^{0\,2}
-0.016\,A^{0\,2}_Q+5.8\,(M_{1/2})^{2}
+0.00\,A^{0}A^{0}_Q\nonumber\\&+&0.028\,A^{0}M_{1/2}
-0.073\,A^{0}_{Q}M_{1/2}.
\end{eqnarray}
\end{itemize}

We have also obtained results for different choices of the initial values of
the Yukawa couplings as can appear in a class of models.  The low
energy results do not change significantly.  For example, with
$h^{0}_{Q}=g_{0}\sqrt{2}$ and 
$h^{0}_{S}=g_0$, the values of the coefficients do not change more than 
$10\%$.

\subsection{Semi-Analytic Solutions}
In the following section we present approximate analytical solutions to the 
RGEs. To solve the RGEs, we first make the approximation that the gauge 
couplings (\ref{gaugecpls3})-(\ref{gaugecpls1pr}) are replaced by their
average values,
\begin{eqnarray}
\label{lisa1}
g_{i}=\frac{1}{2}(g_{i}(M_{Z})+g_{0})\;\; (i=3,2,1,1').
\end{eqnarray}
Similarly, we replace the gaugino masses (\ref{gauginosoln}) with
\begin{eqnarray}
\label{lisa2}
M_{i}=\frac{1}{2}(M_i(M_Z)+M_{1/2})\,.
\end{eqnarray}
This yields the respective values $1.00,\, 0.69,\, 0.59,\,0.58$ for the 
gauge couplings $g_{3},g_{2},g_{1},g_{1}'$, and $2.07M_{1/2},\, 0.91M_{1/2},\,
0.70M_{1/2}$, 
$0.69M_{1/2}$ for the gaugino masses 
$M_{3},\, M_{2},\,M_{1},\,M_{1}'$.
Under these approximations, we can solve the coupled equations 
for the Yukawa couplings by noticing that with the choice of initial conditions,  
$h_{Q}$ remains relatively close to its fixed point
value\footnote{The gauge couplings run, so this is not a fixed point in the exact
sense. However, this approach is valid in the limit that (\ref{lisa1}) holds.},
while $h_{s}$ evolves significantly. The approximate solution is: 
\begin{eqnarray} 
h_{s}^{2}(t)&=&\frac{{\tilde{g}_{S}}^{2}} 
{1-(1-\frac{{\tilde{g}_{S}}^{2}}{{h^{0}_{S}}^{2}}) 
e^{{7{\tilde{g}_{S}}^{2}t}}},\\
h_{Q}^{2}(t)&=&\frac{{\tilde{g}_{Q}}^{2}}{1-
(1-\frac{{\tilde{g}_{Q}}^{2}}{{h^{0}_{Q}}^{2}})
e^{{12{\tilde{g}_{Q}}^{2}t}}},
\end{eqnarray}
in which
\begin{eqnarray}
{\tilde{g}_{S}}^{2}&=&\frac{1}{7}(3g_{2}^{2}-\frac{16}{3}g_{3}^{2}+
\frac{1}{3}g_{1}^{2}+2\rho (2Q_{S}^{2}+2Q_{1}^{2}+Q_{2}^{2}-Q_{U}^2-Q_{Q}^{2})
\,g_{1}'^{2}),\\
{\tilde{g}_{Q}}^{2}&=&\frac{1}{6}(\frac{16}{3}g_{3}^{2}+3g_{2}^{2}+
\frac{13}{15}g_{1}^{2}+
2\rho (Q_{U}^{2}+Q_{Q}^{2}+Q_{2}^{2})\,g_{1}'^{2}-\bar{h}_{S}^{2}),\\
\label{lisa3}
\bar{h}_{S}&=&\frac{1}{2}(h^{0}_{S}+h_{s}(M_{Z})),\\
\label{lisa4}
\bar{h}_{Q}&=&\frac{1}{2}(h^{0}_{Q}+h_{Q}(M_{Z})).
\end{eqnarray}
As a first approximation to solve for the trilinear couplings, we use the averaged 
Yukawa couplings, averaged gaugino masses, and averaged gauge couplings, and 
the $U(1)$ factors are neglected for simplicity.  The equations are then 
solved to yield
\begin{eqnarray}
A_{Q}(t)&=&\alpha_{1Q}e^{\lambda_{1}t}+\alpha_{2Q}e^{\lambda_{2}t}
+\beta_{1Q}e^{\lambda_{1}t}+\beta_{2Q}e^{\lambda_{2}t}-A_{QP},\\
A(t)&=&\alpha_{1S}e^{\lambda_{1}t}+\alpha_{2S}e^{\lambda_{2}t}
+\beta_{1S}e^{\lambda_{1}t}+\beta_{2S}e^{\lambda_{2}t}-A_{SP},
\end{eqnarray}
where the initial condition-dependent $\alpha$ and $\beta$ coefficients are
\begin{eqnarray}
\alpha_{iQ}&=&\alpha_{i}({A}^{0}_{Q}, A^{0}_{S}),\\
\beta_{iQ}&=&\alpha_{i}(A_{QP}, A_{SP}),\\
\alpha_{iS}&=&\alpha_{iQ}\frac{(\lambda_{i}-12\bar{h}_{Q}^{2})}{2\bar{h}_{S}^{2}},\\
\beta_{iS}&=&\beta_{iQ}\frac{(\lambda_{i}-12\bar{h}_{Q}^{2})}{2\bar{h}_{S}^{2}}.
\end{eqnarray}
We have introduced some short-hand notation:
\begin{eqnarray}
\alpha_{1}(A,B)&=&\frac{A(12\bar{h}_{Q}^{2}-\lambda_{2})
+2\bar{h}_{S}^{2}B}{\lambda_{1}-\lambda_{2}},\\
\alpha_{2}(A,B)&=&\frac{A(\lambda_{1}-12\bar{h}_{Q}^{2})
-2\bar{h}_{S}^{2}B}{\lambda_{1}-\lambda_{2}},\\
\lambda_{1,2}&=&6\bar{h}_{Q}^{2}+4\bar{h}_{S}^{2}\pm 
\sqrt{(6\bar{h}_{Q}^{2}-4\bar{h}_{S}^{2})^{2}+12\bar{h}_{Q}^{2}\bar{h}_{S}^{2}},\\
A_{QP}&=&\frac{\frac{128}{3}g^{2}_{3}M_{3}+18g^{2}_{2}M_{2}}{42\bar{h}_{Q}^{2}},\\
A_{SP}&=&\frac{-\frac{32}{3}g^{2}_{3}M_{3}+6g^{2}_{2}M_{2}}{14\bar{h}_{S}^{2}}.
\end{eqnarray}
With the approximations (\ref{lisa1}), (\ref{lisa2}), (\ref{lisa3}), and
(\ref{lisa4}), the fixed point values are $A_{QP}=2.3M_{1/2}$ and
$A_{SP}=-1.8M_{1/2}$. This analysis slightly overestimates the splitting of the
fixed point
values, but shows the tendency for $A_Q(M_Z)$ to be larger than $A(M_Z)$
for
$M_{1/2}$ positive.  

The equations for the trilinear terms can also be solved when the running of the 
$SU(3)$ gauge coupling and gaugino are included.  The others are neglected for 
simplicity, as the $SU(3)$ gauge coupling is dominant.  In this case
the 
solutions are
\begin{eqnarray}
A_{Q}(t)&=&\alpha_{1Q}e^{\lambda_{1}t}+\alpha_{2Q}e^{\lambda_{2}t}
+M_{1/2}f_{Q}(I_{1}(t),I_{2}(t)),\\
A(t)&=&\alpha_{1S}e^{\lambda_{1}t}+\alpha_{2S}e^{\lambda_{2}t}
+M_{1/2}f_{S}(I_{1}(t),I_{2}(t))\, ,
\end{eqnarray}
in which
\begin{eqnarray}
f_{Q}&=&-\frac{16}{9}
\frac{e^{\lambda_{1}t}(12\bar{h}_{Q}^{2}-\lambda_{2})I_{1}(t)+
e^{\lambda_{2}t}(\lambda_{1}-12\bar{h}_{Q}^{2})I_{2}(t)}{\lambda_{1}-\lambda_{2}}\\
f_{S}&=&-\frac{16}{9}\frac{6\bar{h}_{Q}^{2}(e^{\lambda_{1}t}I_{1}(t) 
-e^{\lambda_{2}t}I_{2}(t))}{\lambda_{1}-\lambda_{2}}, 
\end{eqnarray}
and the functions $I_{i}(t)$ are defined by
\begin{eqnarray}
I_{i}(t)=\int_{x=0}^{6g^{2}_{0}\,t}e^{-\frac{\lambda_{i}x}{6g^{2}_{0}}}
\frac {dx}{(1+x)^{2}}. 
\end{eqnarray}
To solve the RGEs for the soft mass-squared parameters, 
only the $SU(3)$ gauge coupling and gaugino are included in the analysis.  
To obtain relatively compact approximate analytical solutions, 
the trilinear couplings are also replaced by their average values:
\begin{eqnarray}
\bar{A}_{Q}=\frac{1}{2}(A^{0}_{Q}+A_{Q}(M_{Z})),\\
\bar{A}=\frac{1}{2}(A^{0}+A(M_{Z}))\, . 
\end{eqnarray}
With these
further approximations, it is useful to consider the solutions for the sums 
defined by
\begin{eqnarray}
\Sigma_{1}&=&m^{2}_{Q}+m^{2}_{U}+m^{2}_{2},\\
\Sigma_{2}&=&m^{2}_{S}+m^{2}_{1}+m^{2}_{2}\, .
\end{eqnarray}
The solutions are given by
\begin{eqnarray}
\label{lisa5}
\Sigma_{1}(t)&=&(\gamma_{1}+\rho_{1})e^{\lambda_{1}t}+(\gamma_{2} 
+\rho_{2})e^{\lambda_{2}t}-\Delta_{1}\, ,\\
\label{lisa6}
\Sigma_{2}(t)&=&(\delta_{1}+\eta_{1})e^{\lambda_{1}t} 
+(\delta_{2}+\eta_{2})e^{\lambda_{2}t}-\Delta_{2}\, , 
\end{eqnarray}
in which
\begin{eqnarray}
\gamma_{i}&=&\alpha_{i}(m^{0\,2}_{Q}+m^{0\,2}_{U}+m^{0\,2}_{2},m^{0\,2}_{S}+
m^{0\,2}_{1}+m^{0\,2}_{2}),\\
\rho_{i}&=&\alpha_{i}(\Delta_{1},\Delta_{2}),\\
\delta_{i}&=&\gamma_{i}\frac{(\lambda_{i}-12\bar{h}_{Q}^{2})}{2\bar{h}_{S}^{2}},\\
\eta_{i}&=&\rho_{i}\frac{(\lambda_{i}-12\bar{h}_{Q}^{2})}{2\bar{h}_{S}^{2}},\\
\Delta_{1}&=&\bar{A}_{Q}^{2}- 
\frac{128}{63}\frac{g^{2}_{3}M_{3}^{2}}{\bar{h}_{Q}^{2}},\\
\Delta_{2}&=&\bar{A}^{2}+
\frac{32}{21}\frac{g^{2}_{3}M_{3}^{2}}{\bar{h}_{S}^{2}}.
\end{eqnarray}
The renormalization group equations for the individual mass-squared parameters
may then be integrated explicitly to yield
\begin{eqnarray}
m^{2}_{1}(t)&=&\frac{5}{7}m^{0\,2}_{1}-\frac{1}{7}m^{0\,2}_{2} 
-\frac{2}{7}m^{0\,2}_{S}+\frac{1}{7}m^{0\,2}_{U}+\frac{1}{7}m^{0\,2}_{Q}
+\frac{1}{7}\Delta_{1}-\frac{2}{7}\Delta_{2}-3.05g^{2}_{3}M_{3}^{2}t\nonumber\\
&+&2\bar{h}_{S}^{2}\{\frac{\eta_{1}+\delta_{1}}{\lambda_{1}}e^{\lambda_{1}t} 
+\frac{\eta_{2}+\delta_{2}}{\lambda_{2}}e^{\lambda_{2}t}\},\\
m^{2}_{2}(t)&=&-\frac{1}{7}m^{0\,2}_{1}+\frac{3}{7}m^{0\,2}_{2}
-\frac{1}{7}m^{0\,2}_{S}-\frac{3}{7}m^{0\,2}_{U}-\frac{3}{7}m^{0\,2}_{Q}
-\frac{3}{7}\Delta_{1}-\frac{1}{7}\Delta_{2}+9.14g^{2}_{3}M_{3}^{2}t\nonumber\\
&+&\frac{2\bar{h}_{S}^{2}(\eta_{1}+\delta_{1})+6\bar{h}_{Q}^{2}(\gamma_{1}+\rho_{1})} 
{\lambda_{1}}\,e^{\lambda_{1}t}
+\frac{2\bar{h}_{S}^{2}(\eta_{2}+\delta_{2})+6\bar{h}_{Q}^{2}(\gamma_{2}+\rho_{2})} 
{\lambda_{2}}\,e^{\lambda_{2}t},\\
m^{2}_{S}(t)&=&-\frac{4}{7}m^{0\,2}_{1}-\frac{2}{7}m^{0\,2}_{2}
+\frac{3}{7}m^{0\,2}_{S}+\frac{2}{7}m^{0\,2}_{U}+\frac{2}{7}m^{0\,2}_{Q}
+\frac{2}{7}\Delta_{1}-\frac{4}{7}\Delta_{2}-6.1g^{2}_{3}M_{3}^{2}t\nonumber\\
&+&4\bar{h}_{S}^{2}\{\frac{\eta_{1}+\delta_{1}}{\lambda_{1}}e^{\lambda_{1}t}
+\frac{\eta_{2}+\delta_{2}}{\lambda_{2}}e^{\lambda_{2}t}\},\\
m^{2}_{U}(t)&=&\frac{2}{21}m^{0\,2}_{1}-\frac{6}{21}m^{0\,2}_{2}
+\frac{2}{21}m^{0\,2}_{S}+\frac{13}{21}m^{0\,2}_{U}-\frac{8}{21}m^{0\,2}_{Q}
+\frac{8}{21}\Delta_{1}+\frac{2}{21}\Delta_{2}-2.54g^{2}_{3}M_{3}^{2}t\nonumber\\
&+&4\bar{h}_{Q}^{2}\{\frac{\rho_{1}+\gamma_{1}}{\lambda_{1}}e^{\lambda_{1}t}
+\frac{\rho_{2}+\gamma_{2}}{\lambda_{2}}e^{\lambda_{2}t}\},\\
m^{2}_{Q}(t)&=&\frac{1}{21}m^{0\,2}_{1}-\frac{2}{21}m^{0\,2}_{2}
+\frac{1}{21}m^{0\,2}_{S}-\frac{4}{21}m^{0\,2}_{U}+\frac{17}{21}m^{0\,2}_{Q}
-\frac{4}{21}\Delta_{1}+\frac{1}{21}\Delta_{2}-6.6g^{2}_{3}M_{3}^{2}t\nonumber\\
&+&2\bar{h}_{Q}^{2}\{\frac{\rho_{1}+\gamma_{1}}{\lambda_{1}}e^{\lambda_{1}t}
+\frac{\rho_{2}+\gamma_{2}}{\lambda_{2}}e^{\lambda_{2}t}\}.
\end{eqnarray}
These solutions are valid in the limit of small initial gaugino masses,
such that their 
contribution to the evolution of the trilinear couplings and the mass squares is 
small.  When this condition is not satisfied, the $SU(3)$ gaugino masses and gauge 
couplings control the evolution of all the parameters, and the 
approximation of neglecting the running of the gaugino masses and gauge couplings 
breaks down.  As stated above, it is possible to incorporate the running of the 
$SU(3)$ gauge coupling and gaugino in solving the equations for the trilinear 
couplings and obtain solutions to these equations that are in better agreement 
with the exact solutions.  This is also possible for the soft mass-squared
parameters, but the solutions are cumbersome and thus do not yield much physical
insight, so they are not presented here.  

In the limit in which the gaugino masses and trilinear couplings are neglected, 
($\Delta_i$, $\rho_i$, and $\eta_i$ are zero), it is possible to use the
semi-analytic expressions to show that with universal initial conditions, the
only soft mass-squared parameter that will run negative is $m^2_2$.  In this
limit, (\ref{lisa5}) and (\ref{lisa6}) approach zero asymptotically.  Therefore,
in the asymptotic limit the appropriate sums of the individual mass-squared
parameters must also approach zero.  Since $H_2$ couples both
to the quarks and the singlet in the superpotential, it has a greater
weight driving it negative in its RGE (\ref{c2eqn}), and it will be
negative at low energies.  The other soft mass-squared parameters have smaller group
theoretical prefactors, and in the asymptotic limit they must be positive to
compensate for the negative value of $m^2_2$.  This indicates that the other soft
mass-squared parameters are necessarily positive at the electroweak scale, as the
asymptotes dominate the low energy  behaviour.  Although the solution of the RGEs
requires a choice of average values of the Yukawa couplings, the asymptotes of the
mass-squared parameters do not depend on the Yukawa couplings; it is only the
group theoretical factors present in the RGEs that lead to this result.  

This also indicates why it
becomes so simple to have $m^2_S$ negative when we add exotics that couple to
the singlet in the superpotential.  This increases the effective group
theoretical factor in the RGE for $m^2_S$, so it is naturally negative at the
electroweak scale for universal boundary conditions.   

\section{Non-Anomalous $U(1)'$}

In this work, we consider the phenomenological consequences of an additional
non-anomalous $U(1)'$ symmetry.  The requirement that the $U(1)'$ symmetry be
anomaly-free severely constrains the $U(1)'$ charge assignments of the theory;
the charges must be chosen so that the $U(1)'$ triangle anomaly and the
mixed anomalies cancel.  Furthermore, we require that the charges forbid an
elementary $\mu$ term $(Q_1+Q_2\ne 0)$ but allow our induced $\mu$ term
$(Q_1+Q_2+Q_S=0)$.  Finally, we require (for models involving light exotic
supermultiplets) that the approximate gauge unification under the standard
model group be respected.  In this appendix, we display two models which
satisfy these constraints and provide ``existence" proofs.  One involves ad hoc
charge assignments for the minimal particle content, and the other is
GUT-motivated and involves exotics.  The construction of realistic
string-derived models is beyond the scope of this paper.

In the model we consider with the MSSM particle
content and one additional singlet, for which approximate gauge unification is
respected, the anomaly constraints are 

\begin{eqnarray}
\label{su3anomaly}
0&=&\sum_i (2{Q_Q}_i+{Q_U}_i+{Q_D}_i),\\
\label{su2anomaly}
0&=&\sum_i(3{Q_Q}_i+{Q_L}_i)+Q_1+Q_2,\\
\label{mixanomaly1}
0&=&\sum_i(\frac{1}{6}{Q_Q}_i+\frac{1}{3}{Q_D}_i 
+\frac{4}{3}{Q_U}_i+\frac{1}{2}{Q_L}_i+{Q_E}_i)+
\frac{1}{2}(Q_1+Q_2),\\
\label{mixanomaly2}
0&=&\sum_i({Q_Q}_i^2+{Q_D}_i^2-2{Q_U}_i^2-{Q_L}_i^2+{Q_E}_i^2)
-Q^2_1+Q^2_2,\\
\label{trianomaly}
0&=&\sum_i(6{Q_Q}_i^3+3{Q_D}_i^3+3{Q_U}_i^3+2{Q_L}_i^3+
{Q_E}_i^3)+2Q^3_1+2Q^3_2+Q^3_S.
\end{eqnarray}
The first four constraints correspond to the
mixed anomalies with $SU(3)$, $SU(2)$, $[U(1)_Y]^2$, and $U(1)_Y$, respectively.
The final equation
is the $U(1)'$ triangle anomaly condition. 

There are also constraints from the requirements of gauge invariance:
\begin{eqnarray}
\label{gauginvu}
{Q_U}_3+{Q_Q}_3+Q_2&=&0,\\
\label{gauginvs}
Q_1+Q_2+Q_S&=&0,
\end{eqnarray}
where (\ref{gauginvu}) and (\ref{gauginvs}) follow from the existence of a Yukawa
interaction for the $t$ quark mass and a term to generate an effective $\mu$
parameter, respectively. We do not require the existence of Yukawa interactions
for leptons ($Q_E+Q_L+Q_1=0$) or down-type quarks ($Q_D+Q_Q+Q_1=0$). This is
consistent with our superpotential (\ref{superpot}), which does not include
Yukawa couplings for these superfields. This implies in general that these
fields must have masses generated by other mechanisms (e.g., higher
dimensional terms in the superpotential and/or extra fields in the model). 
In one of the examples below we obtain that the condition $Q_E+Q_L+Q_1=0$
is automatically satisifed for the third generation, so that the mass of
the tau lepton can be generated by higher dimensional terms. However,
$Q_D+Q_Q+Q_1 \ne 0$ in that model, so that the bottom quark mass (and the masses
of the first two generations) generated by higher dimensional terms would
be suppressed by powers of the $U(1)'$ breaking scale, and are thus too small.
  
We have been able to find examples of charge assignments for our model which
satisfy the anomaly [(\ref{su3anomaly})-(\ref{trianomaly})] and
gauge invariance [(\ref{gauginvu})-(\ref{gauginvs})] constraints. One
simple possibility is the following:
\begin{equation}
\begin{array}{ll} 
{Q_E}_3=Q_2-Q_1, & {Q_L}_3=-Q_2, \vspace{0.1cm}\\
{Q_Q}_3=-\frac{1}{3}Q_1, & Q_S=-(Q_1+Q_2), \vspace{0.1cm}\\
{Q_D}_3=\frac{1}{3}(Q_1+3Q_2), & {Q_U}_3=\frac{1}{3}(Q_1-3Q_2),
\end{array}
\end{equation}
for arbitrary $Q_1$ and $Q_2$, and the first and second families have zero $U(1)'$
charges (other examples with nonzero charges for all three families can easily
be constructed). This choice is consistent with string models where $U(1)'$
charges for quarks and leptons of different families are {\it not equal} in
general.

We now consider the effects of neglecting the $U(1)$
factors (\ref{sumhyp}) and (\ref{sumpr}) in the analysis of the RGEs for the
soft mass parameters.   
It is straightforward to derive the evolution equations for $S_1$ and $S_1'$; 
if the charge assignments are such that the conditions for anomaly cancellation
and gauge invariance of the superpotential [(\ref{su3anomaly})-(\ref{gauginvs})]
are satisfied, one obtains a homogeneous coupled 
system involving only $S_1$, $S_1'$, the $U(1)$ gauge couplings, and 
the $U(1)'$ charges. For universal soft mass-squared parameters at the string scale, 
$S_1$ and $S_1'$ are manifestly zero when the anomaly conditions are
satisfied, and they remain zero from $M_{String}$ to $M_Z$.  When there
are nonuniversal soft mass-squared parameters, $S_1$ and $S_1'$ have nonzero initial 
values. In the semi-analytic approach in which the gauge
couplings are replaced by their average values, it is possible to solve this
coupled system for our 
example of $U(1)'$ charge assignments, and show that the system exponentially
decays.  Therefore, these factors become less important, and neglecting them
in the RGE analysis is well justified.

As an example of a GUT-motivated $U(1)'$, we consider the  $\psi$
\cite{FUTUREZ}, which occurs in the breaking of $E_6$ to
$SO(10) \times U(1)_\psi$. It is not our intention to consider
GUTs per se, but rather to use this as an existence proof of acceptable
$U(1)'$ quantum numbers. The theory will be anomaly-free if the
matter supermultiplets transform according to
\begin{equation}
3 \times 27_L + n(27_L + 27^*_L),
\end{equation}
where $27_L$ and $27^*_L \sim (27_R)^\dagger$ refer to 27-plets of $E_6$.
Since the $27_L$ and $27^*_L$ pairs are vector, any submultiplets can have a
string (or GUT) scale mass and decouple without breaking the $U(1)_\psi$ or
introducing anomalies, and indeed in most string models one expects only parts
of the $27_L + 27^*_L$ to be present in the observable sector.

It is convenient to display the decomposition of the $27_L$ under
the $SU(5) \times U(1)_\psi$ subgroup,
\begin{equation}
27_L \rightarrow (10,1)_L + (5^*,1)_L + (1,1)_L +(5,-2)_L
+ (5^*,-2)_L + (1,4)_L,
\label{decomp}
\end{equation}
where the first and second quantities are the $SU(5)$ multiplet and
 $\sqrt{24} Q_\psi$, respectively.
In (\ref{decomp}), the $(10,1)_L + (5^*,1)_L$ constitutes an ordinary
family,
$(1,1)_L$ and $(1,4)_L$ are standard model singlets, and
$(5,-2)_L
+ (5^*,-2)_L$ are exotic multiplets which form a vector pair under
the standard model gauge group but are chiral under $U(1)_\psi$. In
particular, $(5,-2)_L$ consists of  $D_L$ and $h_2$, where
$D$ is a color-triplet charge $-1/3$ quark and $h_2$ has the standard
model
quantum numbers of the $H_2$. Similarly, $(5^*,-2)_L$ consists
of $\bar{D}_L$ and $h_1$, where $h_1$ has the quantum number of either
the $H_1$ or a lepton doublet.

Any of the three $h_1$'s and three
$h_2$'s have the appropriate quantum numbers to be the MSSM Higgs
doublets.
Furthermore, the $(1,4)_L$ could be the singlet $S$, with the two Yukawa
couplings in (\ref{superpot}) allowed by $U(1)_\psi$. An exotic $h_D \hat{S}
\hat{D}
\hat{\bar{D}}$ coupling, as in (\ref{superpotex}), is also allowed. Hence, a
model consisting
of three 27-plets has most of the ingredients needed to display the
considerations of this paper, albeit with additional singlets and
$(5,-2)_L + (5^*,-2)_L$ pairs.

The model as such is not consistent with the observed
approximate gauge unification. The two extra $(5,-2)_L + (5^*,-2)_L$ pairs
and the singlets do not affect the standard model gauge unification at
one-loop. However, the $D$ and $\bar{D}$ associated with the two Higgs
doublets destroy the unification, and they cannot be made superheavy
without
breaking the $U(1)_\psi$ and also introducing anomalies in the effective
low-energy
theory.

Gauge unification can be restored without introducing
anomalies by adding a single $27_L + 27^*_L$ pair, and assuming, for
example, that
only the Higgs-like doublets $h_2$ and $h_3$ associated with the
$(5,-2)_L$ (from $27_L$) and $(5^*,+2)_L$ (from $27^*_L$) remain
in the observable sector. The $h_2$ is equivalent to the
$h_2$'s from the other 27-plets, while the $h_3$ is similar
to the $h_1$ multiplets, except that it has the opposite  $Q_\psi$.
The $h_3$ is not a candidate for the $H_1$, because its $Q_\psi$
would not allow the Yukawa interactions in (\ref{superpot}) needed
to generate an effective $\mu$ (an elementary $\mu$ is allowed
by $U(1)_\psi$ is this case) or the
effective Yukawa interactions (e.g., generated by
higher-dimension terms in the superpotential) for the down-type quarks
and electrons. Thus, in this model the Higgs multiplets (or at least
$H_1$)
are not associated with the extra $27_L + 27^*_L$, although the
latter are needed for gauge unification. This is not an ad hoc assumption,
but a consequence of the allowed couplings; the model actually has
eight Higgs-like doublets, 4 $h_2$'s, 3 $h_1$'s, and one $h_3$. Assuming
positive soft mass squares at the Planck scale, the only fields to
actually acquire VEVs will be those which have the necessary
Yukawa interactions in (\ref{superpot}) and possibly (\ref{superpotex}), i.e.,
an $h_1$ and $h_2$ pair. 


\vskip2.mm
\begin{references}
\bibitem{PRESENTZ}{See the Appendix
of M. Cveti\v c and P. Langacker,
Mod. Phys. Lett. {\bf 11A}, 1247 (1996),
and references therein.}
\bibitem{FUTUREZ}{For a review see, {\it e.g.,} 
M. Cveti\v c and S. Godfrey, in Proceedings of  {\it Electro-weak Symmetry 
Breaking
and beyond the Standard model}, eds. T. Barklow, S Dawson, H. Haber and J.
Siegrist (World Scientific 1995), hep-ph/9504216, and references therein.}
\bibitem{CDFLIM}{K. Maeshima,
Proceedings of the 28th
International Conference on High Energy Physics
(ICHEP'96), Warsaw, Poland, July 25-31, 1996.}
\bibitem{LEPTOPHOBIC}{P. Chiappetta et al., Phys. Rev.
{\bf D54}, 789 (1996);
G. Altarelli et al., Phys. Lett. {\bf B375}, 292 (1996);
K.S. Babu, C. Kolda, and  J. March-Russell,
Phys. Rev. {\bf D54}, 4635 (1996);
P. Frampton, M. Wise, and B. Wright, 
Phys. Rev. {\bf D54}, 5820 (1996);
K. Agashe et al., Phys. Lett. {\bf B385}, 218 (1996);
V. Barger, K. Cheung, and P. Langacker, 
Phys. Lett. {\bf B381}, 226 (1996); J. Rosner, Phys. Lett. {\bf B387}, 
113 (1996).}
\bibitem{LEPRB}{The current situation is summarized in the
joint report of the LEP Collaborations, LEP Electroweak
Working Group, and SLD Heavy Flavor Group, CERN-PPE/96-183.}
\bibitem{MUPROB}{J. E. Kim and H. P. Nilles, Phys. Lett.
{\bf B138}, 150 (1984).}
\bibitem{SY} D. Suematsu and Y. Yamagishi, Int. J. Mod. Phys. {\bf A10}, 4521
(1995). 
\bibitem{CL} M. Cveti\v c and P. Langacker, Phys. Rev. {\bf D54}, 3570 (1996),
and Mod. Phys. Lett. {\bf 11A}, 1247 (1996). 
\bibitem{shrock}
V. Jain and R. Shrock, Phys. Lett. {\bf B352}, 83 (1995)
and ITP-SB-95-22, hep-ph/9507238;
 Y. Nir, Phys. Lett. {\bf 354}, 107
(1995).
\bibitem{LEPTONASY}
{M. Fukugita and T. Yanagida, Phys. Lett. {\bf B174},
45 (1986);  Phys. Rev. {\bf D42}, 1285 (1990);
P. Langacker, R. D. Peccei and T. Yanagida, Mod.
Phys. Lett. {\bf A1}, 541 (1986).}
\bibitem{COSTRINGS}{R. Brandenberger, A.-C. Davis, and
M. Rees, Phys. Lett. {\bf B349}, 329 (1995), and references therein.}
\bibitem{LYKKEN}{J.D. Lykken, Preprint FERMILAB-CONF-96-344-T, hep-ph/9610218.}
\bibitem{ABK}{I. Antoniadis, C.
Bachas and C. Kounnas, Nucl. Phys. {\bf B289},
 87 (1987); H. Kawai, D. Lewellen and S.H.-H. Tye, Phys. Rev. Lett. {\bf
57},  1832 (1986); Phys. Rev. {\bf D34}, 3794 (1986).}
\bibitem{NAHE}{I. Antoniadis, J. Ellis, J. Hagelin and D. Nanopoulos,
Phys. Lett. {\bf B231}, 65 (1989).}
\bibitem{FARA}{A. Faraggi, D.V. Nanopoulos and K. Yuan, Nucl. Phys. {\bf B335},
347 (1990); A. Faraggi, Phys Lett. {\bf B278}, 131 (1992).}
\bibitem{CHL}{S. Chaudhuri, S.-W. Chung, G. Hockney and J. Lykken,
Nucl. Phys. {\bf B456}, 89 (1995);
S. Chaudhuri,  G. Hockney and J. Lykken,
Nucl. Phys. {\bf B469}, 357 (1996).}
\bibitem{GM}{G. F. Giudice and A. Masiero, Phys. Lett. {\bf B206}, 480 (1988). 
See also J. A. Casas and C. Mu\~ noz, Phys. Lett.{\bf B306}, 288 (1993).}
\bibitem{NMSSM} {P. Fayet, Nucl. Phys. {\bf B90}, 104 (1975);
H.-P. Nilles, M. Srednicki and D. Wyler, Phys. Lett. {\bf B120}, 346 (1983);
J.-M. Fr\`ere, D.R.T. Jones and S. Raby, Nucl. Phys. {\bf B222}, 11 (1983);
J.-P. Derendinger and C.A. Savoy, Nucl. Phys. {\bf B237}, 307 (1984);
L. Durand and J.L. L\'opez, Phys. Lett. {\bf B217}, 463 (1989);
M. Drees, Int. J. Mod. Phys. {\bf A4}, 3645 (1989);
J. Ellis et al., Phys. Rev. {\bf D39},  844 (1989);
S. F. King and P. L. White, Phys. Rev. {\bf D52}, 4183 (1995);
U. Ellwanger, M. Rausch de Traubenberg and C.A. Savoy, 
Nucl. Phys. {\bf B492}, 21 (1997) and references therein.}
\bibitem{Faraggi} See e.g., A.E. Faraggi. Phys.Lett. {\bf B377}, 43(1996).
\bibitem{oneloopmin}{G. Gamberini, G. Ridolfi and F. Zwirner, Nucl. Phys. {\bf
B331}, 331 (1990).} 
\bibitem{KLS} {A. Kusenko, P. Langacker and G. Segr\`e, Phys. Rev. {\bf D54}, 
 5824 (1996); A. Kusenko and P. Langacker, Phys. Lett. {\bf B391},  29 (1997). }
\bibitem{HABSH} H.E. Haber and M. Sher, Phys. Rev. D35 2206 (1987);
M. Drees, Phys. Rev. {\bf D35}, 2910 (1987).
\bibitem{XG} J.R. Espinosa and M. Quir\'os, Phys. Lett. {\bf B279}, 92 (1992)
and Phys.Lett. {\bf B302}, 51 (1993); G. Kane, C. Kolda, J.D. Wells, Phys. Rev.
Lett. {\bf 70}, 2686 (1993); D. Comelli and C. Verzegnassi, Phys. Rev. {\bf D47},
764  (1993) and Phys. Lett. {\bf B303}, 277 (1993). 
\bibitem{radcor}
Y.~Okada, M.~Yamaguchi and T.~Yanagida, Prog. Theor. Phys. Lett.
{\bf 85}, 1  (1991)  and Phys. Lett. {\bf B262}, 54 (1991);
J.~Ellis, G.~Ridolfi and F.~Zwirner, Phys. Lett. {\bf B257}, 83 (1991);
H.E.~Haber and R.~Hempfling, Phys. Rev. Lett. {\bf 66}, 1815 (1991);
R.~Barbieri and M.~Frigeni, Phys. Lett. {\bf B258}, 395 (1991).  
\bibitem{NTL} J. Kamoshita, Y. Okada and M. Tanaka, Phys. Lett. {\bf B328},
67 (1994); D. Comelli and J.R. Espinosa, Phys. Lett. {\bf B388}, 793 (1996).
\bibitem{lep} Special CERN particle physics seminar on physics results from the
LEP run at $\sqrt{s}=172$ GeV (25 Feb. 1997) by the four LEP collaborations; and
G. Alexander et al. (Opal Collaboration), Phys.Lett. {\bf B377}, 181 (1996).
\bibitem{BG} M. Pietroni, Nucl. Phys. {\bf B402}, 27 (1993).
\bibitem{CCB}{ J.A. Casas, A. Lleyda and C. Mu\~noz, Nucl.Phys.
{\bf B471}, 3 (1996) and references therein.}
\end{references}
\end{document}